\documentclass[a4paper,fleqn,usenatbib]{mnras}

\usepackage{newtxtext,newtxmath}
\usepackage[T1]{fontenc}
\usepackage{ae,aecompl}
\usepackage{subfigure}

\usepackage{graphicx}
\usepackage{float}
\usepackage{placeins} 
\usepackage{paralist}

\newcommand{\HI}{H\,{\sc i}}

\newcommand{\kms}{km\,s$^{-1}$}

\newcommand{\Htwo}{H$_{2}$}
\newcommand{\Ts}{$T_{s}$}

\title[Measuring the temperature of the Riegel-Crutcher cloud]{Calibrating the HISA temperature: Measuring the temperature of the Riegel-Crutcher cloud}

\author[H. D\'{e}nes et al.]{
H. D\'{e}nes,$^{1,2}$\thanks{E-mail: Helga.Denes@csiro.au (HD)}
N. M. McClure-Griffiths,$^{2}$
J. M. Dickey,$^{3}$
J. R. Dawson,$^{1,4}$
C. E. Murray,$^{5,6}$
\\
$^{1}$Australia Telescope National Facility, CSIRO Astronomy and Space Science, P.O. Box 76, Epping, NSW 1710, Australia\\
$^{2}$Research School of Astronomy and Astrophysics, Australian National University, Canberra, ACT 2611, Australia\\
$^{3}$University of Tasmania, School of Maths and Physics, Private Bag 37, Hobart, TAS 7001, Australia\\
$^{4}$Department of Physics and Astronomy and MQ Research Centre in Astronomy, Astrophysics and Astrophotonics,\\ Macquarie University, NSW 2109, Australia\\
$^{5}$Department of Astronomy, University of Wisconsin, Madison, 475 N Charter Street, Madison, WI 53706, USA\\
$^{6}$Space Telescope Science Institute, 3700 San Martin Drive, Baltimore, MD 21218\\
}

\date{Accepted XXX. Received YYY; in original form ZZZ}

\pubyear{2016}

\begin{document}
\label{firstpage}
\pagerange{\pageref{firstpage}--\pageref{lastpage}}
\maketitle

\begin{abstract}
	\HI\ self absorption (HISA) clouds are clumps of cold neutral hydrogen (\HI) visible in front of warm background gas, which makes them ideal places to study the properties of the cold atomic component of the interstellar medium (ISM). The Riegel-Crutcher (R-C) cloud is the most striking HISA feature in the Galaxy. It is one of the closest HISA clouds to us and is located in the direction of the Galactic Centre, which provides a bright background. High-resolution interferometric measurements have revealed the filamentary structure of this cloud, however it is difficult to accurately determine the temperature and the density of the gas without optical depth measurements. In this paper we present new \HI\ absorption observations with the Australia Telescope Compact Array (ATCA) against 46 continuum sources behind the Riegel-Crutcher cloud to directly measure the optical depth of the cloud. We decompose the complex \HI\ absorption spectra into Gaussian components using an automated machine learning algorithm. We find 300 Gaussian components, from which 67 are associated with the R-C cloud ($0 < v_{LSR} < 10$ \kms, FWHM $< 10$ \kms). Combining the new \HI\ absorption data with \HI\ emission data from previous surveys we calculate the spin temperature and find it to be between 20 and 80 K. Our measurements uncover a temperature gradient across the cloud with spin temperatures decreasing towards positive Galactic latitudes. We also find three new OH absorption lines associated with the cloud, which support the presence of molecular gas.
\end{abstract}




\section{Introduction}

Theoretical models (e.g. \citealt{Field1969, McKee1977, Wolfire2003}) predict that the neutral hydrogen (\HI) in the interstellar medium (ISM) is in two phases - a warm and a cold phase. The warm neutral medium (WNM) has typical temperatures of $\sim 10^{4}$ K and the cold neutral medium (CNM) has typical temperatures $< 300$ K. These two phases are predicted to be in thermal equilibrium throughout most of the Galaxy (\citealt{McKee1977, Wolfire2003}) and play an important role in regulating star formation. Observations support the two-phase nature of the neutral medium (e.g. \citealt{Garwood1989, Liszt1993, Kulkarni1987}). The WNM is mostly detected in emission and the CNM in absorption. \HI\ emission-absorption studies find that typically 20 - 50 \% of the \HI\ mass is cold and largely missed by emission studies (e.g. \citealt{Dickey1982, Heiles2003_2, Strasser2004,Dickey2009}). Observational constraints on the physical properties are important for understanding how much of the \HI\ is in the cold phase, when it transitions into molecular gas, and which heating and cooling processes influence it. By combining \HI\ emission and absorption measurements we can directly measure the spin temperature ($T_{s}$) and column density of CNM and WNM structures along the line of sight. However, this method provides only very limited information on the spatial structure of the cold gas. 

\HI\ self-absorption (HISA) clouds are unique places where we can study the structure and the physical properties of the CNM at the same time. HISA arises when cold \HI\ gas is located in front of warm \HI. If the two gas components have the same velocity and the cold gas has a lower excitation temperature compared to the background \HI\ brightness temperature, the cold gas appears in absorption against the warm background gas. Large sky surveys, such as the Canadian Galactic Plane Survey (CGPS; \citealt{Taylor2003}) and the Southern Galactic Plane Survey (SGPS; \citealt{McClure-Griffiths2005}), showed that HISA is prevalent throughout the Galaxy and probes the spatial structure of the cold \HI\ (e.g., \citealt{Gibson2000, Gibson2005, Kavars2005}). Several studies suggest that HISA clouds are transitional clouds where either the atomic \HI\ gas is transitioning into \Htwo\ molecular gas via cooling, or the molecular \Htwo\ gas is dissociating into \HI\ (e.g. \citealt{Minter2001, Kavars2003, Kavars2005}). This theory is supported by the wide range of molecular gas associated with HISA clouds. Some HISA clouds have a significant fraction of molecular gas (e.g. \citealt{Baker1979, Liszt1981, Jacq1988}), but others do not have any (e.g. \citealt{Garwood1989, Gibson2005}). \cite{Kavars2005} found that $\sim$60 \% of the HISA clouds in the SGPS have associated CO emission. However, a challenge in interpreting HISA is that variations in the background emission make it difficult to accurately determine the temperature and optical depth of the HISA. A way to overcome this challenge is to measure the \HI\ optical depth directly from \HI\ absorption measurements against bright continuum background sources. By combining the \HI\ emission and absorption measurements we can derive spin temperatures for the cold gas.

To better understand the physical conditions inside cold \HI\ clouds, we mapped the optical depth and spin temperature structure of one of the most prominent HISA clouds, the Riegel-Crutcher (R-C) cloud. The R-C cloud is one of the best studied HISA clouds, located 125 $\pm$ 25 pc from us \citep{Crutcher1984} in the direction of the Galactic Centre (GC) \citep{Crutcher1973}. The \HI\ emission toward the centre of the Galaxy provides an ideal bright background against which to image the CNM. The R-C cloud was discovered by \cite{Heeschen1955} and fully mapped later by \cite{Riegel1969}, who found that the cloud extends $\sim 40^{\circ}$ of Galactic longitude and $\sim 10^{\circ}$ of latitude. It has been suggested that the cloud could be part of a large extended structure at the wall of the Local Bubble, called Lindblad's Feature A \citep{Lindblad1973}. The cloud covers a velocity range of $\sim$10 \kms\ centred at $v_{LSR} = 5$ \kms\ \citep{Riegel1969, Montgomery1995}. Based on background stellar observations, its thickness is estimated to be 1-5 pc \citep{Crutcher1974}. Observations with the Australia Telescope Compact Array (ATCA) revealed a highly filamentary structure aligned with the local magnetic field \citep{McClure-Griffiths2006, Clark2014}. Previous studies also found molecular gas associated with the cloud. \cite{Riegel1969} and \cite{Crutcher1973} observed OH absorption against several lines of sight (LOS) towards the R-C cloud. Overlapping $^{12}$CO emission was detected by \cite{Dame2001} at the positive latitude region of the cloud. In addition to the detected molecular gas, carbon recombination lines (CRRL) at 327 MHz were also observed by \cite{Roshi2011} in the direction of the cloud. \cite{Roshi2011} concluded, that based on their models, the electron temperature of the cloud is between 40-60 K and the CII line cooling is the dominant process in the cloud suggesting that the R-C cloud is in the early stages of molecular formation.

Based on the HISA feature in the \HI\ emission data, the excitation or spin temperature of the R-C cloud was estimated to be $\sim 40$ K (\citealt{Riegel1969, Montgomery1995, McClure-Griffiths2006}) which is consistent with the results of the CRRL lines. However, the spin temperatures derived from the HISA observation alone have a large uncertainty due to the degeneracy of the optical depth and the spin temperature in the radiative transfer equations. To break this degeneracy and to gain a better understanding of the temperature structure of the cloud, we observed \HI\ and OH absorption against 46 continuum sources distributed across the whole R-C cloud. We compare the new absorption data to the \HI\ emission data from the Galactic All Sky Survey (GASS, \citealt{McClure-Griffiths2009, Kalberla2010, Kalberla2015}) and the SGPS Galactic Centre Survey (SGPS GC; \citealt{McClure-Griffiths2012}) to measure the spin temperature and column density distribution across the cloud. 

This paper is structured the following way: in Section \ref{sec:Data} we describe our ATCA observations. In Section ~\ref{sec:Methods} we discuss the radiative transfer equations and the Gaussian decomposition of the \HI\ absorption data. In Section~\ref{sec:Results} we discuss the derived optical depths, spin temperatures and column densities. In Section~\ref{sec:Discussion} we discuss the presence of associated molecular gas and compare our results to \HI\ absorption studies in the literature. Finally, we summarise our results in Section~\ref{sec:Summary}.  

\FloatBarrier
\section{Data}
\label{sec:Data}

In this section we describe our source selection, the ATCA observations and the archival \HI\ emission data that we use in this paper. 

\subsection{HI absorption measurements with ATCA}
\label{sec:ATCA_data}

To measure the optical depth and spin temperature across the R-C cloud we targeted the brightest compact continuum sources in the direction of the HISA cloud. Our aim was an optical depth sensitivity of $\sigma_{\tau} \sim 0.1$, which depends on the flux density of the sources: $\sigma_{\tau} = \sigma_{S}/S$, where $S$ is the background source flux density and $\sigma_{S}$ is the observational sensitivity. To achieve this we selected background continuum sources from the NVSS catalogue \citep{NVSS} based on the following criteria: 
\begin{inparaenum} 
	\item integrated 1.4 GHz flux $\geq$ 200 mJy
	\item and unresolved in NVSS (diameter $\leq$ 20'').
\end{inparaenum}
This selection resulted in 47 target sources.

We obtained high-resolution synthesis \HI\ line observations for the sample with the ATCA. We used a single pointing for each source with the 1.5C antenna configuration, observing each target for ~100 minutes. Observations were carried out in May-June 2015. Details of the observations are given in Tab.~\ref{tab:observations}. The ATCA is a radio-interferometer consisting of six 22 m dishes, creating 15 baselines in a single configuration. While five antennas (CA01-CA05) are reconfigurable along a 3 km long east-west track (and a 214 m long, north-south spur), one antenna (CA06) is fixed at a distance of 3 km from the end of the track creating the longest baselines. 

We observed each source hourly for 10 minutes in a 12 hour observing run, which gave sufficient \textit{uv} coverage for imaging. For each 12 hour observing session we observed the ATCA flux and bandpass calibrator PKS 1934-638 for 30 minutes at the start and at the end of the observation. We used the 1M-0.5k correlator configuration on the Compact Array Broad-band Backend (CABB; \citealt{Wilson2011}) with a 3 MHz wide zoom band\footnote{The 3 MHz wide zoom band consists of 5 concatenated 1 MHz zoom bands, each with 2048 channels overlapped by 50 \% to obtain a flat bandpass.} divided into 6145 channels. This gives a velocity resolution of 0.103 \kms. We used two zoom bands for the \HI\ observations centred at 1417 and at 1420 MHz. The 1417 MHz band was used for bandpass calibration. 

Data reduction was carried out with the {\sc Miriad} software package \citep{Miriad}. We used the standard ATCA primary calibrator PKS 1934-638 for bandpass and amplitude calibration and the brightest continuum source observed each day for phase calibration (NVSS J172920-234535, J174713-192135, J172836-271236, J175233-223012, J175151-252359, J174713-192135, J175114-323538, J174716-191954). The bandpass calibration is not trivial because all potential bandpass calibrators, including PKS 1934-638, show strong \HI\  absorption near $v_{LSR}=0$ \kms\ (See Fig.~\ref{fig:1934-638}), which is close to the velocity of the R-C cloud. We tested two different methods for bandpass calibration: 

\begin{enumerate}
	\item Flagging the \HI\ absorption from the calibrator's data and interpolating across the flagged channels to derive  the  bandpass solution.  The absorption affects $\sim$ 150-200 channels in the middle of the spectra. This method relies on the assumption that the bandpass is flat throughout the zoom band, which is a fair assumption for CABB data.
	\item An alternative option is to use frequency-switching to calibrate the bandpass. In this case we used two zoom bands, one centred at 1420 MHz and second one centred at 1417 MHz, where both bands have the same number of channels. This assumes that the bandpass is the same in the two neighbouring zoom bands. 
\end{enumerate}

Employing both methods we found that the two bandpass solutions were very similar, such that the difference between the solutions is smaller than the measured RMS of the PKS 1934-638 data. However, we did find that after calibration there was a consistently net positive inflection in the bandpass derived via the first method. We therefore chose to proceed with the frequency-switched bandpass solution for all data described below.  

After the calibration we made data cubes for all sources using natural weighting with three different velocity channel samplings: 0.2 \kms, 0.824 \kms\ and 0.826 \kms. The $0.2$ \kms\ resolution was created to have a high velocity resolution cube, while the $0.824$ and $0.826$ \kms\ cubes were created to to match the velocity resolution of the \HI\ emission datasets (described below). We made several different sets of data cubes 
\begin{inparaenum} 
	\item using all 15 baselines, 
	\item excluding baselines with antenna 6 (longest baselines ranging from 3 km to 4.5 km) and 
	\item excluding the shortest baseline (76.5 meter, between antennas 4-5). 
\end{inparaenum} 

Our use of the 1.5 k array ensured that we did not resolve the targeted continuum sources, while simultaneously providing good spatial filtering of the \HI\ emission towards most of the observed sources, because the fluctuations in \HI\ emission at our sensitivity are on large angular scales. However, very close to the GC bright emission fluctuations are also observed on smaller scales. To accurately measure \HI\ absorption we need to filter out any contributions from \HI\ emission. These contribute both positively and negatively to the interferometric spectrum, effectively increasing the measured noise. Excluding the shortest baseline from the data cubes provides additional spatial filtering of the \HI\ emission. We found that we obtained the best signal to noise when we exclude the shortest baseline and all baselines with antenna 6 as well.

\begin{figure}
	\centering
	\includegraphics[width=84mm]{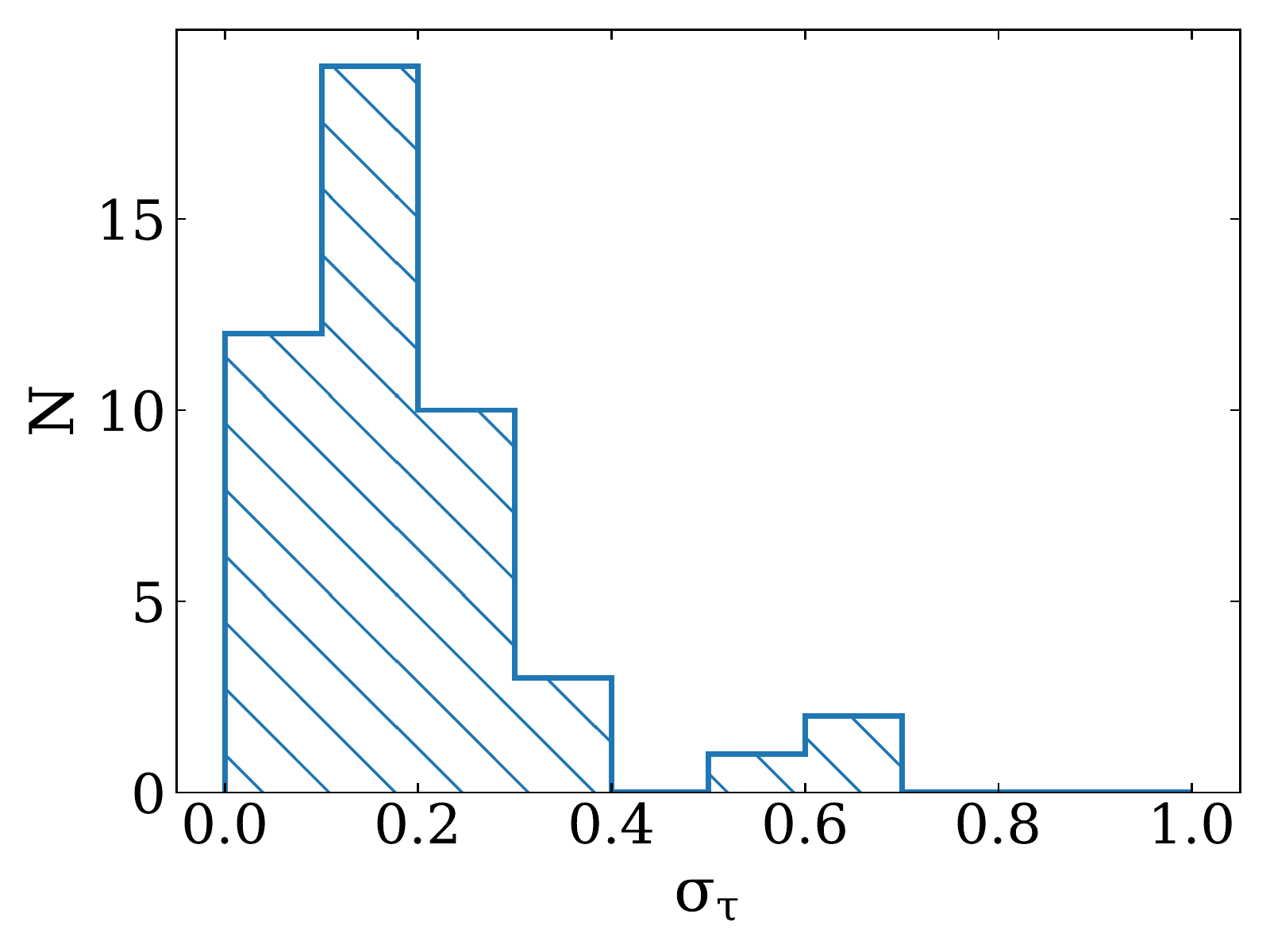}
	\caption{Sensitivity ($\sigma_{\tau}$) histogram of the \HI\ absorption measurements. The majority of our sources (41/47) have optical depth sensitivity $\sigma_{\tau}$ < 0.3.}
	\label{fig:sigma_tau}
\end{figure}

We made continuum images for all of our data cubes and extract the \HI\ spectra at the position of the 1.4 GHz peak flux. From these spectra we derive the \HI\ absorption spectra the following way: 
\begin{equation}
	e^{-\tau} = \frac{T_{B}}{T_{src}},
\end{equation}
where $T_{B}$ is the measured brightness temperature and $T_{src}$ is the brightness temperature of the continuum source. We use a linear fit to the line free channels to measure $T_{src}$. We find that despite the source selection, we resolve or partially resolve 9 sources. This means that some of the sources have a poorer optical depth sensitivity than planned. Fig.~\ref{fig:sigma_tau} shows the optical depth sensitivity of our sample. The majority of the sources (41/47) have optical depth sensitivity $\sigma_{\tau}$ < 0.3. In addition to the resolved sources, the fainter sources toward the GC have a high RMS and low optical depth sensitivity. Although we observed 47 sources, we excluded NVSS J174007-284203, the closest source to the GC, because of a high noise level.  In the remainder of this paper we discuss data towards 46 continuum sources. 

\begin{table*}
	\centering
	\caption{Summary of ATCA observations. $S_{1.4}$ are 1.4 GHz fluxes from the NVSS catalogue \citep{NVSS}.}
	\label{tab:observations}
	\begin{tabular}{l c c c c c c c c}
		\hline
		NVSS name& R.A. & Decl. &  l & b & $S_{1.4}$ & Time & Synthesized Beam & $\sigma_{\tau}$   \\
		& [hh:mm:ss] & [dd:mm:ss] & [$^\circ$] & [$^\circ$] & [mJy] & [hours] & [arcsec$^{2}$] & \\
		\hline
		NVSS J172829-284610   & 17:28:28 &-28:45:50 & -1.8738 & 3.2555 & 585.3 & 1.81 & 51 $\times$ 22 & 0.17 \\
		NVSS J172836-271236   & 17:28:36 &-27:12:36 & -0.5563 & 4.0941 & 530.4 & 1.63 & 60 $\times$ 21 & 0.07 \\
		NVSS J172908-265751   & 17:29:08 &-26:57:51 & -0.2845 & 4.1296 & 364.1 & 1.64 & 60 $\times$ 21 & 0.21 \\
		NVSS J172920-234535   & 17:29:20 &-23:45:35 & 2.432 & 5.8467 & 1825 & 1.81 & 61 $\times$ 22 & 0.04 \\
		NVSS J173107-245703   & 17:31:07 &-24:57:03 & 1.6506 & 4.8577 & 284.8 & 1.81 & 58 $\times$ 22 & 0.14 \\
		NVSS J173133-264015   & 17:31:33 &-26:40:15 & 0.2574 & 3.8392 & 236.1 & 1.69 & 59 $\times$ 21 & 0.18 \\
		NVSS J173203-285516   & 17:32:03 &-28:55:16 & -1.5699 & 2.5169 & 213.4 & 1.81 & 51 $\times$ 22 & 0.3 \\
		NVSS J173205-242651   & 17:32:05 &-24:26:51 & 2.1956 & 4.945 & 597.2 & 1.64 & 66 $\times$ 21 & 0.08 \\
		NVSS J173252-223511   & 17:32:52 &-22:35:21 & 3.8664 & 5.7991 & 353.2 & 1.48 & 80 $\times$ 20 & 0.11 \\
		NVSS J173524-251036   & 17:35:24 &-25:10:26 & 1.9877 & 3.9175 & 258.6 & 1.64 & 64 $\times$ 21 & 0.21 \\
		NVSS J173620-283552   & 17:36:20 &-28:35:42 & -0.7887 & 1.9026 & 210.8 & 1.64 & 74 $\times$ 27 & 0.58 \\
		NVSS J173713-224734   & 17:37:13 &-22:47:34 & 4.2289 & 4.8441 & 209.9 & 1.65 & 70 $\times$ 21& 0.19 \\
		NVSS J173718-260426   & 17:37:18 &-26:04:26 & 1.4579 & 3.0762 & 283.1 & 1.81 & 56 $\times$ 22 & 0.14 \\
		NVSS J173722-223000   & 17:37:22 &-22:30:00 & 4.498 & 4.968 & 317.7 & 1.48 & 80 $\times$ 19 & 0.14 \\
		NVSS J173753-254642   & 17:37:53 &-25:46:42 & 1.7778 & 3.1242 & 217.2 & 1.81 & 56 $\times$ 22 & 0.24 \\
		NVSS J173806-262443   & 17:38:06 &-26:24:43 & 1.2669 & 2.7455 & 309.7 & 1.81 & 55 $\times$ 22 & 0.14 \\
		NVSS J173811-204411   & 17:38:13 &-20:42:41 & 6.1012 & 5.7415 & 342.4 & 1.64 & 77 $\times$ 21 & 0.12 \\
		NVSS J173850-221918   & 17:38:50 &-22:19:18 & 4.8301 & 4.7751 & 244.9 & 1.65 & 72 $\times$ 21&  0.19 \\
		NVSS J173939-205505   & 17:39:41 &-20:53:35 & 6.1259 & 5.3569 & 544.1 & 1.64 & 77 $\times$ 21 & 0.07 \\
		NVSS J174007-284203   & 17:40:07 &-28:42:03 & -0.4324 & 1.1463 & 219.1 & 1.81 & 67 $\times$ 30 & 0.74 \\
		NVSS J174202-271311   & 17:42:02 &-27:13:11 & 1.0472 & 1.5731 & 204.2 & 1.64 & 77 $\times$ 27 & 0.61 \\
		NVSS J174224-203729   & 17:42:26 &-20:35:59 & 6.7168 & 4.9604 & 434.5 & 1.64 & 77 $\times$ 21 & 0.1 \\
		NVSS J174317-305819   & 17:43:18 &-30:58:29 & -1.9982 & -0.6363 & 357.5 & 1.81 & 48 $\times$ 22 & 0.13 \\
		NVSS J174343-182838   & 17:43:45 &-18:26:58 & 8.7203 & 5.8132 & 568.3 & 1.64 & 87 $\times$ 21 & 0.06 \\
		NVSS J174351-261059   & 17:43:51 &-26:10:59 & 2.1435 & 1.7718 & 229.6 & 1.64 & 80 $\times$ 27 & 0.41 \\
		NVSS J174423-311636   & 17:44:24 &-31:16:46 & -2.135 & -0.9956 & 424.6 & 1.81 & 47 $\times$ 22 & 0.1 \\
		NVSS J174513-315104   & 17:45:14 &-31:51:04 & -2.533 & -1.446 & 406.6 & 1.81 &  46 $\times$ 22& 0.39 \\
		NVSS J174618-193006   & 17:46:18 &-19:30:16 & 8.1553 & 4.7619 & 300.2 & 1.64 & 82 $\times$ 21 & 0.2 \\
		NVSS J174637-182629   & 17:46:39 &-18:24:49 & 9.1081 & 5.2404 & 268.7 & 1.63 & 87 $\times$ 21 & 0.15 \\
		NVSS J174709-295802   & 17:47:12 &-29:58:12 & -0.7073 & -0.8216 & 269.9 & 1.81 & 63 $\times$ 30 & 0.3 \\
		NVSS J174713-192135   & 17:47:13 &-19:21:35 & 8.3906 & 4.6467 & 2276.5 & 1.64 & 83 $\times$ 21 & 0.02 \\
		NVSS J174716-191954   & 17:47:16 &-19:19:54 & 8.4192 & 4.6536 & 1286.2 & 1.7 & 79 $\times$ 21 & 0.02 \\
		NVSS J174748-312315   & 17:47:50 &-31:23:15 & -1.8502 & -1.6766 & 540.7 & 1.93 & 44 $\times$ 24 & 0.19 \\
		NVSS J174831-324102   & 17:48:32 &-32:41:22 & -2.8834 & -2.4753 & 398 & 1.79 & 45 $\times$ 22 & 0.24 \\
		NVSS J174832-225211   & 17:48:31 &-22:52:21 & 5.5312 & 2.5801 & 282.8 & 1.65 & 71 $\times$ 21 & 0.42 \\
		NVSS J174915-200033   & 17:49:15 &-20:00:33 & 8.0755 & 3.9058 & 344.2 & 1.64 & 80 $\times$ 21 & 0.29 \\
		NVSS J174931-210847   & 17:49:31 &-21:08:37 & 7.1301 & 3.2676 & 355.4 & 1.64 & 76 $\times$ 21 & 0.27 \\
		NVSS J175104-235215   & 17:51:04 &-23:52:05 & 4.9686 & 1.5693 & 267.2 & 1.58 & 57 $\times$ 24 & 0.24 \\
		NVSS J175114-323538   & 17:51:14 &-32:35:38 & -2.514 & -2.9188 & 646.8 & 1.83 & 45 $\times$ 23 & 0.1 \\
		NVSS J175151-252359   & 17:51:51 &-25:23:59 & 3.745 & 0.6349 & 1223.6 & 1.67 & 57 $\times$ 22 & 0.04 \\
		NVSS J175157-240425   & 17:51:57 &-24:04:25 & 4.8976 & 1.2915 & 220.9 & 1.64 & 54 $\times$ 25 & 0.29 \\
		NVSS J175218-210508   & 17:52:18 &-21:04:56 & 7.513 & 2.7416 & 242.5 & 1.8 & 69 $\times$ 22 & 0.2 \\
		NVSS J175233-223012   & 17:52:33 &-22:30:12 & 6.32 & 1.9719 & 1936.5 & 1.78 & 64 $\times$ 22&  0.02 \\
		NVSS J175427-235235   & 17:54:27 &-23:52:25 & 5.358 & 0.8988 & 267.7 & 1.48 &  58 $\times$ 23& 0.27 \\
		NVSS J175526-223211   & 17:55:26 &-22:32:11 & 6.6287 & 1.3809 & 225.6 & 1.81 & 64 $\times$ 22 & 0.1 \\
		NVSS J175548-233322   & 17:55:48 &-23:33:22 & 5.7906 & 0.7936 & 439.8 & 1.53 & 60 $\times$ 23 & 0.14 \\
		NVSS J175727-223901   & 17:57:26 &-22:39:01 & 6.7646 & 0.9197 & 541.5 & 1.64 & 66 $\times$ 21.65 & 0.29 \\
		\hline
	\end{tabular}
\end{table*}

\subsection{HI emission data}

We use \HI\ emission data from two sources: the GASS \citep{McClure-Griffiths2009, Kalberla2010, Kalberla2015} and from the SGPS GC; \citep{McClure-Griffiths2012}. GASS maps the Galactic \HI\ emission across the whole Southern sky ($\delta \leq 1^{\circ}$), including the whole area of the R-C cloud around the GC. GASS has a spatial resolution of ~16', a velocity resolution of 0.826 \kms\ and an RMS brightness temperature noise of 57 mK. 

The SGPS GC data cover 100 deg$^{2}$ around the GC ($-5^{\circ} \leq l \leq 5^{\circ}, -5^{\circ} \leq b \leq 5^{\circ}$) and encompass most of the R-C cloud. The SGPS GC \HI\ data cubes are constructed from interferometric data observed with the ATCA in several different array configurations, combined with Parkes single dish data (for details see \citealt{McClure-Griffiths2012}). These data have higher spatial resolution, $\theta = 145^{\prime\prime}$, but worse RMS brightness temperature sensitivity, $\sim$2 K, compared to GASS. The velocity resolution is 0.824 \kms. 

We extract spectra from the GASS cubes at the same position as from the ATCA data cubes. For the SGPS GC data cube we average spectra from four positions 145'' away from the ATCA positions. This is to avoid contamination of the emission spectra with remaining absorption toward the continuum source. When comparing the 4 point average spectra and the spectra at the continuum sources position we find that there is only a small, 3 - 8 K (3 - 6 \% of the peak flux) difference between the two. 

\subsection{OH data}

In addition to the \HI\ observation we observed all continuum sources in the two main OH lines (1665 and 1667 MHz) with the ATCA. We centred the two OH zoom bands on 1665 and 1667 MHz, with a bandwidth of 2 MHz divided into 4097 channels, \footnote{The 2 MHz wide zoom band consists of 3 concatenated 1 MHz zoom bands, each with 2048 channels overlapped by 50 \% to obtain a flat bandpass.} which gives a velocity resolution of 0.088 \kms. 

We reduced the OH data in the same way as the \HI\, using PKS 1934-638 for bandpass and amplitude calibration and the brightest continuum source observed each day for phase calibration (NVSS J172920-234535, J174713-192135, J172836-271236, J175233-223012, J175151-252359, J174713-192135, J175114-323538, J174716-191954). After the calibration we made OH data cubes for all sources using natural weighting and 0.26 \kms\ velocity channels. We extracted OH spectra for each source at the same position as the \HI\ spectra.

\section{Methods}
\label{sec:Methods}

In this section we describe our methodology for calculating spin temperatures, our decomposition of the \HI\ absorption spectra into Gaussian components, and our method to reconstruct the \HI\ emission without the HISA.  

\subsection{Deriving the spin temperature of the cold \HI}

\begin{figure*}
	\centering
	\subfigure{\includegraphics[width=84mm]{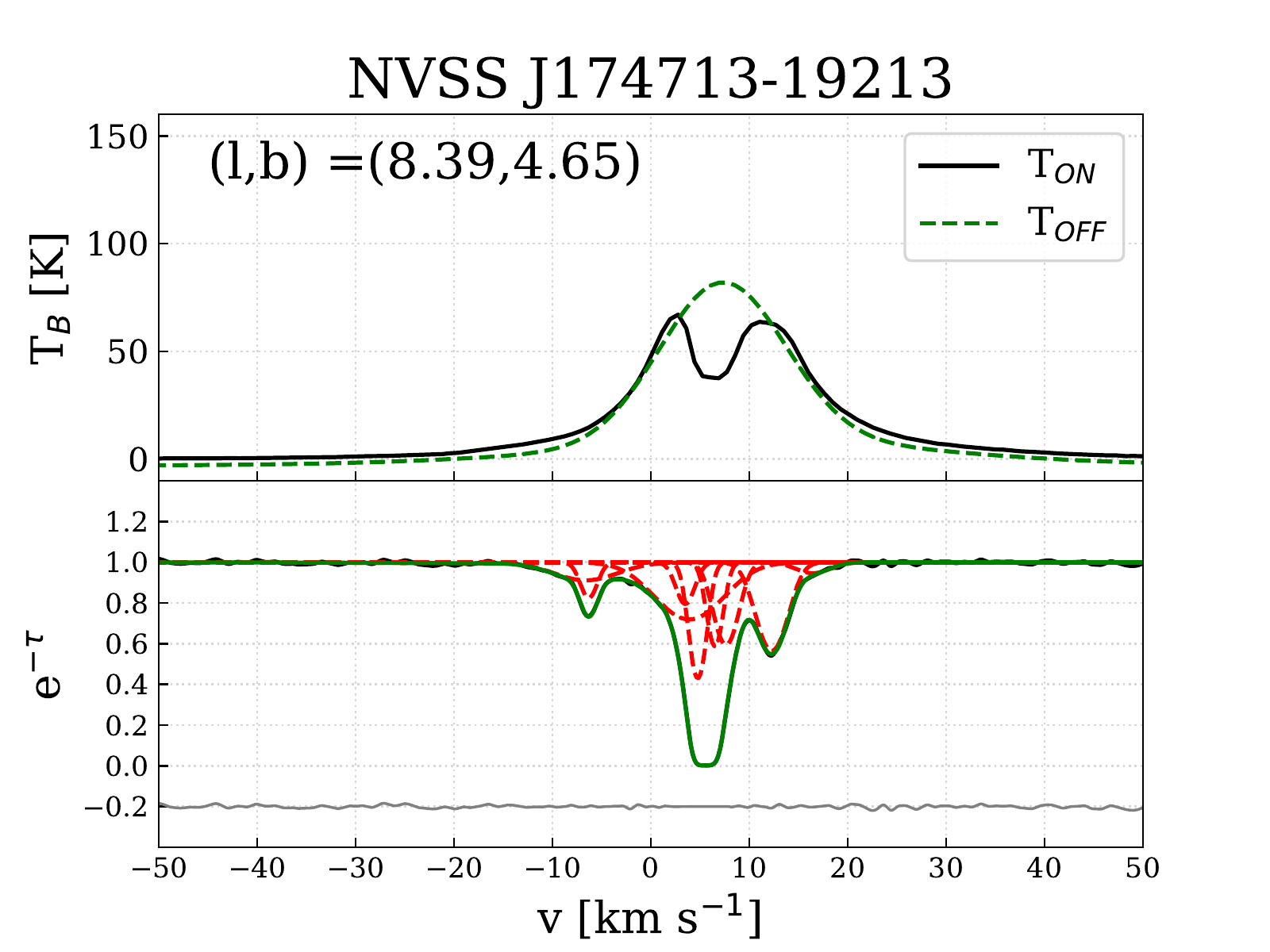}}
	\hfill
	\subfigure{\includegraphics[width=84mm]{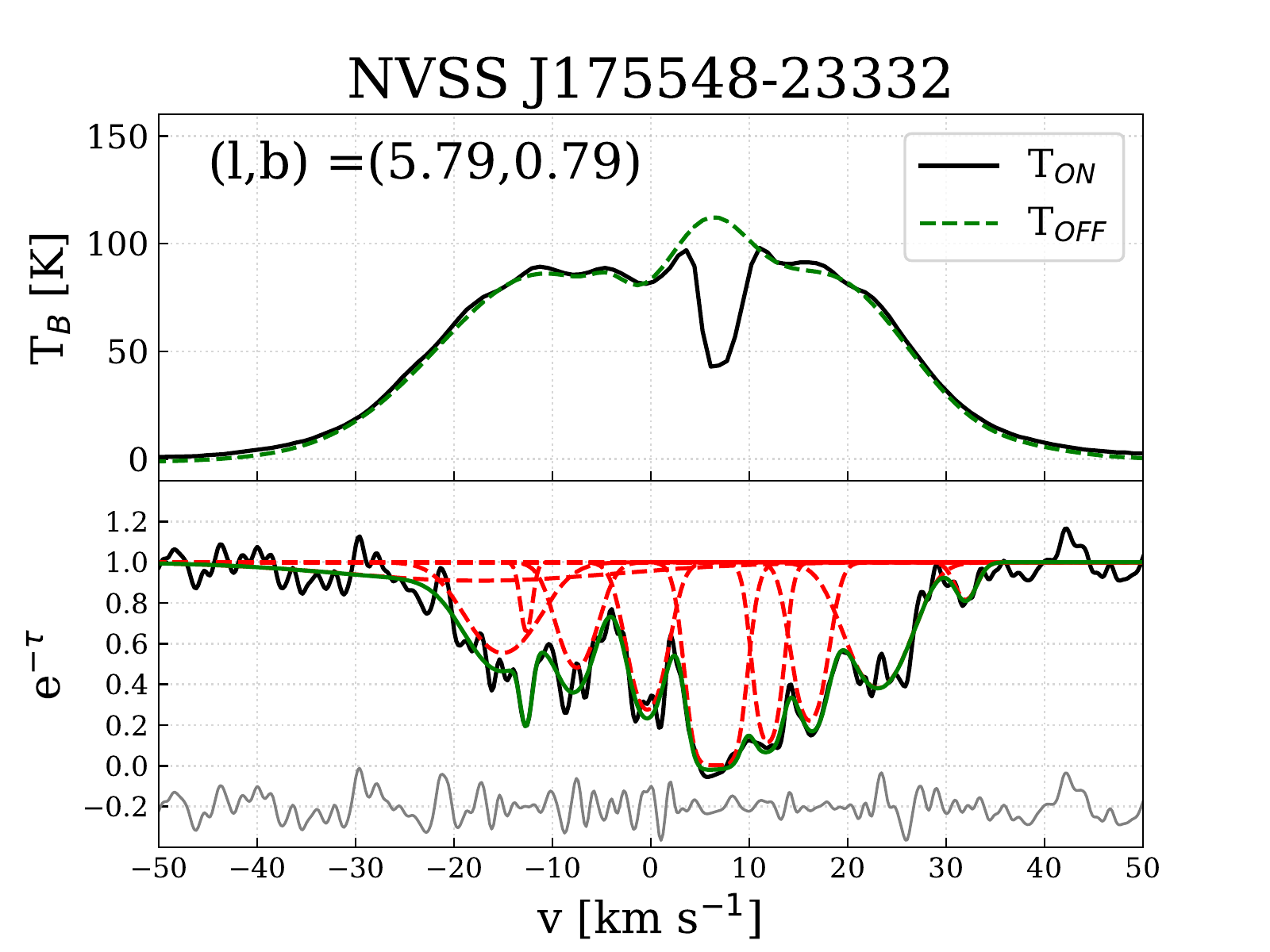}}
	\hfill
	\caption{Example \HI\ emission and absorption spectra of sources NVSS J174713-192135 ($l,b$ = 8.39, 4.64) and NVSS J175548-233322 ($l,b$ = 5.78, 0.79), with $\sigma_{\tau}$ 0.02 and 0.14 respectively. Top panel: the solid black line shows the emission spectra ($T_{ON}$) extracted from GASS and the dashed line shows the modelled $T_{OFF}$ spectra. Bottom panel: the black line shows the ATCA absorption spectra, the green line is the fitted Gaussian model and the red dashed lines are the individual components of the fit. The grey line is the residual from the absorption model shifted to -0.2.}
	\label{fig:example_spectra}
\end{figure*}

There are several methods that are used to calculate spin temperatures for the cold \HI\ gas. Each of these methods have different assumptions, advantages and disadvantages, and are generally suitable for different science goals. For a detailed summary of the various methods of deriving \Ts\ see \cite{Dickey2003} and \cite{Murray2015}. In this work we are investigating the spin temperature of a HISA cloud that is seen towards the GC and has a large angular extent on the sky (more than 100 square degrees). This means that the \HI\ spectra are a complex mixture of various, overlapping emission components and at least one deep self-absorption feature. The \HI\ absorption spectra measured with the ATCA are composed of several strong, blended absorption lines. Fig.~\ref{fig:example_spectra} shows two typical emission and absorption spectra pairs towards the R-C cloud (NVSS J174713-192135 and NVSS J175548-233322). The strong HISA feature at $\sim$ 5 \kms\ is evident in both emission spectra. Fig.~\ref{fig:example_spectra} also illustrates that several of the absorption lines from the ATCA data have corresponding emission features in the GASS data, but it is only the absorption lines at $\sim 5$ \kms\ which appear as HISA. 

\begin{figure}
	\centering
	\includegraphics[width=84mm]{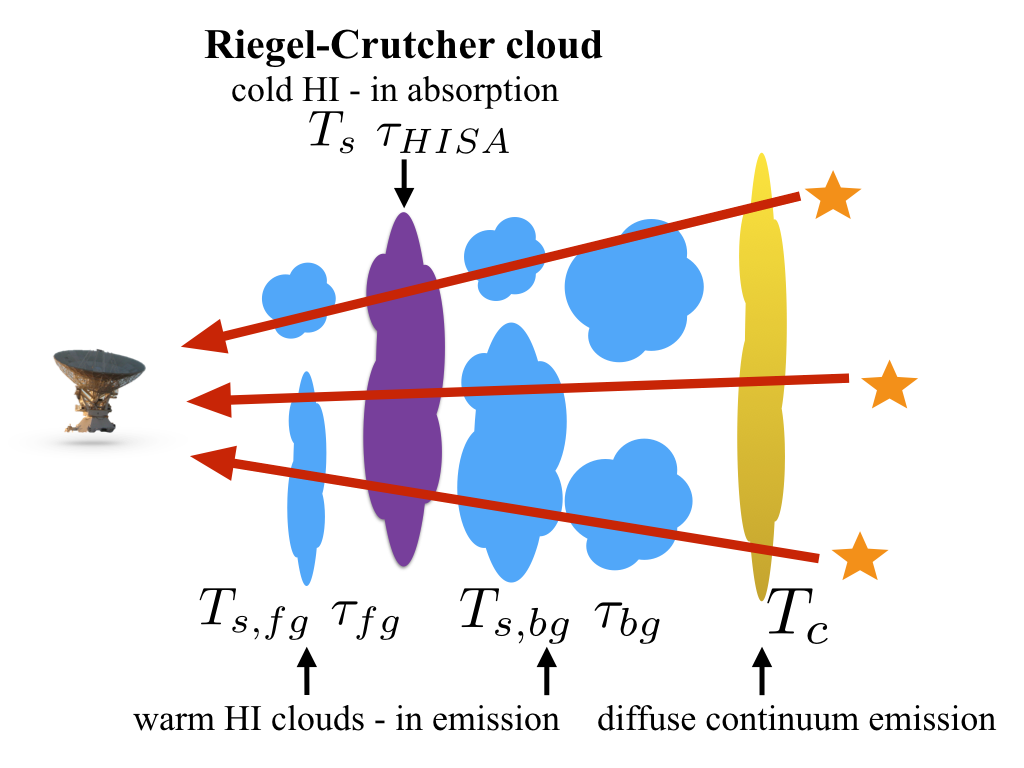}
	\caption{Illustration of the Riegel-Crutcher cloud observations. We are observing the cold gas from the RC cloud in HI absorption and we are observing the background and foreground warmer HI clouds in emission. }
	\label{fig:RC-cloud-illustration}
\end{figure}

Based on these spectral properties and the requirement that we are specifically interested in the HISA cloud, we chose to calculate the spin temperatures with the method presented in \cite{Gibson2000} and \cite{Kavars2003}. This model assumes a four component ISM model to solve the radiative transfer equation and derives the spin temperature as a function of optical depth. The four component model consists of warm \HI\ in the foreground and background of the HISA cloud, cold \HI\ within the HISA cloud, and diffuse continuum emission in the background (see Fig.~\ref{fig:RC-cloud-illustration} for a schematic diagram). This model describes the observed brightness temperature ($T_{B}$) in the following way:
\begin{equation}
	\begin{split}
		T_{B} = T_{s,fg} (1-e^{-\tau_{fg} }) + T_{s} (1-e^{-\tau_{HISA}})e^{-\tau_{fg}} \\
		+ T_{s,bg} (1-e^{-\tau_{bg}})e^{-(\tau_{fg} +\tau_{HISA})} + T_{c}e^{-(\tau_{fg} +\tau_{HISA} +\tau_{bg})},
	\end{split}
\end{equation}
where \Ts\ and $\tau_{HISA}$ are the temperature and optical depth of the HISA components, $T_{s,fg}$, $T_{s,bg}$, $\tau_{fg}$ and $\tau_{bg}$ are the temperature and optical depth of the foreground and background emission and $T_{c}$ is the brightness temperature of the diffuse continuum background. Except for $T_{c}$, all of the above variables are a function of velocity. For simplicity we omit $(v)$ from the equations. We also assume that all the diffuse continuum radiation is in the background. 

From this ISM model we can expresses the HISA component as:
\begin{equation}
	T_{ON} - T_{OFF}= (T_{s} - pT_{OFF} - T_{C} e^{-\tau_{bg}}) e^{-\tau_{fg}}(1-e^{-\tau_{HISA}}),
	\label{eq:HISA}
\end{equation}
where $T_{ON}$ is the \HI\ emission spectra with the self-absorption ($T_{B}$) and $T_{OFF}$ is the emission spectra without self-absorption (see Fig~\ref{fig:example_spectra}). The  fraction of background to foreground emission is $p = T_{s,bg}(1-e^{-\tau_{bg}}) / T_{OFF}$. Because the R-C cloud is located at the wall of the Local Bubble we can assume that a relatively small fraction of the warm \HI\ emitting gas is in the foreground and that most of it is in the background. This is also supported by the 3D gas and dust maps of the local environment around the Sun (e.g. \citealt{Vergely2010, Lallement2014}). We assume that no more than 10 \% of the \HI\ emission is in the foreground of the R-C cloud and use p = 0.9 for all \Ts\ calculations in this paper. Varying p between 1 and 0.9 decreases the average spin temperature of the cloud by $\sim$ 10 K. This corresponds to a 10 K uncertainty for all our derived temperatures.  

Because of the large angular extent of the R-C cloud we cannot estimate $T_{OFF}$ from nearby spectra, hence we need to interpolate over the HISA feature in the emission spectra. To estimate $T_{OFF}$ we reconstruct the emission spectra by fitting multiple-component Gaussian profiles (see Section~\ref{interpollating_GASS}). For the diffuse continuum emission ($T_{c}$) we use values extracted from the CHIPASS ZOA map\footnote{http://www.atnf.csiro.au/research/CHIPASS/} \citealt{Calabretta2014}. CHIPASS is a 1.4 GHz continuum map of the Southern sky ($\delta < +25^{\circ}$) based on data observed for the \HI\ Parkes All-Sky Survey (HIPASS) and the \HI\ Zone of Avoidance (HIZOA) survey. For our sample the range of $T_{c}$ is between a few K and 22 K (Fig.~\ref{fig:Tc_histogram}). For simplicity we assume that $\tau_{fg}$ and $\tau_{bg}$ are negligible, corresponding to warm, optically thin \HI. We note that the background of the R-C cloud is very complex and $\tau_{bg}$ may not be negligible in this region. If $\tau_{bg}$ varies between 0.1 and 1, then \Ts\ can change by 0.5 - 10 K. We discuss the uncertainty associated with $\tau_{bg}$ in section~\ref{sec:uncertainties} in more detail. After these assumptions, the spin temperature can be expressed as:

\begin{equation}
	T_{s} = \frac{ T_{ON} - T_{OFF}}{1-e^{-\tau_{HISA}}} + T_{c} + pT_{OFF}. 
	\label{eq:Ts}
\end{equation}
It is clear from this that \Ts\ can only be derived as a function of $\tau_{HISA}$. This degeneracy limits studies of HISA unless extra information about either \Ts\ or $\tau_{HISA}$ can be employed. Often authors have made assumptions about \Ts\ based on the width of the \HI\ absorption line, which gives a crude limit on \Ts\ (e.g. \citealt{McClure-Griffiths2006}). In this paper we break the degeneracy by directly measuring $\tau_{HISA}$ from the \HI\ absorption spectra toward continuum sources measured with the ATCA. With this additional step we can effectively ``calibrate'' the HISA spin temperature with the \HI\ continuum absorption.  
We present our derived spin temperatures in Section~\ref{par:Ts}.

\subsection{Gaussian decomposition}
\newcommand{\vc}[3]{\overset{#2}{\underset{#3}{#1}}}

The ATCA \HI\ absorption spectrum ($e^{-\tau}$) is a composition of several \HI\ clouds ($\tau_{i}$) along the LOS from which the HISA ($\tau_{HISA}$) is only one or two components. The $\tau$ spectra can be described as the sum of Gaussian components (e.g. \citealt{Heiles2003_1, Murray2017}). To decompose the absorption spectra into Gaussian components we use the Autonomous Gaussian Decomposition algorithm (AGD; {\sc GaussPy}) from \cite{Lindner2015}. AGD uses computer vision and machine learning algorithms to decompose complicated spectra and provide a multi-component Gaussian model. The algorithm uses a two step approach by first providing initial guesses for the Gaussian parameters and then performing a least squares fit of the model to the data. The initial guesses for the number, location and width of the components are estimated from the derivatives of the spectra with respect to velocity. The location is determined from the maxima in the second and fourth derivatives and the width is determined from the value of the second derivative at the location. AGD is a flexible algorithm that can be tuned to the properties of different data sets, e.g. the signal to noise of the data set and the number and expected width of the Gaussian components. Derivative spectroscopy is frequently used to study spectral line properties (e.g. \citealt{Krco2008, Krco2010}).

AGD can be fine-tuned primarily through the use of the $\alpha$ parameters, which control the balance between smoothness and data fidelity in the solution by suppressing the noise when decomposing the spectra. Setting these parameters to an optimal value is essential to avoid classifying noise as a Gaussian component and yet also finding all the weak signal components. The ideal $\alpha$ parameters vary between different data sets, depending on the noise level of the spectra, the width of the components and on how blended the components are. The best $\alpha$ values can be determined with machine learning, by ``training'' AGD with a representative data set, where we know the underlying distribution of Gaussian components. 

We created a set of artificial absorption spectra with the following parameters: 800 spectra, with 3-15 Gaussian components each, a varying RMS level (0.01-0.32), amplitude between the RMS level and 1, and FWHM randomly distributed between 4 to 50 channels (for a channel width of 0.2 \kms\ this corresponds to lines with FWHM 0.8 - 10 \kms). These parameters are representative of our ATCA absorption spectra. We use these spectra to determine the best $\alpha_{1}$ and $\alpha_{2}$ values to decompose our spectra with AGD (for details see \citealt{Lindner2015}). We found the following values: $\alpha_{1} = 2.5$ and $\alpha_{2} = 5.97$ using a signal to noise threshold of 5.   

We note that because our data have a higher noise level than 21-SPONGE for which AGD was developed, and because several of our LOS have saturated \HI\ lines, we have to employ a two-step ``hybrid'' method to decompose the absorption spectrum more accurately.
\begin{inparaenum} 
	\item We decompose the $e^{-\tau}$ spectra into Gaussian components with \textsc{GaussPy},
	\item then we use the results of the decomposition as an initial guess for $\tau_{i}$ and fit the following function to the \HI\ absorption spectrum:
\end{inparaenum}	
\begin{equation}
	f(v) = e^{-\vc{\Sigma}{n}{i=1}\tau_{i}(v)}.
	\label{eq:absorption_fit}
\end{equation} 
Here $n$ is the number of clouds in the LOS. Converting the $e^{-\tau}$ spectra into $\tau$ amplifies the noise substantially and data points where $e^{-\tau} < 0 $ (due to the noise) cannot be converted. Using this hybrid fitting method allows us to avoid these issues and adequately fit saturated absorption lines in the presence of noise.

The bottom panels of Fig.~\ref{fig:example_spectra} show two examples of the decomposition with AGD. NVSS J174713-192135 is one of the brightest sources in our sample with $\sigma_{\tau}$  = 0.02 and NVSS J175548-233322 is an average source with $\sigma_{\tau}$  = 0.14. The black line shows the observed spectra (Gaussian smoothed), the green line is the fitted model and the red dashed lines are the individual $e^{-\tau(v)}$ components. We also show the GASS \HI\ emission spectra and the interpolated ($T_{OFF}$) spectra in the upper panel of Fig.~\ref{fig:example_spectra}. All decomposed spectra are presented in Appendix~\ref{Appendix:HIspectra}.

\subsection{Interpolating the \HI\ emission data }
\label{interpollating_GASS}

The strong HISA feature and the complicated structure of the \HI\ emission towards the GC makes it difficult to determine the ``off-source'' brightness temperature ($T_{OFF}$). To overcome this, we fitted the GASS and the SGPS GC spectra with a combination of negative and positive Gaussian components, which we identify with AGD. 

\begin{figure}
	\centering
	\includegraphics[width=8.4cm]{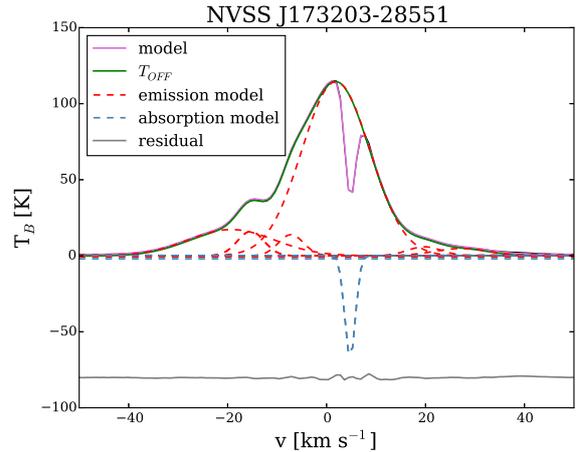}
	\caption{An example GASS spectra towards NVSS J173203-28551 (black). The modelled spectra is shown with a pink line. The interpolated $T_{OFF}$ is shown with a solid green line and the individual Gaussian components of our model are shown with dashed lines, red for emission and blue for absorption. The residual between the GASS spectra and our model is shown in grey.} 
	\label{fig:HI_emission}
\end{figure}

We trained the algorithm to fit the \HI\ emission data of GASS and SGPS. For the training we constructed spectra with the following parameters: 1000 spectra, with 5-15 Gaussian components each, a varying RMS level, amplitude between 100-1000 times the RMS level and FWHM randomly distributed between 4 to 50 channels. The resulting $\alpha_{1}$ and $\alpha_{2}$ parameters are 9.01 and 5.39. We determine the position and width of the HISA feature based on the second derivative of the spectra and mask the absorption. Then we fit the emission and the HISA feature separately with \textsc{GaussPy} and use the fitted components as the initial guess parameters for a composite model, which we fit to the GASS and SGPS GC spectra. We extract the positive emission components from the final fitted model and combine them into $T_{OFF}$. Fig.~\ref{fig:HI_emission} illustrates this method with the \HI\ emission spectrum towards NVSS J173203-285516, where the pink line shows the composite model, the red and blue dashed lines show the individual emission and absorption components, and the green line is the reconstructed $T_{OFF}$ spectrum. The observed spectrum is under the pink line in black and the residuals of the fit are shown in grey.    

This method gives a reasonable $T_{OFF}$ model for most lines-of-sight. We estimate the uncertainty of the fitted model by comparing the maximum of the emission spectra and the maximum of the $T_{OFF}$ model. This is a conservative approach that gives $T_{OFF}$ errors on the order of 10 K. The average $T_{OFF}$ uncertainty for GASS data is 8 K and 13 K for SGPS GC. 

\section{Results}
\label{sec:Results}

\subsection{Optical depth}

Inspecting the \HI-absorption spectra we find that all LOS have multiple blended absorption components (See Fig.~\ref{fig:example_spectra} and Appendix~\ref{Appendix:HIspectra}). To disentangle which absorption components belong to the R-C cloud, we decompose the absorption spectra using our hybrid fitting method. We find between 1-12 components for the individual LOS, with an average of 6 components. Altogether we find 300 Gaussian components for 46  LOS. From these 85 have $\tau > 1$. 

We select the components associated with the R-C cloud based on the following criteria: $2 < v_{LSR} < 10$ \kms\ and FWHM $< 10$ \kms, assuming that all the components of the cloud are relatively cold with narrow line widths. Based on this, 67 Gaussian components are associated with the R-C cloud. Of these 53 have $\tau_{HISA} > 0.5$ and 43 have $\tau_{HISA} > 1$. In Fig.~\ref{fig:example_spectra} we show two typical absorption spectra, with the fitted model (green) and the individual fitted components (red). For comparison we also plot the \HI\ emission spectra from GASS with the $T_{OFF}$ model (green dashed line) and the individual Gaussian components (red dashed lines). We present all of the spectra with the fitted models in Appendix~\ref{Appendix:HIspectra}. 

\begin{figure*}
	\centering
	\includegraphics[width=160mm]{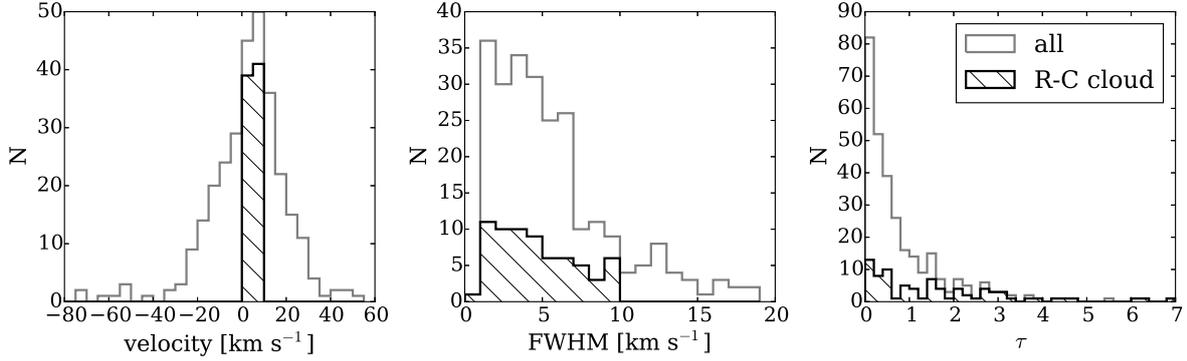}
	\caption{Parameters of the fitted Gaussian components to the \HI\ absorption spectra: velocity of the components, derived optical depth and FWHM. All components of the sample are shown in the grey histogram and the R-C cloud components are highlighted in the black hatched histogram.}
	\label{fig:RC-histograms}
\end{figure*}

We present the Gaussian parameters of the decomposition in Tab.~\ref{tab:parameters} and in Fig.~\ref{fig:RC-histograms}. We note that Tab.~\ref{tab:parameters} only contains the Gaussian parameters of components associated with the R-C cloud. Some of our low signal-to-noise spectra do not have components matching the criteria and so not all sources from Tab.~\ref{tab:observations} have components in Tab.~\ref{tab:parameters}. Fig.~\ref{fig:RC-histograms} shows the distribution of all Gaussian parameters - position, optical depth, FWHM - of the whole sample in grey and the line parameters associated with the R-C cloud ($0 < v_{LSR} < 10$ \kms, FWHM $< 10$ \kms) in black hatched.  

\begin{figure*}
	\centering
	\subfigure{\includegraphics[width=84mm]{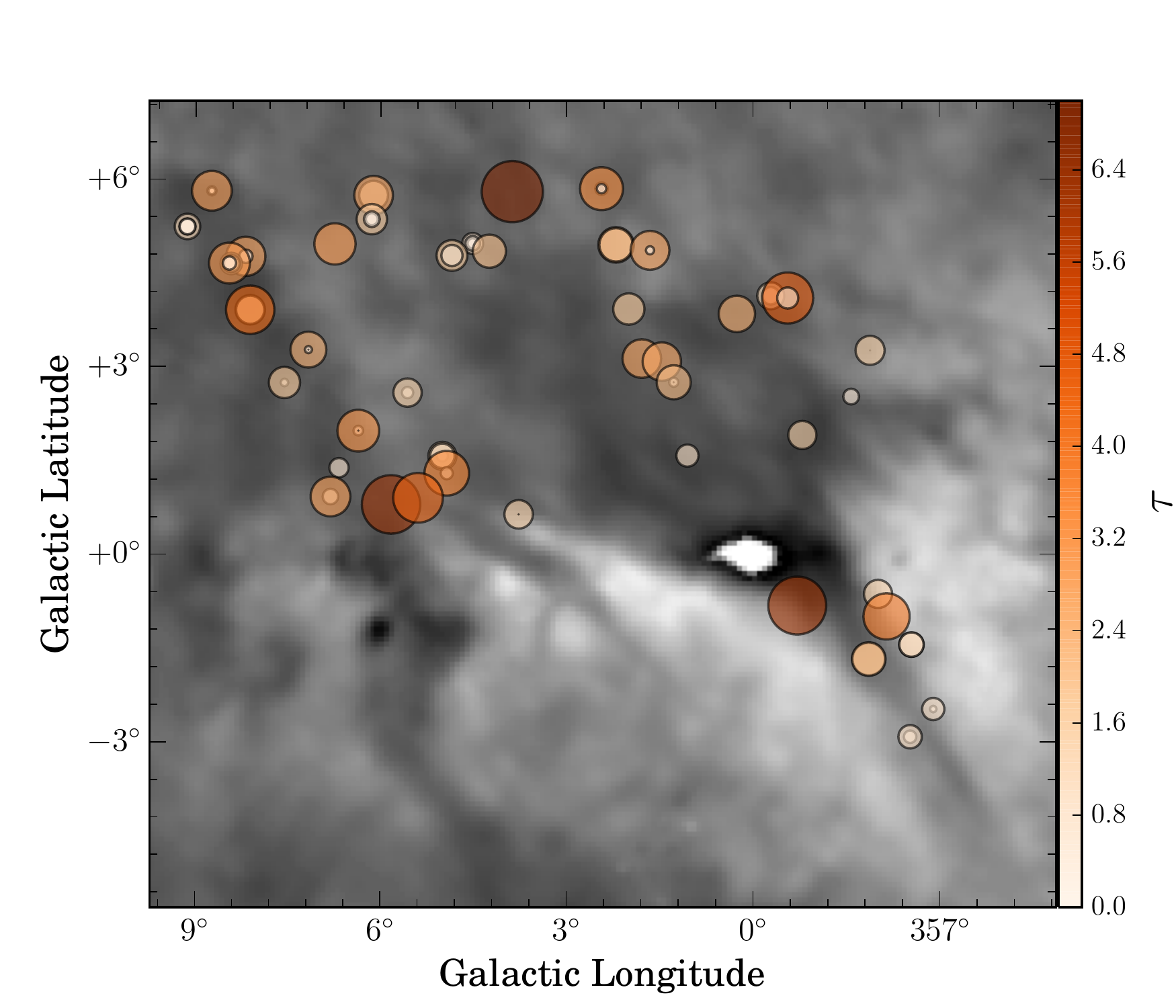}}
	\hfill
	\subfigure{\includegraphics[width=84mm]{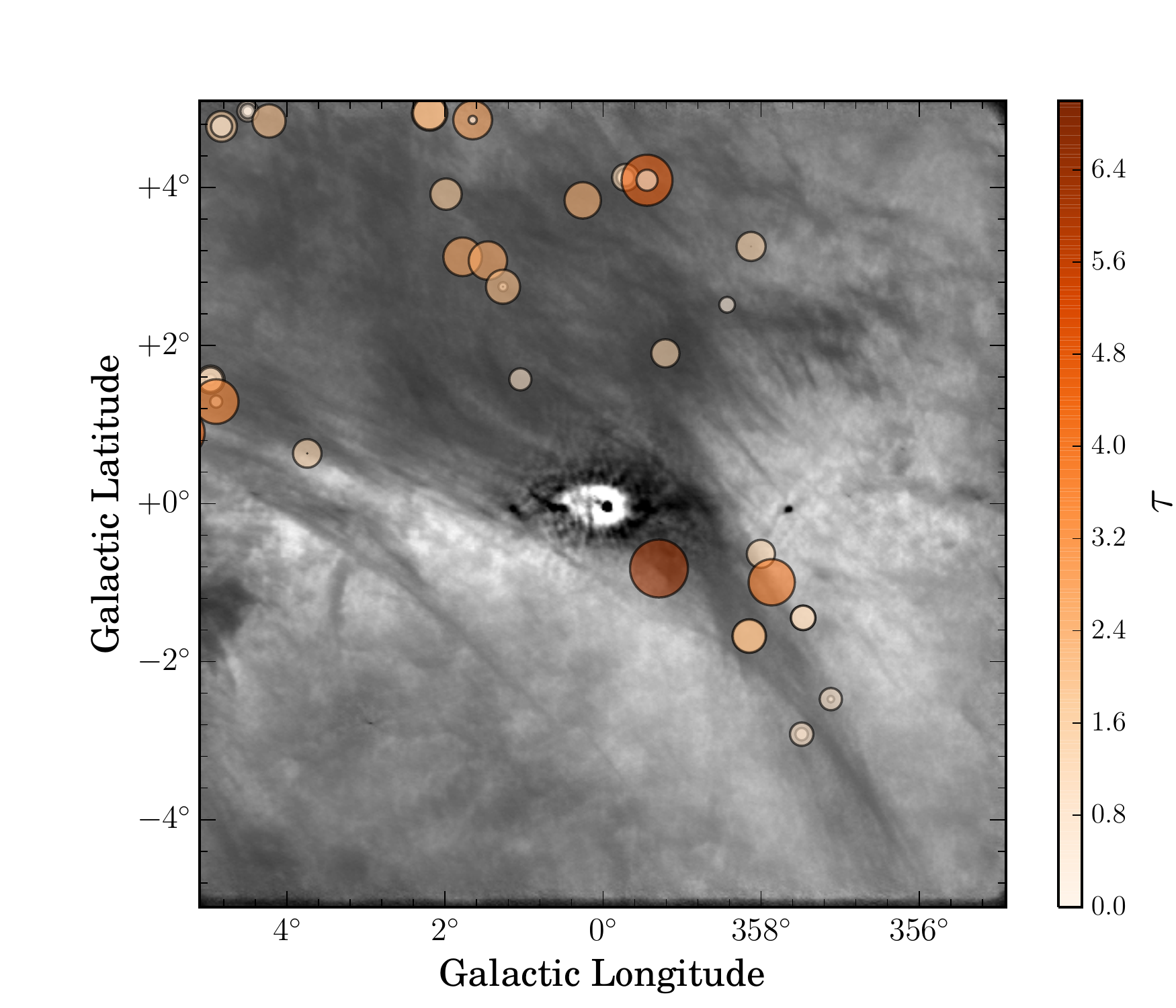}}
	\hfill
	\caption{GASS (left, $v = 5.2$ \kms) and SGPS GC (right, $v = 5.7$ \kms) \HI\ intensity maps of the R-C cloud. The circles show the position of the observed continuum background sources scaled and coloured by the optical depth. Most  LOS have two Gaussian components between $0 < v < 10$ \kms and FWHM $< 10$ \kms.}
	\label{fig:optical-depths}
\end{figure*}

Fig.~\ref{fig:optical-depths} shows the optical depth distribution across the R-C cloud over-plotted on the GASS and SGPS GC \HI\ intensity maps. The symbol sizes are colour coded and scaled with the optical depth from each Gaussian component. Some LOS have multiple Gaussian components associated with the cloud, indicating internal sub-structure. \cite{Montgomery1995} found that the R-C cloud has two main components at $v_{LSR} = 5$ and $v_{LSR} = 7$ \kms. We also find that several sight lines have more than one component associated with the R-C cloud. We find most components at 5 \kms\ and do not see a peak in the distribution at 7 \kms. This may be due to the fact that we are only sampling 46 LOS across the whole cloud. 

\begin{table*}
	\centering
	\caption{Line parameters from the Gaussian decomposition associated with the R-C cloud (0 < $v_{LSR}$ < 10 \kms, FWHM $< 10$ \kms) and calculated T$_{k,max}$, T$_{s}$ and N(\HI).  See full version in Appendix~\ref{Appendix:Tables}.}
	\label{tab:parameters}
	\begin{tabular}{l c c c c c c c c}
		\hline
		name & $\tau_{HISA}$ & $v_{LSR}$ & FWHM & T$_{k,max}$ & T$_{s,HISA}$ GASS & T$_{s,HISA}$ SGPS & N(\HI) GASS & N(\HI) SGPS   \\
		& & [\kms] & [\kms] & [K] & [K] & [K]  & [10$^{20}$ cm$^{-2}$]  & [10$^{20}$ cm$^{-2}$]  \\
		\hline
		NVSS J172920-234535   & 0.29 $\pm$ 0.16 & 3.2 $\pm$ 0.2 & 2.4 $\pm$ 0.2 & 127 $\pm$ 1 & - $\pm$ - & - $\pm$ - & - $\pm$ - & - $\pm$ - \\
		NVSS J172920-234535   & 3.38 $\pm$ 0.31 & 6.7 $\pm$ 0.2 & 4.7 $\pm$ 0.2 & 485 $\pm$ 1 & 46 $\pm$ 54 & - $\pm$ - & 14.48 $\pm$ 0.05 & - $\pm$ - \\
		NVSS J172920-234535   & 0.16 $\pm$ 0.02 & 0.4 $\pm$ 0.3 & 3.4 $\pm$ 0.6 & 254 $\pm$ 7 & - $\pm$ - & - $\pm$ - & - $\pm$ - & - $\pm$ - \\
		NVSS J173205-242651   & 2.33 $\pm$ 0.43 & 2.2 $\pm$ 0.2 & 3.0 $\pm$ 0.2 & 192 $\pm$ 1 & 72 $\pm$ 4 & 67 $\pm$ 1 & 9.75 $\pm$ 0.11 & 9.08 $\pm$ 0.1 \\
		NVSS J173753-254642   & 2.73 $\pm$ 0.21 & 3.4 $\pm$ 0.2 & 9.8 $\pm$ 0.3 & 2098 $\pm$ 2 & 35 $\pm$ 24 & 41 $\pm$ 30 & 18.46 $\pm$ 0.04 & 21.38 $\pm$ 0.04 \\
		\hline
	\end{tabular}
\end{table*}

\subsection{Spin temperature and column density}
\label{par:Ts}

\begin{figure}
	\centering
	\includegraphics[width=84mm]{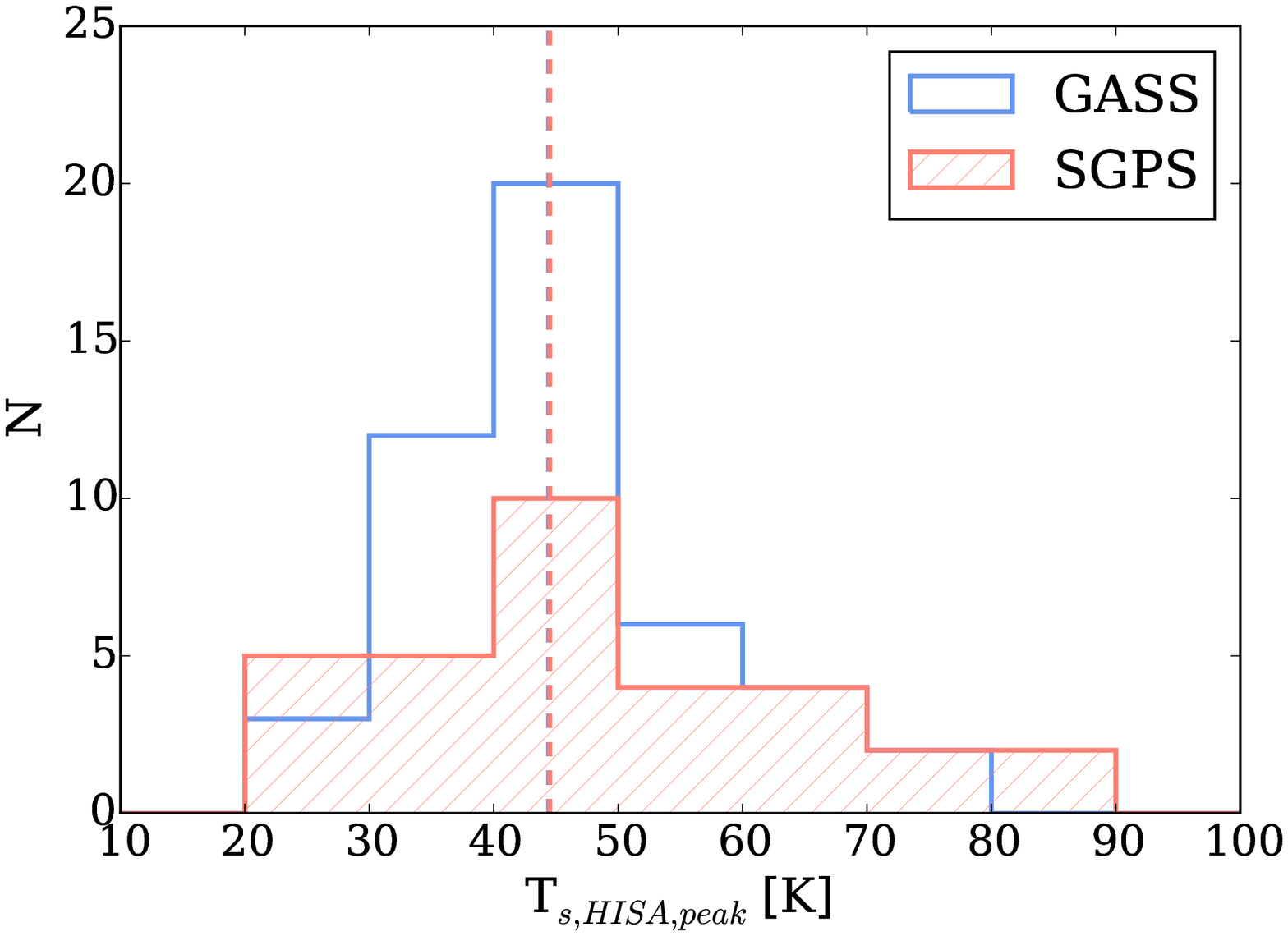}
	\caption{Distribution of \Ts$_{,HISA,peak}$. The blue histogram shows temperatures calculated with the GASS data and the red histogram shows temperatures calculated with SGPS data. The dashed lines show the median values.}
	\label{fig:Ts_peak}
\end{figure}

We derive \Ts\ using Equation~\ref{eq:Ts} in two different ways. 
\begin{inparaenum} 
	\item We calculate \Ts$_{,HISA}$ for the individual Gaussian components of the absorption spectra in the range of the R-C cloud ($0 < v_{LSR} <10$ \kms, FWHM $< 10$\kms) and 
	\item we  calculate \Ts\ at the minimum point of the HISA feature (\Ts$_{,HISA,peak}$). 
\end{inparaenum}	
The difference between the two approaches is that the first method estimates the temperature for different velocity components, while the second method derives a single harmonic mean temperature at the given position. We consider the second method as a better representation of \Ts\ because it does not depend on the uncertainties of the Gaussian decomposition. Fig.~\ref{fig:Ts_peak} compares the distribution of \Ts$_{,HISA,peak}$ for the GASS (blue) and SGPS GC (red) data. Both histograms peak at 50 K, but the SGPS GC data has a longer tail towards warmer temperatures. This is due to a systematic offset in brightness temperature between the two surveys, which we will further discuss in Section~\ref{sec:uncertainties}.

We find that \Ts$_{,HISA,peak}$ varies across the cloud between 20-80 K, with a median \Ts$_{,HISA,peak}$ of 48 $\pm 7$ K for GASS and 64 $\pm 9$ K for SGPS GC. This is slightly higher compared to previous studies which found temperatures around $\sim$40 K \citep{Crutcher1984, Montgomery1995, McClure-Griffiths2006}. The range of spin temperatures is also slightly larger compared to the ones estimated by \cite{McClure-Griffiths2006}, who found temperatures between 30-65 K. However, we directly measure the optical depth with the ATCA \HI\ absorption spectra which make our results better constrained compared to previous studies.

\begin{figure*}
	\centering
	\includegraphics[width=160mm]{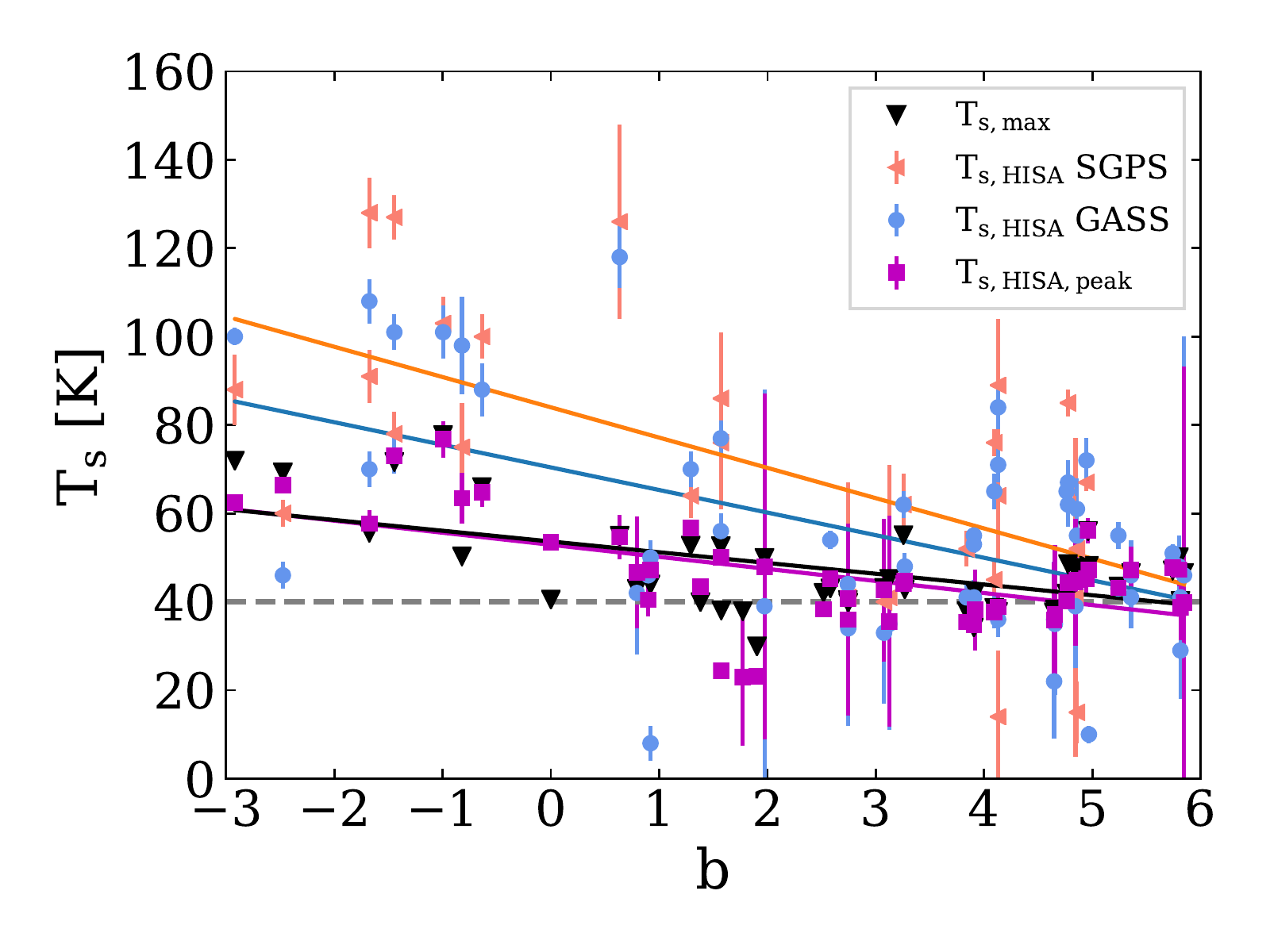}
	\caption{Spin temperatures as a function of Galactic latitude. Blue circles show \Ts\ calculated with GASS data and red triangles show \Ts\ calculated with SGPS GC data. The black triangles show T$_{B}$ at the tip of the HISA in the GASS data for each LOS, which are lower limits for \Ts. The purple squares are \Ts$_{,HISA,peak}$ calculated at the tip of the HISA for each LOS. The coloured lines are linear fits to the data to show the decrease of \Ts\ towards higher Galactic latitudes. }
	\label{fig:Ts_latitude}
\end{figure*}

In Fig.~\ref{fig:Ts_latitude} we show the distribution of \Ts\ with Galactic latitude. We mark \Ts$_{,HISA}$ from GASS with blue circles and \Ts$_{HISA}$ from SGPS GC with red triangles (see also Tab.~\ref{tab:parameters}). We find an offset between the GASS and the SGPS GC results, with systematically lower temperatures based on the GASS data. We mark \Ts$_{,HISA,peak}$ from the GASS data with magenta squares (see also Tab.~\ref{tab:Ts_ridge}). We only show the GASS data because it has better brightness temperature sensitivity compared to SGPS GC. For comparison we also included the brightness temperature of the GASS data at the minimum point of the HISA (black triangles), which is an upper limit on the spin temperature of the cold gas. We note that in some cases \Ts$_{,HISA,peak}$ has a lower temperature compared to \Ts$_{,HISA}$, this is because several LOS have 2 separate Gaussian components associated with the R-C cloud and that the Gaussian decomposition underestimates the optical depth in some cases. We discuss this further in section ~\ref{sec:uncertainties}. The coloured lines in Fig.~\ref{fig:Ts_latitude} are linear fits to the data, showing the trend of decreasing \Ts\ towards higher Galactic latitudes. This temperature gradient is steepest for \Ts$_{,HISA}$ derived for individual Gaussian components, however it also has the largest scatter and a Pearson's r of -0.5 (for the GASS data). The temperature gradient is present in all the data and $T_{s,HISA,peak}$ and $T_{s,max}$ show a slightly stronger correlation with a Pearson's r of -0.6. 

\begin{figure*}
	\centering
	\subfigure{\includegraphics[width=84mm]{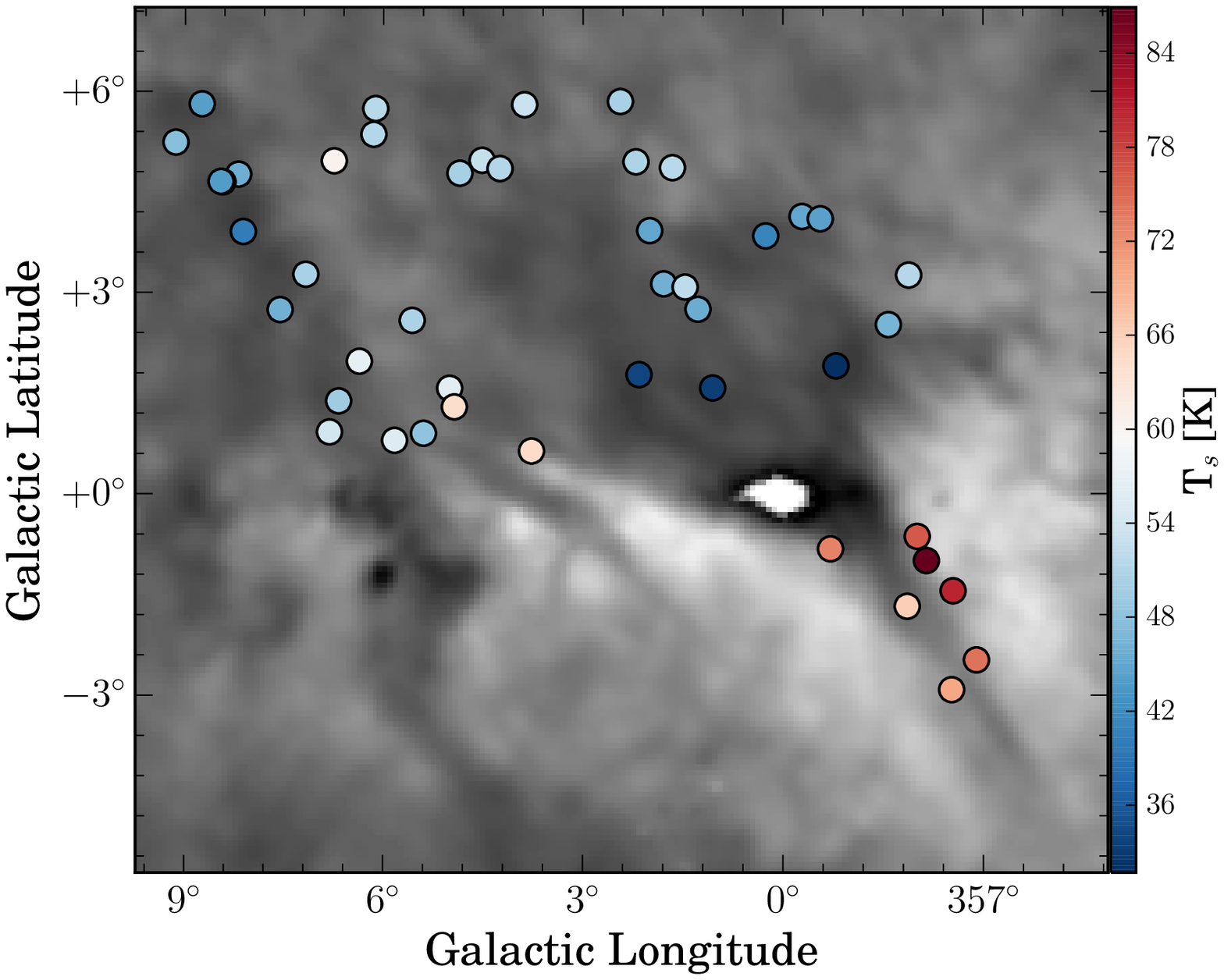}}
	\hfill
	\subfigure{\includegraphics[width=84mm]{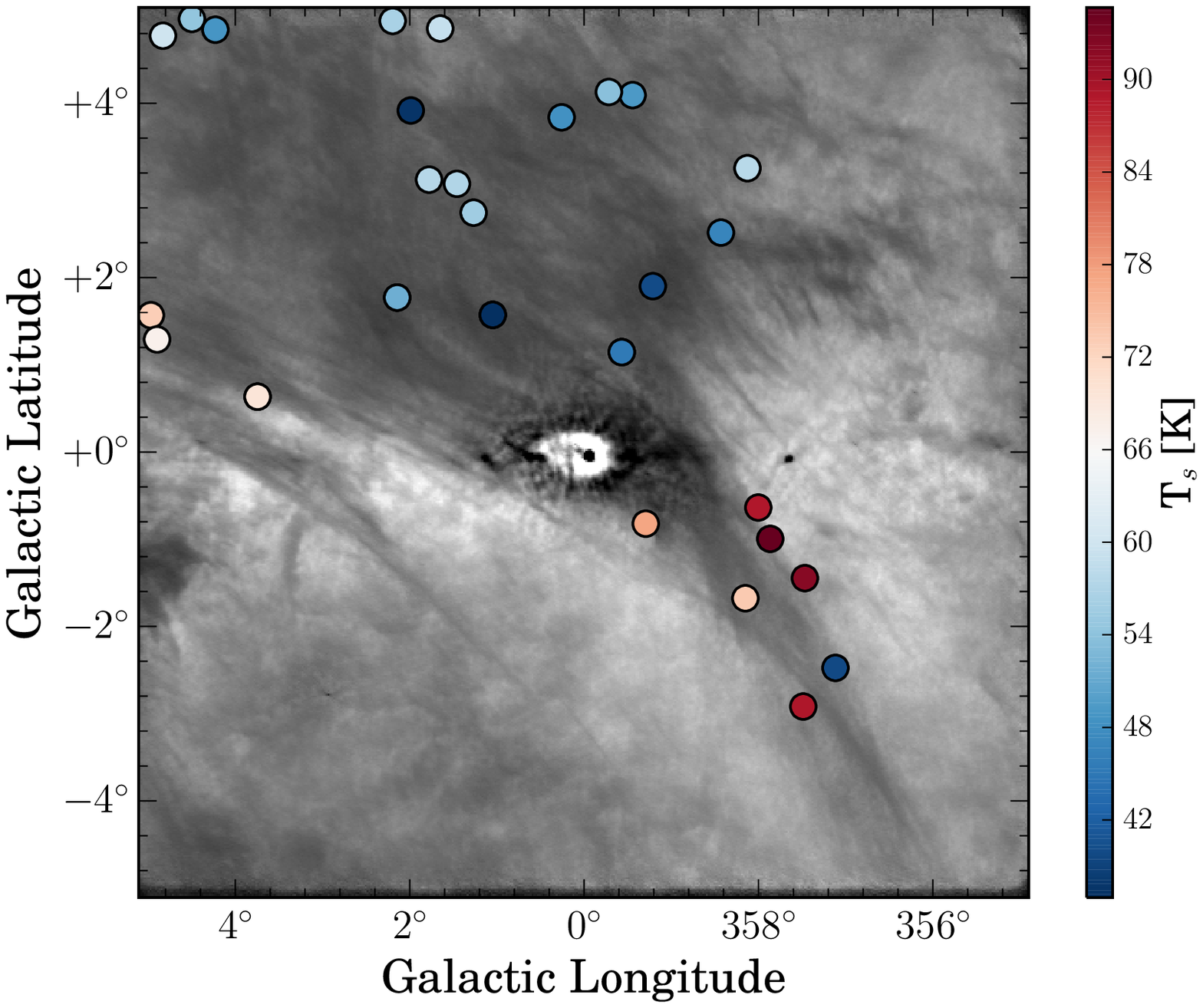}}
	\hfill
	\caption{\Ts$_{,HISA,peak}$ for each  LOS over plotted on the \HI\ intensity map of the R-C cloud ($v = 5.2$ and $5.7$ \kms for the GASS and the SGPS GC respectively). The left side shows \Ts$_{,HISA,peak}$ calculated from GASS data, the right side \Ts$_{,HISA,peak}$ calculated from SGPS  GC data.}
	\label{fig:Ts}
\end{figure*}

In Fig.~\ref{fig:Ts} we show the temperature distribution (\Ts$_{,HISA,peak}$) with colour coded circles over-plotted on the \HI\ intensity map from GASS and SGPS GC. The temperature gradient is also evident here, with temperatures decreasing towards higher latitudes in the cloud. The coldest temperatures are between $6^{\circ} < l < 9^{\circ}, 1^{\circ} < b < 3^{\circ}$ (30 - 50 K) of the cloud and warmer temperatures are in the filamentary tip of the cloud (50 - 80 K). 

\begin{table*}
	\centering
	\caption{\Ts\ calculated at the minimum point ($v_{min}$, T$_{min}$) of the HISA feature in the \HI\ emission spectra for both GASS and SGPS data. See full table in Appendix~\ref{Appendix:Tables}}
	\label{tab:Ts_ridge}
	\begin{tabular}{l c c c c c c c c}
		\hline
		name & l & b & v$_{min}$ (GASS) & T$_{s,max}$  & T$_{s,HISA,peak}$  & v$_{min}$ (SGPS) & T$_{s,max}$  & T$_{s,HISA,peak}$   \\
		& [$^{\circ}$] & [$^{\circ}$] &  [\kms] & (GASS) [K] & (GASS) [K] & [\kms]  & (SGPS) [K]  & (SGPS)  [K]  \\
		\hline
		NVSS J172829-284610   & -1.8738 & 3.2555 & 4.4 & 55.11 & 51 $\pm$ 1 & 4.9 & 66.06 & 57 $\pm$ 7 \\
		NVSS J172836-271236   & -0.5563 & 4.0941 & 4.4 & 38.65 & 44 $\pm$ 1 & 4.1 & 44.05 & 49 $\pm$ 2 \\
		NVSS J172908-265751   & -0.2845 & 4.1296 & 4.4 & 37.75 & 44 $\pm$ 3 & 4.1 & 44.66 & 53 $\pm$ 16 \\
		NVSS J172920-234535   & 2.432 & 5.8467 & 5.3 & 46.58 & 50 $\pm$ 58 & - & - & - $\pm$ - \\
		NVSS J173107-245703   & 1.6506 & 4.8577 & 3.6 & 46.52 & 51 $\pm$ 1 & 4.1 & 56.83 & 59 $\pm$ 1 \\
		\hline
	\end{tabular}
\end{table*}

The column density of the \HI\ absorption can be calculated with
\begin{equation}
	N(HI) = 1.823 \times 10^{18} T_{s} \int \tau(\nu) d\nu .
\end{equation}
We calculate the column density for each Gaussian component associated with the R-C cloud ($0 < v_{LSR} < 10$ \kms, FWHM $<10$ \kms). The column density for each component is
\begin{equation}
	N(HI) = 1.064 \times 1.823 \times 10^{18} \times T_{s,HISA} \times \tau_{HISA} \times \Delta v,
\end{equation}
where $T_{s,HISA}$ is the spin temperature, $\tau_{HISA}$ is the peak optical depth, and  $\Delta v$ is the FWHM in \kms\ of each component. 

\begin{figure}
	\centering
	\includegraphics[width=84mm]{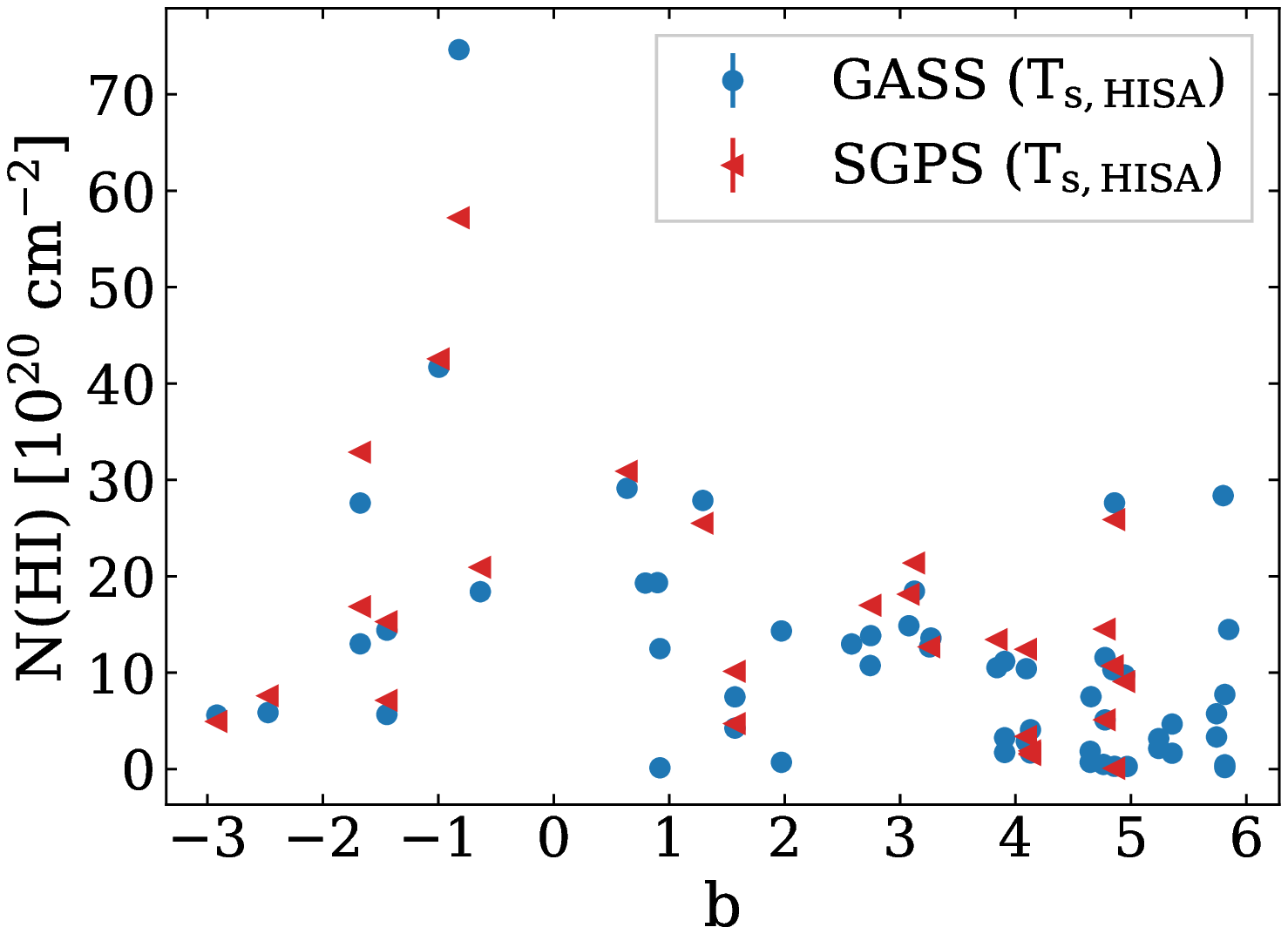}
	\caption{\HI\ column density as a function of Galactic latitude. The blue circles show N(HI) calculated from \Ts$_{,HISA}$ GASS and the red circles show N(HI) calculated from \Ts$_{,HISA}$ SGPS GC.}
	\label{fig:NHI_latitude}
\end{figure}

We find an average column density of 11.1 $\pm 0.2 \times 10^{20}$ cm$^{-2}$ for GASS and 15.3 $\pm 0.3 \times 10^{20}$ cm$^{-2}$ for SGPS GC, which is higher than the 2 $\pm 1.4 \times 10^{20}$ cm$^{-2}$ found by \cite{McClure-Griffiths2006}. The reason for the difference is that we have more accurate measurements of the optical depth for this study. \cite{McClure-Griffiths2006} assumed \Ts=40 K for the whole cloud which resulted in $\tau\sim 2.5$. In contrast to that approach we use optical depths derived from the Gaussian decomposition and spin temperatures (\Ts$_{HISA}$ GASS) ranging between 10 - 120 K. Furthermore, \cite{McClure-Griffiths2006} found that the positive latitude region of the cloud has higher N(\HI) which we do not see in our data (Fig.~\ref{fig:NHI_latitude}). This is due to the fact the \cite{McClure-Griffiths2006} assumed a constant spin temperature throughout the cloud, where we find that the spin temperature decreases towards positive latitudes. 

\subsection{Sources of uncertainty and errors}
\label{sec:uncertainties}

We take the following sources of uncertainties into account when calculating the spin temperature:
\begin{enumerate}
	\item The RMS noise of the observations. This is the smallest source of uncertainty for \Ts.
	\item Errors in estimating $T_{OFF}$. We estimate the uncertainty by comparing the maximum of the emission spectra and the maximum of the $T_{OFF}$ model. This is a conservative approach that gives errors on the order of 10K. 
\end{enumerate}

The errors of the line parameters in Tab.~\ref{tab:parameters} ($\tau_{HISA}, v_{LSR}$, FWHM) are based on the errors calculated by {\sc GaussPy}, the errors on \Ts\ and N(HI) are propagated errors. 

Additional sources of uncertainty:
\begin{enumerate}	
	\item \HI\ emission fluctuations on a small scale can introduce errors because the spatial scale probed by the \HI\ emission data (16') is much larger compared to the \HI\ absorption data ($\sim$60''). To investigate this issue, we compared \Ts\ for the closest absorption pair in our sample (NVSS J174713-19213 and NVSS J174716-19195), separated by 1.7',  and found that the derived spin temperatures (\Ts$_{,HISA,peak}$) agree well ($42 \pm 14$ K and $43 \pm 17$ K). This suggests that our data are not too sensitive to arcminute scale variations.
	\item Comparing \Ts$_{,HISA,peak}$ derived with SGPS GC and GASS data we find that SGPS GC gives on average 6 K higher temperatures. This is due to the 1.2 scaling factor between GASS and SGPS GC. Since SGPS GC data are a combination of interferometric and single dish data, the single dish $T_{B}$ needs to be scaled to match the interferometric data. Because of this, and the fact that the GASS data has better brightness sensitivity, we use the results from the GASS data to derive our main conclusions.
	\item The fraction of the foreground \HI\ emission compared to the background \HI\ emission (p) can introduce some errors. Allowing a 10\% variation in p gives a 10 K uncertainty for \Ts.
	\item If $\tau_{bg}$ is not negligible, the spin temperature is calculated the following way:
	\begin{equation}
		T_{s} = \frac{ T_{ON} - T_{OFF}}{1-e^{-\tau_{HISA}}} + T_{c}e^{-\tau_{bg}} + pT_{OFF}. 
		\label{eq:Ts_2}
	\end{equation}
	$\tau_{bg}$ only appears in the term for the diffuse continuum emission. If $\tau_{bg}$ varies between 0.1 and 1, \Ts\ can decrease by 0.5 - 10 K.  
	\item Spectral baseline errors are due to difficulties in data calibration. For the \HI\ absorption data the problem is the strong absorption line in the bandpass calibrator between 0 - 10 \kms (see Appendix~\ref{Appendix:1934}). This bandpass uncertainty is negligible compared to the other sources of errors. For the \HI\ emission data, the main difficulty is the strong continuum emission around the GC. The uncertainty from this is on the order of a few K.  
	\item Decomposing the \HI\ absorption spectra has inherent uncertainty as there is no unique solution for decomposing spectra with multiple blended components. Because of the location of the R-C cloud, in the Galactic plane and in front of the GC, these data are especially susceptible. \cite{Murray2017} investigated the recovery rate of the AGD algorithm on simulated \HI\ spectra. They found that AGD has a very good recovery rate for simple LOS, e.g. high latitude spectra (99\% for $|b| > 50^{\circ}$), performs well for moderately complex spectra (67\% for $20^{\circ} < |b| < 50^{\circ}$) and accurately recovers 53 \% of the components of complex, Galactic plane sources ($4^{\circ} < |b| < 20^{\circ}$). All our sources fall into the last category which means that AGD recovers approximately 50 \% of the underlying Gaussian distribution towards the R-C cloud. Considering the complexity of the spectra in this region and the sensitivity of our observations this is a reasonable recovery rate. We find that in some cases AGD fits one broad Gaussian to spectra, which are likely composed of two or three components based on visual inspection. This means that AGD will underestimate the number of components and their optical depths. This would then lead to higher spin temperatures and kinetic temperatures. However, the advantage of AGD over other decomposition methods is that it does not require any subjective initial guesses and the decomposition allows us to compare our results with studies such as the Millennium \citep{Heiles2003_1} and the Perseus \citep{Stanimirovic2014} survey data in a quantitative way. To avoid systematic bias from the line decomposition, we also calculate \Ts\ without decomposing the spectra (\Ts$_{HISA,peak}$) and consider these our most accurate temperatures for this region.
\end{enumerate}

\newpage

\subsection{OH absorption lines}

We detected four OH absorption line candidates at $v_{LSR} \sim 3 - 7$ \kms\ towards three sources: NVSS J172920-23453,  NVSS J173850-22191, and NVSS J175526-22321 (Fig.~\ref{fig:OH_spectra} and ~\ref{fig:OH_spectra2}). We detect lines towards NVSS J172920-23453 in both frequencies, 1665 and 1667 MHz. Our detection criteria is $\tau_{OH, max} > 3 \sigma_{tau,OH}$, where $\tau_{OH, max}$ is the maximum of the OH optical depth spectra and $\sigma_{tau,OH}$ is the optical depth sensitivity. Typical $\tau_{OH}$ are much smaller than $\tau_{HI}$ (e.g. \citealt{Colgan1989, Liszt1996, Li2018}). \cite{Liszt1996} found $\tau_{OH}$ between 0.006 and 0.27 for the 1667 MHz line, in regions where $\tau_{HI}$ was between 0.06 and 3.16. For our observations $\sigma_{\tau OH} $ is between 0.01 and 0.6. We note that our OH detections are weak and would need follow up observations to better constrain the optical depth and the column density. To estimate the OH optical depth and column density we fitted a single Gaussian to each spectrum. The fitted parameters are presented in Tab.~\ref{tab:OHparameters}. We found optical depths between 0.05 $\pm 0.01$ to 0.23$\pm 0.07$, where the errors are based on the RMS of the spectra and the spectral resolution (0.26 \kms). 

We calculate OH column densities with the following equation:
\begin{equation}
	N(OH) = 1.064 \times 2.24 \times 10^{14} T_{ex}(1667)\tau_0(1667) \Delta v, 
\end{equation}

\begin{equation}
	N(OH) = 1.064 \times 4.07  \times 10^{14} T_{ex}(1665)\tau_0(1665) \Delta v, 
\end{equation}
where $T_{ex}$ is the excitation temperature of the gas in K, $\tau_0$ is the peak optical depth and $\Delta v$ is the FWHM in \kms\ assuming Gaussian line profiles. We assume $T_{ex}$ is between 3-20 K - based on typical measurements from the literature (e.g. \citealt{Dickey1981, Colgan1989, Liszt1996}) - and calculate the N(OH) for these two values (Tab.~\ref{tab:OHparameters}). The OH column density towards NVSS J172920-23453 calculated from the 1665 MHz and the 1667 MHz line agrees well. NVSS J172920-23453 is one of the brightest continuum sources in our sample. 

In addition, we detected two un-associated OH absorption lines in two different  LOS at $v \sim 50$ \kms\ and $v \sim 160$ \kms. We also detected an OH maser in the field of view of NVSS J174513-31510 at 1667 MHz. 

\begin{table*}
	\centering
	\caption{Fitted Gaussian parameters to the OH lines. For calculating N(OH) we assume that $T_{ex}$ is between 3 - 20 K.}
	\label{tab:OHparameters}
	\begin{tabular}{l c c c c c c c c}
		\hline
		name & l & b & Frequency  & amplitude & $\tau$ & v & FWHM & N(OH)\\
		& [$^{\circ}$]  &[$^{\circ}$]  & [MHz]  & & & [\kms] & [\kms]   & [$10^{14}$ cm$^{-2}$]\\
		\hline
		NVSS J172920-23453 & 2.43 & 5.85 & 1665   & 0.05 $\pm$ 0.01 & 0.05 $\pm$ 0.01 & 7.0 $\pm$ 0.3 & 1.5 $\pm$ 0.3 &1.0 - 6.5 \\
		NVSS J173850-22191 & 4.83 & 4.77 & 1665   & 0.23 $\pm$ 0.07 & 0.27 $\pm$ 0.07 & 3.9 $\pm$ 0.3 & 1.8 $\pm$ 0.3 & 6.4 - 42.8\\
		NVSS J172920-23453 & 2.43 & 5.85 & 1667   & 0.05  $\pm$ 0.01 & 0.05 $\pm$ 0.01 & 6.4 $\pm$ 0.3 & 2.7 $\pm$ 0.3 & 1.0 - 6.5\\
		NVSS J175526-22321 & 6.63 & 1.38 & 1667   & 0.11  $\pm$ 0.03 & 0.12 $\pm$ 0.03 & 7.7 $\pm$ 0.3 & 1.2 $\pm$ 0.3 & 1.1 - 7.1\\
		\hline
	\end{tabular}
\end{table*}

\section{Discussion}
\label{sec:Discussion}

The R-C cloud is a cold \HI\ structure at the edge of the Local Bubble. According to theories of its formation (e.g. \citealt{Weaver1979, Heiles1998}) a collection of massive stars and supernovae swept out the material from the Local Bubble and stretched the magnetic field along the wall of the cavity. The cold material swept up at the edge of the bubble ended up entrained in the strong magnetic field of the wall and produced the filamentary structure that was observed by \cite{McClure-Griffiths2006}. Since HISA clouds are theorised to be transitional clouds between purely atomic and molecular clouds, the R-C cloud may also be a location where the ISM is transitioning from an atomic to a molecular phase. In this section we discuss the presence of associated molecular gas and compare our results with the literature.

\subsection{Associated molecular gas}

\cite{Kavars2005} found that 60\% of HISA clouds in the SGPS have associated molecular gas. The R-C cloud is one of the largest HISA complexes in the Galaxy and as our results show, has a substantial amount of high optical depth ($\tau > 1$) cold gas. Furthermore, our results in Fig.~\ref{fig:Ts_latitude} and ~\ref{fig:Ts} reveal a temperature gradient across the cloud, varying from $\sim 30$ K in the upper left of the cloud (base) to almost 80 K at the lower right (tip) of the cloud. This might suggest that the presence or absence of molecular gas would correlate with the temperature variations, with molecular gas more likely at the base of the cloud than at the tip. Indeed, the R-C cloud has a significant amount of associated OH and $^{12}$CO gas. \cite{Crutcher1973} found several OH absorption lines across the cloud, indicating the presence of substantial molecular gas. This is further supported by detected $^{12}$CO emission at $l \sim 5^{\circ}, ~b \sim 6^{\circ}, ~v \sim 5$ \kms \citep{Dame2001}. 

\begin{figure}
	\centering
	\includegraphics[width=84mm]{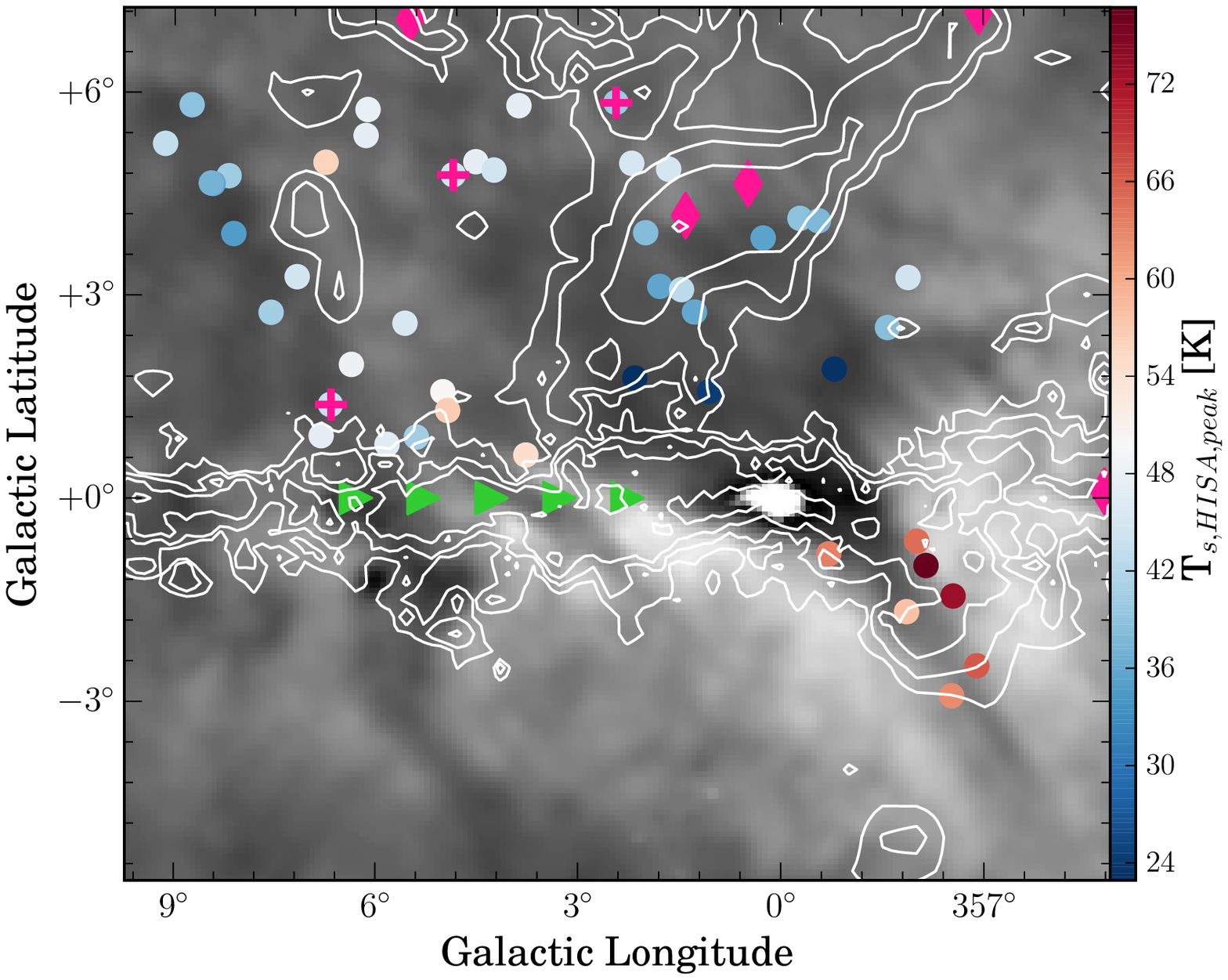}
	\caption{GASS \HI\ intensity map of the R-C cloud ($v_{LSR}=5.2$). White contours show the integrated $^{12}$CO intensity between 2.6 - 7.8 \kms\ \citep{Dame2001}. Contour levels are 2, 4, 8, 16, 32 K \kms. The pink diamonds show detected OH lines by \citep{Crutcher1973} and the pink crosses show our OH line detections, the green triangles show the CRRL lines from \citep{Roshi2011}. The red and blue circles show the measured spin temperatures (\Ts$_{HISA,peak}$) for reference.}
	\label{fig:CO_map}
\end{figure}

Fig.~\ref{fig:CO_map} shows the integrated $^{12}$CO intensity\footnote{https://www.cfa.harvard.edu/rtdc/CO/CompositeSurveys/} between $2.6 < v_{LSR} < 7.8$ \kms\ overlaid on the GASS \HI\ intensity map. We also marked the location of the OH lines detected by \cite{Crutcher1973} with pink diamonds, our OH line candidates with pink crosses and the detected CRRL lines with green triangles \citep{Roshi2011}. For reference we also plot \Ts$_{HISA,peak}$ as red and blue circles. Fig.~\ref{fig:CO_map} shows significant $^{12}$CO emission in the Galactic plane, which is expected, and it also shows an elongated feature across the base of the cloud. The OH detections from \cite{Crutcher1973} and our strongest OH line candidate (NVSS J172920-23453 at $l,b = 2.43^{\circ}, 5.85^{\circ}$) overlap with the $^{12}$CO emission.    

The Southern Parkes Large-Area Survey (SPLASH; \citealt{Dawson2014}) has detected significant OH emission and absorption throughout the R-C cloud. The SPLASH survey covers approximately the same region of the R-C cloud that is covered by the SGPS GC cube ($b \sim 5^{\circ}$). Approximately 30\% of the LOS we observed have OH lines associated with the R-C cloud in the SPLASH data. These lines mostly overlap with the $^{12}$CO emission. One of our OH absorption line candidates (NVSS J175526-22321;  $l,b = 6.63^{\circ}, 1.38^{\circ}$) overlaps with the SPLASH data, which confirms our detection. We will further discuss the SPLASH OH detections for this region in an upcoming publication.

Assuming that the R-C cloud was formed by the Local Bubble pushing out and piling up the \HI\ in a sheet-like structure, which we observe as an \HI\ self-absorption cloud, makes it a plausible site for triggered molecular cloud formation similar to the edges of other supershells (e.g. \citealt{Fukui2010, Dawson2011, Ehlerova2016, Dawson2013} and references within). 

\subsection{Comparison to the literature}

\begin{figure*}
	\centering
	\subfigure{\includegraphics[width=84mm]{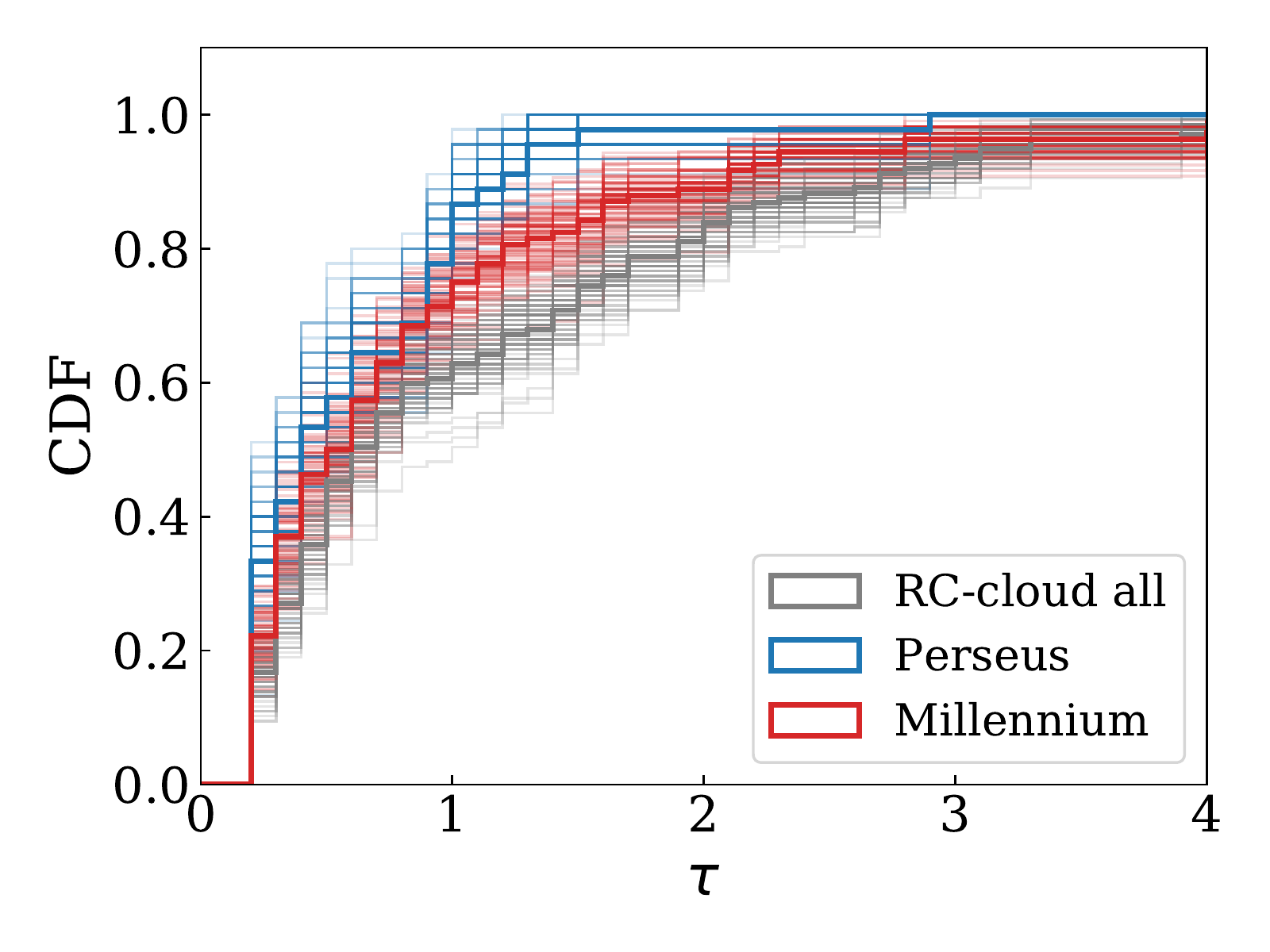}}
	\hfill
	\subfigure{\includegraphics[width=84mm]{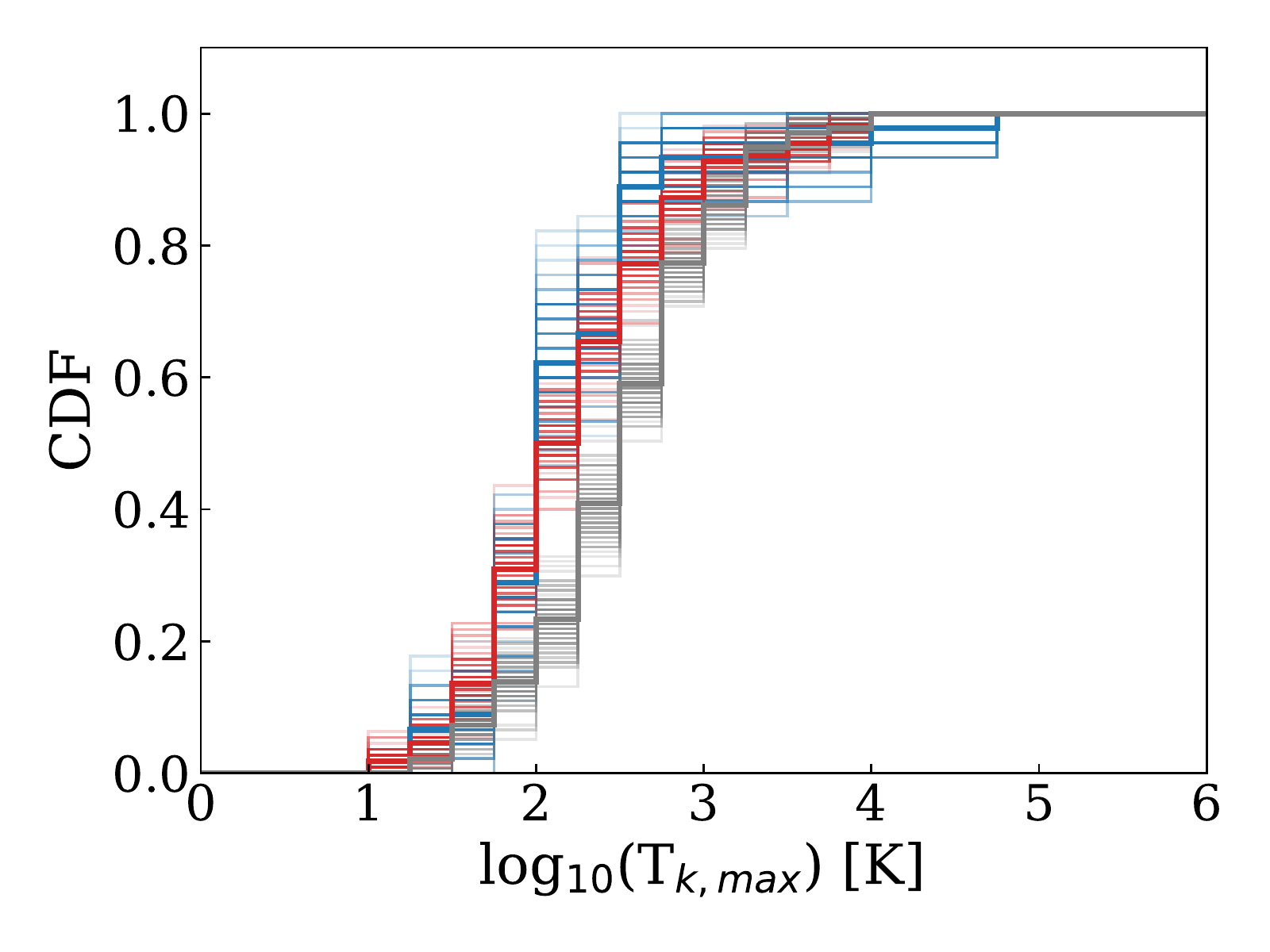}}
	\hfill
	\caption{Comparison of optical depth and maximum kinetic temperature ($T_{k, max}$) with data from the Millennium survey (red) and the Perseus molecular cloud (blue). To illustrate the uncertainties of the distributions we plot a 100 bootstrap samples for each dataset as lighter-shaded lines.}
	\label{fig:RC-histograms_2}
\end{figure*}

In this section we briefly compare our results with Galactic \HI\ absorption datasets from the Millenium survey \citep{Heiles2003_1, Heiles2003_2} and the Perseus \citep{Stanimirovic2014} molecular cloud. Fig.~\ref{fig:RC-histograms_2} compares the optical depth and the maximum kinetic temperature (T$_{k,max}$) of our sample (grey, including all Gaussian components from the decomposition) with data from the Millenium survey (red) and data from around the Perseus molecular cloud (blue). The Millennium survey consists of 79 continuum sources mostly towards high Galactic latitudes observed with the Arecibo telescope (for details see  \citealt{Heiles2003_1}). The Perseus sample has 26 continuum sources distributed around the molecular cloud observed with Arecibo \citep{Stanimirovic2014}. Both surveys have an optical depth sensitivity of 0.002 per 1 \kms\ channel. The Perseus sample is similar to our sample in the sense that it is also probing a region that is expected to have a significant amount of cold ISM, whereas the Millennium survey does not have a specific target region. This gives us the opportunity to compare the cold gas fraction of a HISA cloud, where atomic gas may be transitioning into molecular gas, to a molecular cloud, where this transition already happened and to the general distribution of the ISM. 

\begin{figure}
	\centering
	\includegraphics[width=84mm]{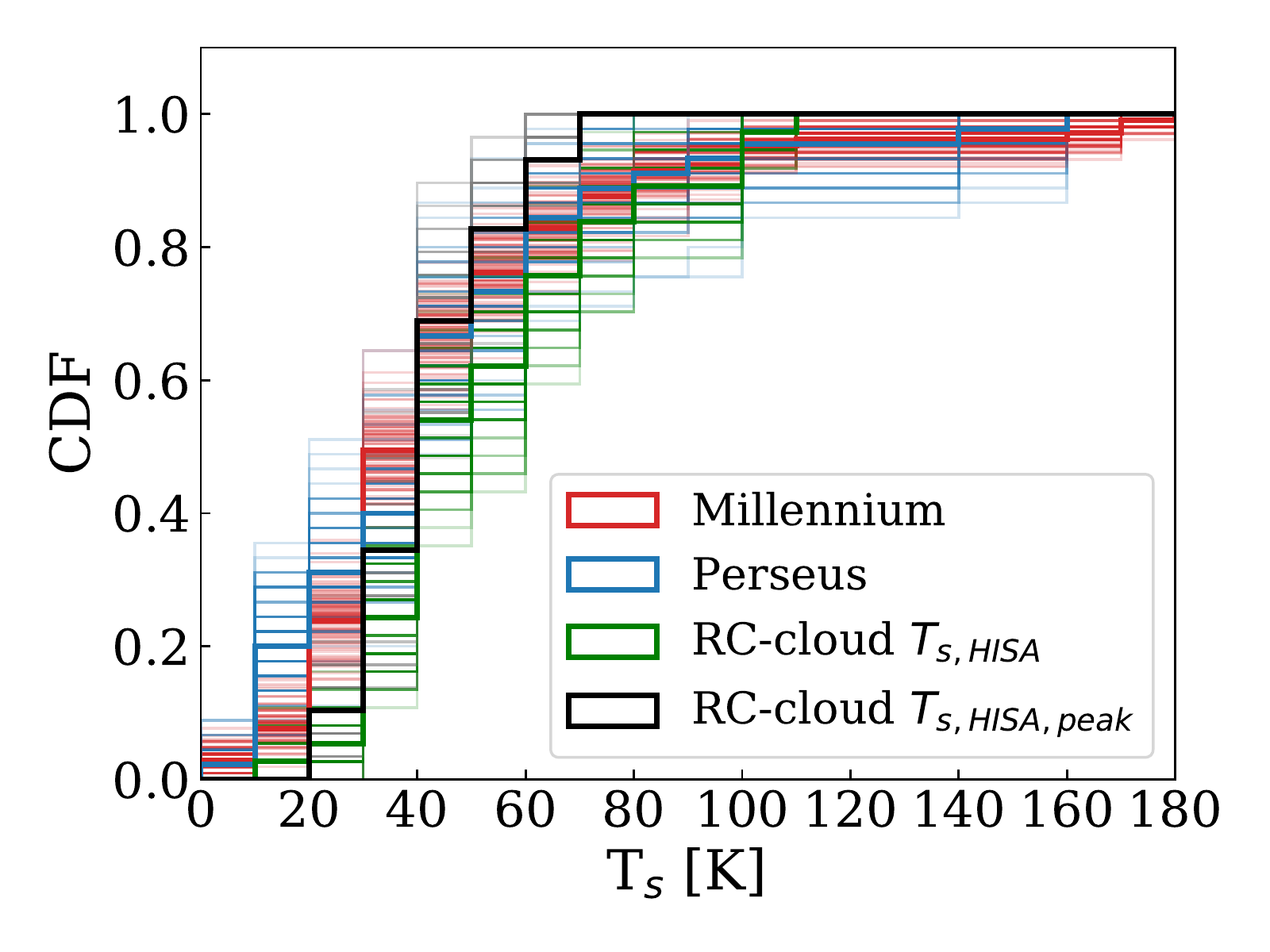}
	\caption{CDF for \Ts$_{,HISA,peak}$ (black) and \Ts\ (green). Data from the Perseus molecular cloud and the Millennium survey are in red and blue for comparison. To illustrate the uncertainties of the distributions we plot a 100 bootstrap samples for each dataset as lighter-shaded lines.}
	\label{fig:Ts_histogram}
\end{figure}

Considering that we had a different observing strategy compared to the Millennium and the Perseus surveys our data have a much poorer optical depth sensitivity $0.01 < \sigma{_\tau} < 0.66$ per 0.2 \kms\ channel. To fairly compare our data to the literature we only include Gaussian components in Fig.~\ref{fig:RC-histograms_2} and ~\ref{fig:Ts_histogram} which have $\tau > 0.2$ and $\sigma_{\tau} < 0.2$. The optical depth cumulative density function (CDF) in Fig.~\ref{fig:RC-histograms_2} shows that we detect more components with high optical depths, especially compared to Perseus. The highest optical depth around Perseus is 2.9, whereas for our sample and for Millennium $\tau_{max} \sim 7$. 

The second panel of Fig.~\ref{fig:RC-histograms_2} compares the maximum kinetic temperature ($T_{k,max}$) distribution of our data with results from the Millennium (red) and Perseus (blue) surveys. We derive $T_{k,max}$ for each Gaussian component the following way:
\begin{equation}
	T_{k,max} = m_{H}/(8k_{B} ln(2)) \times \Delta v^{2} = 21.866 \times \Delta v^{2},
\end{equation}
where $m_{H}$ is the hydrogen mass, $k_{B}$ is Boltzmann's constant, and $\Delta v$ is the FWHM in \kms. The lighter-shaded lines in Fig.~\ref{fig:RC-histograms_2} show the uncertainty of the distribution with 100 bootstrapped samples. The T$_{k,max}$ distribution of our sample shows fewer components between $10^{2}$ - $10^{3}$ K compared to the Perseus and Millennium data. This translates into our sample containing a higher fraction of broader lines with a median FWHM of 4 \kms\ compared to the other two surveys median of 3 \kms. This may be the result of the strong line blending in the direction of the R-C cloud. 

We find that overall the R-C cloud has a much narrower spin temperature range and a significantly lower median spin temperature (44 K for \Ts$_{HISA,peak}$ and 49 K for \Ts$_{HISA}$ based on GASS data), compared to the median \Ts\ found in the Millennium survey (114 K) and in Perseus (115 K). However, if we only include components from Millenium and Perseus with $\tau >0.2$ (Fig.~\ref{fig:Ts_histogram}) in our comparison, we find that the \Ts\ distribution of all three samples is very similar. With this cut the median spin temperature of the Millennium survey is 42 K and the Perseus survey is 44 K.


If we compare our results to other HISA measurements, the R-C cloud seems relatively ``warm''. \cite{Gibson2000} find that \Ts\ for HISA clouds in the CGPS is between 7 - 20 K and \cite{Kavars2005} find \Ts\ ranging between 6 - 41 K for HISA clouds in the SGPS survey. \cite{Kavars2003} analysed a HISA cloud in detail from the SGPS survey and found \Ts\ between 20 - 25 K and also found associated $^{12}$CO with the cloud. These values are lower than the R-C cloud \Ts$_{HISA,peak}$ range of 20 - 80 K. There can be several reasons for this difference, one of the biggest uncertainties is the assumed fraction of foreground \HI\ emission. A 10 \% difference in the foreground can result in $\sim$10 K temperature difference. Typical assumptions on the foreground are 25 - 50 \%. In the case of the R-C cloud we know that the cloud is approximately at the edge of the Local Bubble and that there should not be much gas in the foreground. For this reason, we assume a 10\% foreground fraction for all our calculations. Another key difference is that we have better constraints on the optical depth with 46 LOS where we directly measure $\tau$ which we use to ``calibrate'' the HISA spin temperature.  \cite{Kavars2003} found a likely range of $\tau$ between 0.06 and 2.3, whereas we measure $\tau$ between 0.09 and 6.8. The temperature and optical depth difference could also indicate that the two clouds are at different evolutionary stages. 

It has been suggested that the R-C cloud may be the wall of the Local Bubble (e.g. \citealt{Weaver1979, Heiles1998}), which makes it a similar structure to the walls of Galactic supershells. HISA features with high optical depths ($\tau > 1$) and low spin temperatures (10 - 40 K) were also observed in the walls of Galactic supershells (e.g. \citealt{Knee2001, Moss2012}), providing further support to the idea that the R-C cloud could be part of the wall of the Local Bubble. 

Our \Ts$_{,HISA,peak}$ temperatures are in agreement with the results found by \cite{Roshi2011} who derived an electron temperature between 40 - 60 K for the R-C cloud from Carbon RRRLs. If we assume that the CII gas is co-located with the \HI, then the electron temperature should be the same as the spin temperature. We find very similar temperatures ($\sim$50 K) close to the region where the CRRL lines were measured (see Fig.~\ref{fig:Ts} and \ref{fig:CO_map}). 

\section{Summary and Conclusions}
\label{sec:Summary}

HISA clouds are ideal places to study the cold neutral component of the ISM. They are common across the Galaxy, abundantly detected in large-scale \HI\ surveys such as GASS \citep{McClure-Griffiths2006}, the Canadian Galactic Plain Survey (CGPS; \citealt{Gibson2000, Gibson2005}), and SGPS  \citep{Kavars2003, Kavars2005}. These surveys provide data to study the location and spatial structure of these clouds. However, determining the physical properties of the cold gas, such as the optical depth, spin temperature and column density, is uncertain. Here we present a method to calibrate the HISA spin temperature by directly measuring the optical depth with \HI\ absorption measurements against continuum sources. 

We present new high-resolution \HI\ absorption observations with the ATCA towards 46  LOS across the Riegel-Crutcher cloud. We use the AGD algorithm \citep{Lindner2015} to decompose the strongly blended \HI\ absorption spectra towards the R-C cloud and find overall 300 Gaussian components, from which 67 are associated with the R-C cloud ($0 < v_{LSR} < 10$ \kms, FWHM $< 10$ \kms). We find high optical depths ($\tau > 1$) throughout the cloud. 

We calculate spin temperatures for the cloud for the individual Gaussian components (\Ts$_{,HISA}$) and for the peak of the HISA feature (\Ts$_{,HISA,peak}$) by using a four-component ISM model. For \Ts$_{,HISA,peak}$ we find temperatures ranging from 20 - 80 K with a median of 48 $\pm 7$ K for GASS and 64 $\pm 9$ K for SGPS GC. These median temperatures are higher compared to the estimated 40 K in previous studies (\citealt{Crutcher1984, Montgomery1995, McClure-Griffiths2006}).  However, our estimates should be more accurate given the direct measurements of the optical depth. We find a clear temperature gradient across the R-C cloud. The spin temperature is decreasing towards the high-latitude base of the R-C cloud. 

We compared our spin temperature results from the Gaussian decomposition and the \Ts\ calculations to results from the Millennium Survey \citep{Heiles2003_1} and the Perseus molecular cloud \citep{Stanimirovic2014}. The R-C cloud seems to have more high optical depth components, with slightly broader line widths. We also found that if we take the sensitivity limitations of our survey into account, then the spin temperature distribution of the three samples is very similar.  

There is evidence for associated molecular gas with the R-C cloud. \cite{Crutcher1973} found OH absorption in the region and \cite{Dame2001} detected $^{12}$CO emission across the high latitude part of the cloud. We add to the detected molecular gas with four new OH absorption lines. 

We are planing to use the method presented in this paper in future Galactic \HI\ absorption surveys and map the \Ts\ distribution across a large range of \HI\ self-absorption clouds. One such an upcoming survey is the Galactic Australian SKA Pathfinder Survey (GASKAP; \citealt{GASKAP}) with the Australian Square Kilometre Array Pathfinder (ASKAP). GASKAP will map the plane of the Milky Way and the Magellanic system in \HI\ and OH lines in high-spatial and high-frequency resolution. It will provide \HI\ absorption measurements against hundreds of continuum sources. 

\section*{Acknowledgements}

We would like to thank the anonymous referee for the useful comments and suggestions, which helped us to greatly improve the paper. The Australia Telescope Compact Array and the Parkes telescope are part of the Australia Telescope National Facility which is funded by the Commonwealth of Australia for operation as a National Facility managed by CSIRO. N.M.-G. acknowledges the support of the Australian Research Council through Future Fellowship FT150100024. We thank Patrick Jacobson for his help in the original planning for this project. This research made use of {\sc Astropy}, a community-developed core {\sc Python} package for Astronomy \citep{Astropy}.  http://www.astropy.org




\bibliographystyle{mnras}
\bibliography{RC-cloud_paper} 


\FloatBarrier
\appendix
\section{PKS 1934-638}
\label{Appendix:1934}

\begin{figure}
	\centering
	\includegraphics[width=84mm]{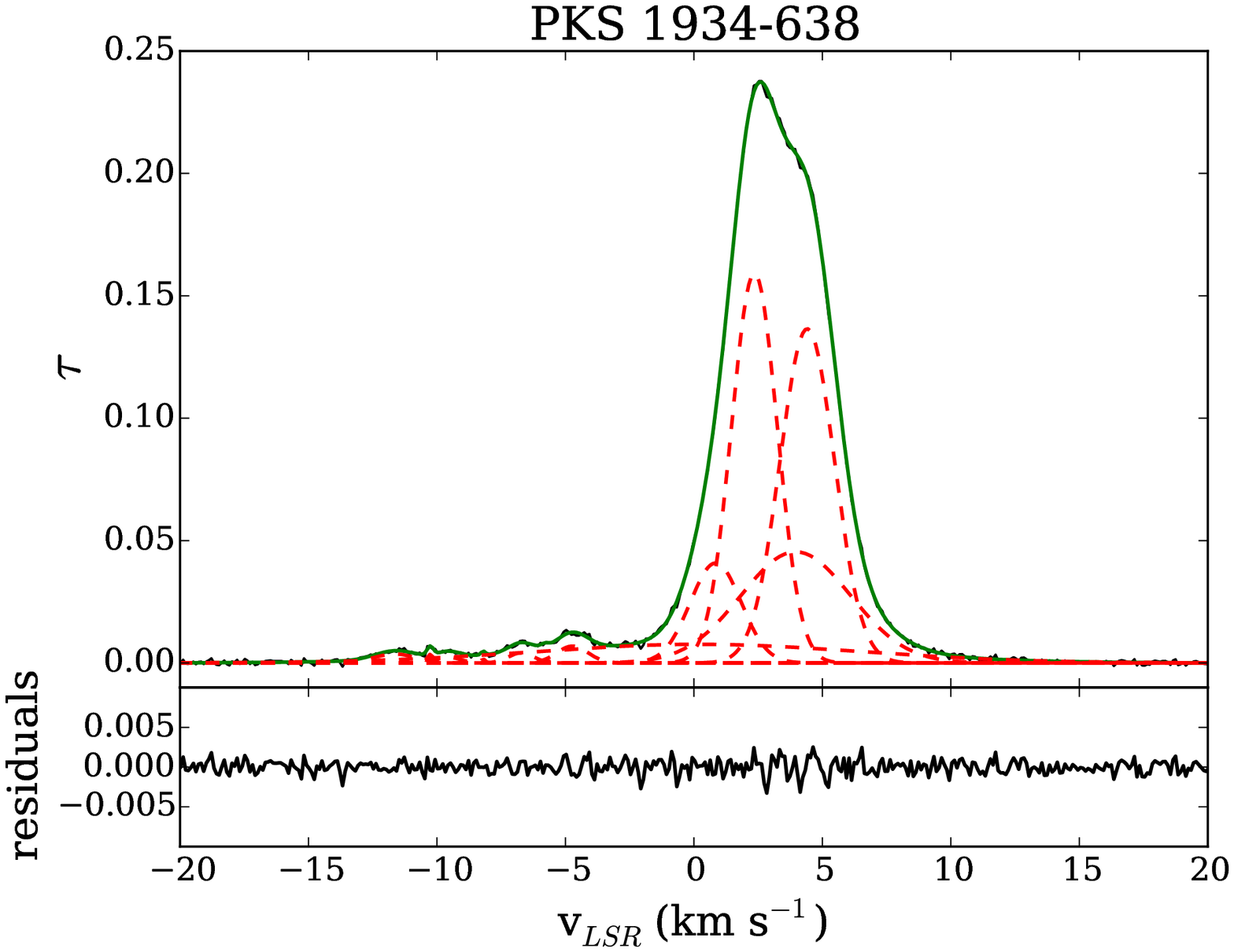}
	\caption{Optical depth spectrum of PKS 1934-638 (total integration time: 15.44 hours, $\sigma_{\tau} = 8 \times 10^{-4}$ per 0.1 \kms\ per channel). Red dashed lines show the Gaussian components found by {\sc GaussPy} and the green line shows the fitted model to the data.}
	\label{fig:1934-638}
\end{figure}

PKS 1934-638 is the main source used for ATCA 16 cm bandpass calibration. However PKS 1934-638 has a strong absorption line between 0 and 10 \kms, which is exactly the velocity range where the R-C cloud is located. In addition to the well known absorption line, there are also absorption features between -15 to 0 \kms. Fig.~\ref{fig:1934-638} shows the optical depth spectrum of PKS 1934-638 (total integration time: 15.44 hours, $\sigma_{\tau} = 8 \times 10^{-4}$ per 0.1 \kms\ per channel). We decomposed the optical depth spectra of PKS 1934-638 with {\sc GaussPy} following \cite{Lindner2015} ($\alpha_{1} = 1.12$ and $\alpha_{2} = 2.73$).  We present the results of the decomposition in Tab.~\ref{tab:1934_parameter} and Fig.~\ref{fig:1934-638}. 

\begin{table}
	\centering
	\caption{Line parameters from the Gaussian decomposition for PKS 1934-638 ($l, b = 332.7461^{\circ}, -29.3893 ^{\circ}$) and calculated T$_{k,max}$.  }
	\label{tab:1934_parameter}
	\begin{tabular}{c c c c}
		\hline
		$\tau$ & $v_{LSR}$ & FWHM & T$_{k,max}$   \\
		& [\kms] & [\kms] & [K]  \\
		\hline
		0.16 $\pm$ 0.014 & 2.36 $\pm$ 0.07 & 2.1 $\pm$ 0.09 & 96.53 $\pm$ 0.17 \\
		0.137 $\pm$ 0.016 & 4.41 $\pm$ 0.06 & 2.49 $\pm$ 0.09 & 135.27 $\pm$ 0.19 \\
		0.044 $\pm$ 0.019 & 4.05 $\pm$ 0.75 & 4.96 $\pm$ 0.48 & 537.1 $\pm$ 4.97 \\
		0.043 $\pm$ 0.025 & 0.87 $\pm$ 0.44 & 2.36 $\pm$ 0.47 & 122.12 $\pm$ 4.89 \\
		0.008 $\pm$ 0.001 & 0.37 $\pm$ 0.39 & 14.52 $\pm$ 0.69 & 4610.19 $\pm$ 10.36 \\
		0.007 $\pm$ 0.001 & -4.73 $\pm$ 0.04 & 1.42 $\pm$ 0.14 & 44.06 $\pm$ 0.44 \\
		0.004 $\pm$ 0.001 & -6.74 $\pm$ 0.06 & 1.18 $\pm$ 0.19 & 30.38 $\pm$ 0.77 \\
		0.004 $\pm$ 0.001 & -11.64 $\pm$ 0.08 & 1.9 $\pm$ 0.23 & 79.35 $\pm$ 1.2 \\
		0.003 $\pm$ 0.001 & -10.26 $\pm$ 0.03 & 0.28 $\pm$ 0.08 & 1.7 $\pm$ 0.15 \\
		0.002 $\pm$ 0.001 & -8.2 $\pm$ 0.05 & 0.21 $\pm$ 0.13 & 1.01 $\pm$ 0.35 \\
		0.001 $\pm$ 0.001 & -5.8 $\pm$ 0.1 & 0.35 $\pm$ 0.26 & 2.72 $\pm$ 1.52 \\
		\hline
	\end{tabular}
\end{table}				

\section{Diffuse continuum emission}
\label{Appendix:Tc}

\begin{figure}
	\centering
	\includegraphics[width=84mm]{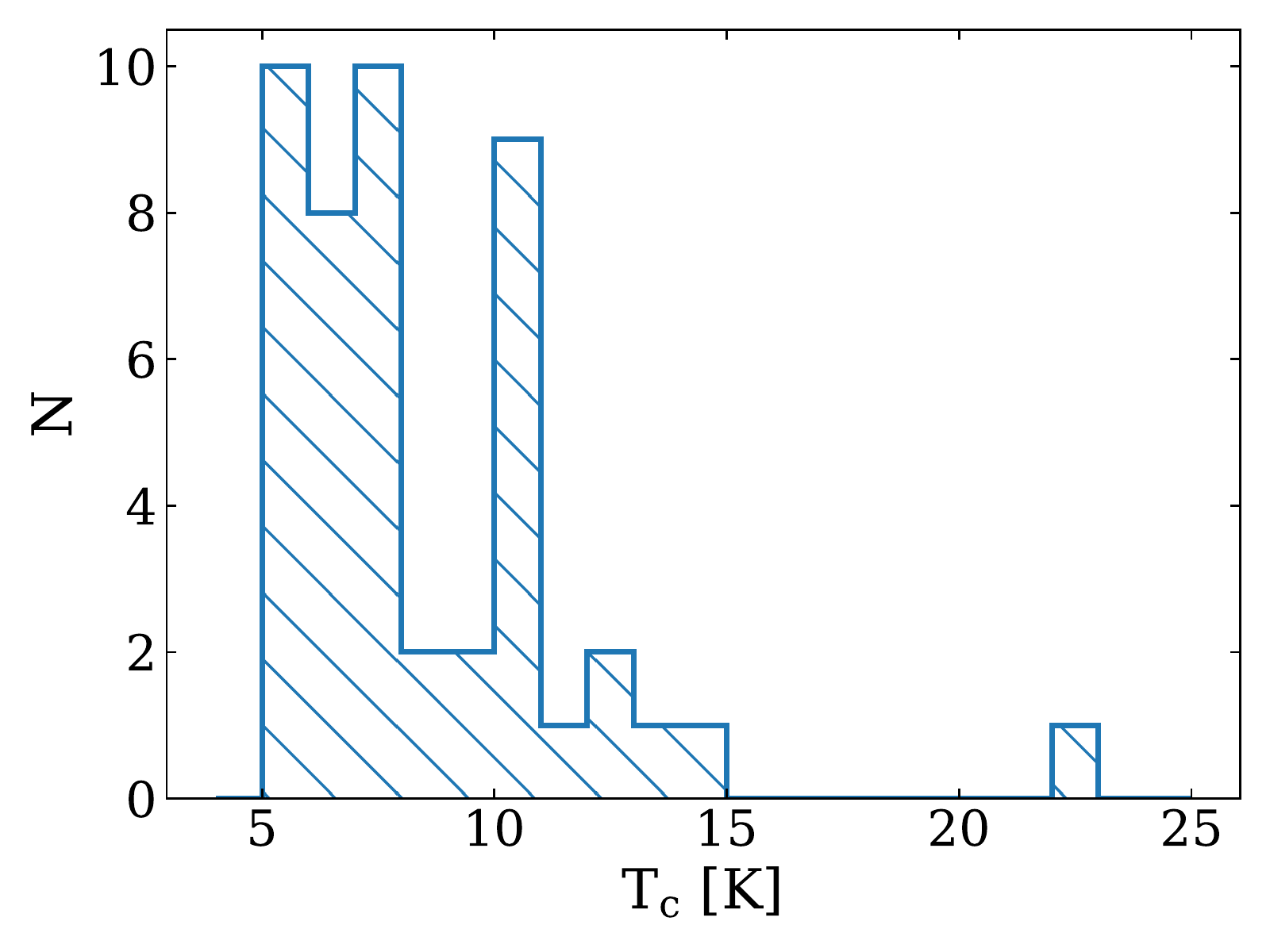}
	\caption{Histogram of the $T_{c}$ distribution in the LOS of our sources.}
	\label{fig:Tc_histogram}
\end{figure}

\begin{figure}
	\centering
	\includegraphics[width=84mm]{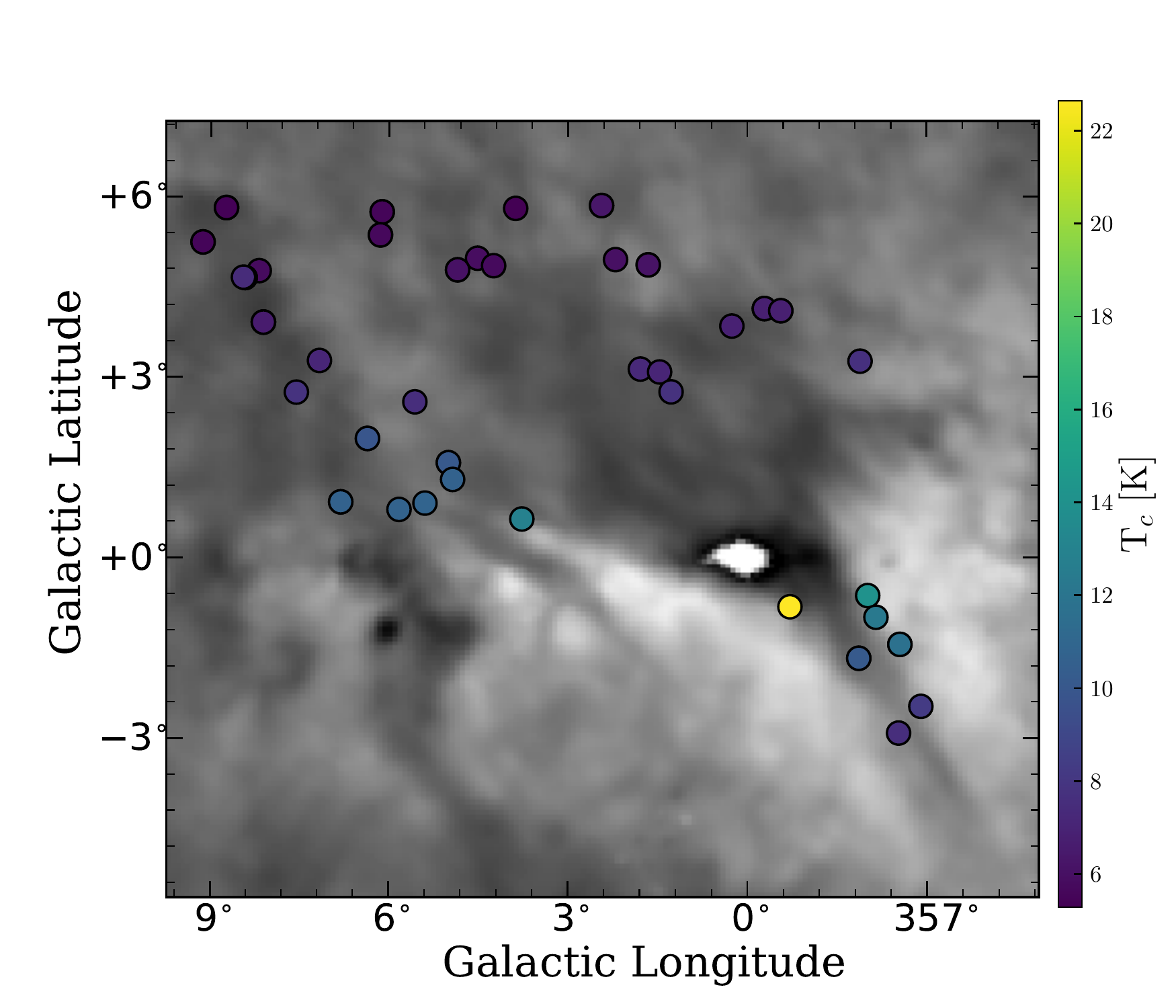}
	\caption{$T_{c}$ distribution overlayed on the GASS \HI\ intensity map ($v_{LSR}=5.2$) of the R-C cloud. $T_{c}$ decreases with Galactic latitude and all LOS with $|b| > 2^{\circ}$ have $T_{c} <$ 8K.}
	\label{fig:Tc_distribution}
\end{figure}

Fig.~\ref{fig:Tc_histogram} shows the distribution of the diffuse continuum emission ($T_{c}$) extracted from CHIPASS at the positions of our sources. Except for one source, all $T_{c}$ values range between 5 - 15 K. Fig.~\ref{fig:Tc_distribution} displays the distribution of $T_{c}$ overlayed on the GASS \HI\ intensity map. $T_{c}$ smoothly decreases with distance from the Galactic Plain and the highest $T_{c}$ (22 K) is at the closest position to the GC.

\section{Tables}
\label{Appendix:Tables}

\begin{table*}
	\centering
	\caption{Line parameters from the Gaussian decomposition in the velocity range of the R-C cloud (0 < $v_{LSR}$ < 10 \kms) and calculated T$_{k,max}$, T$_{s}$ and N(\HI).  }
	\label{tab:Ts}
	\begin{tabular}{l c c c c c c c c}
		\hline
		name & $\tau_{HISA}$ & $v_{LSR}$ & FWHM & T$_{k,max}$ & T$_{s,HISA}$ GASS & T$_{s,HISA}$ SGPS & N(\HI) GASS & N(\HI) SGPS   \\
		& & [\kms] & [\kms] & [K] & [K] & [K]  & [10$^{20}$ cm$^{-2}$]  & [10$^{20}$ cm$^{-2}$]  \\
		\hline
NVSS J172920-234535   & 0.29 $\pm$ 0.16 & 3.2 $\pm$ 0.2 & 2.4 $\pm$ 0.2 & 127 $\pm$ 1 & - $\pm$ - & - $\pm$ - & - $\pm$ - & - $\pm$ - \\
NVSS J172920-234535   & 3.38 $\pm$ 0.31 & 6.7 $\pm$ 0.2 & 4.7 $\pm$ 0.2 & 485 $\pm$ 1 & 46 $\pm$ 54 & - $\pm$ - & 14.48 $\pm$ 0.05 & - $\pm$ - \\
NVSS J172920-234535   & 0.16 $\pm$ 0.02 & 0.4 $\pm$ 0.3 & 3.4 $\pm$ 0.6 & 254 $\pm$ 7 & - $\pm$ - & - $\pm$ - & - $\pm$ - & - $\pm$ - \\
NVSS J173205-242651   & 2.33 $\pm$ 0.43 & 2.2 $\pm$ 0.2 & 3.0 $\pm$ 0.2 & 192 $\pm$ 1 & 72 $\pm$ 4 & 67 $\pm$ 1 & 9.75 $\pm$ 0.11 & 9.08 $\pm$ 0.1 \\
NVSS J173753-254642   & 2.73 $\pm$ 0.21 & 3.4 $\pm$ 0.2 & 9.8 $\pm$ 0.3 & 2098 $\pm$ 2 & 35 $\pm$ 24 & 41 $\pm$ 30 & 18.46 $\pm$ 0.04 & 21.38 $\pm$ 0.04 \\
NVSS J173107-245703   & 2.77 $\pm$ 0.26 & 2.9 $\pm$ 0.2 & 9.2 $\pm$ 0.2 & 1847 $\pm$ 1 & 55 $\pm$ 1 & 52 $\pm$ 1 & 27.62 $\pm$ 0.05 & 25.88 $\pm$ 0.05 \\
NVSS J173107-245703   & 0.12 $\pm$ 0.03 & 7.3 $\pm$ 0.2 & 1.9 $\pm$ 0.5 & 80 $\pm$ 6 & 61 $\pm$ 7 & 15 $\pm$ 7 & 0.28 $\pm$ 0.02 & 0.07 $\pm$ 0.01 \\
NVSS J173718-260426   & 2.71 $\pm$ 0.44 & 4.9 $\pm$ 0.7 & 8.5 $\pm$ 0.6 & 1564 $\pm$ 8 & 33 $\pm$ 16 & 40 $\pm$ 1 & 14.87 $\pm$ 0.17 & 18.14 $\pm$ 0.2 \\
NVSS J173252-223511   & 6.88 $\pm$ 0.75 & 4.3 $\pm$ 0.2 & 4.4 $\pm$ 0.2 & 416 $\pm$ 0 & 48 $\pm$ 7 & - $\pm$ - & 28.37 $\pm$ 0.1 & - $\pm$ - \\
NVSS J173722-223000   & 0.41 $\pm$ 0.04 & 4.4 $\pm$ 0.2 & 3.3 $\pm$ 0.3 & 245 $\pm$ 2 & 10 $\pm$ 1 & - $\pm$ - & 0.27 $\pm$ 0.1 & - $\pm$ - \\
NVSS J174618-193006   & 0.34 $\pm$ 0.12 & 1.0 $\pm$ 0.2 & 1.1 $\pm$ 0.3 & 27 $\pm$ 2 & 65 $\pm$ 1 & - $\pm$ - & 0.48 $\pm$ 0.05 & - $\pm$ - \\
NVSS J173713-224734   & 2.02 $\pm$ 0.35 & 4.9 $\pm$ 0.2 & 6.6 $\pm$ 0.3 & 966 $\pm$ 2 & 39 $\pm$ 14 & 41 $\pm$ 36 & 10.29 $\pm$ 0.08 & 10.72 $\pm$ 0.08 \\
NVSS J173850-221918   & 1.74 $\pm$ 0.33 & 2.5 $\pm$ 0.4 & 5.1 $\pm$ 0.4 & 560 $\pm$ 3 & 67 $\pm$ 4 & 85 $\pm$ 2 & 11.57 $\pm$ 0.17 & 14.52 $\pm$ 0.21 \\
NVSS J173850-221918   & 0.82 $\pm$ 0.21 & 7.7 $\pm$ 0.7 & 5.2 $\pm$ 0.6 & 581 $\pm$ 8 & 62 $\pm$ 4 & 62 $\pm$ 2 & 5.1 $\pm$ 0.16 & 5.11 $\pm$ 0.16 \\
NVSS J174713-192135   & 0.84 $\pm$ 0.69 & 4.8 $\pm$ 0.2 & 1.9 $\pm$ 0.2 & 78 $\pm$ 0 & 22 $\pm$ 13 & - $\pm$ - & 0.7 $\pm$ 0.04 & - $\pm$ - \\
NVSS J174713-192135   & 0.52 $\pm$ 0.36 & 7.6 $\pm$ 0.4 & 2.8 $\pm$ 0.6 & 168 $\pm$ 8 & - $\pm$ - & - $\pm$ - & - $\pm$ - & - $\pm$ - \\
NVSS J174713-192135   & 0.33 $\pm$ 0.13 & 3.9 $\pm$ 2.2 & 7.8 $\pm$ 2.3 & 1328 $\pm$ 115 & 36 $\pm$ 13 & - $\pm$ - & 1.84 $\pm$ 0.22 & - $\pm$ - \\
NVSS J172829-284610   & 1.55 $\pm$ 0.01 & 6.6 $\pm$ 0.2 & 6.8 $\pm$ 0.2 & 1013 $\pm$ 0 & 62 $\pm$ 1 & 62 $\pm$ 6 & 12.67 $\pm$ 0.1 & 12.68 $\pm$ 0.01 \\
NVSS J172908-265751   & 0.57 $\pm$ 0.3 & 8.5 $\pm$ 0.2 & 2.1 $\pm$ 0.3 & 98 $\pm$ 2 & 71 $\pm$ 3 & 64 $\pm$ 15 & 1.67 $\pm$ 0.14 & 1.5 $\pm$ 0.12 \\
NVSS J172908-265751   & 0.57 $\pm$ 0.17 & 0.9 $\pm$ 0.2 & 1.9 $\pm$ 0.3 & 79 $\pm$ 2 & 84 $\pm$ 5 & 89 $\pm$ 15 & 1.78 $\pm$ 0.08 & 1.88 $\pm$ 0.09 \\
NVSS J172908-265751   & 1.34 $\pm$ 0.16 & 4.9 $\pm$ 0.3 & 4.3 $\pm$ 1.1 & 406 $\pm$ 24 & 36 $\pm$ 3 & 14 $\pm$ 15 & 4.08 $\pm$ 0.12 & 1.64 $\pm$ 0.05 \\
NVSS J173203-285516   & 0.46 $\pm$ 0.06 & 5.6 $\pm$ 0.2 & 3.3 $\pm$ 0.4 & 242 $\pm$ 4 & - $\pm$ - & - $\pm$ - & - $\pm$ - & - $\pm$ - \\
NVSS J172836-271236   & 4.77 $\pm$ 0.26 & 3.3 $\pm$ 0.2 & 3.0 $\pm$ 0.2 & 191 $\pm$ 0 & 38 $\pm$ 4 & 45 $\pm$ 4 & 10.41 $\pm$ 0.01 & 12.43 $\pm$ 0.02 \\
NVSS J172836-271236   & 0.82 $\pm$ 0.02 & 7.9 $\pm$ 0.2 & 2.8 $\pm$ 0.2 & 169 $\pm$ 0 & 65 $\pm$ 4 & 76 $\pm$ 2 & 2.87 $\pm$ 0.01 & 3.39 $\pm$ 0.01 \\
NVSS J173133-264015   & 2.43 $\pm$ 0.4 & 6.3 $\pm$ 0.4 & 5.5 $\pm$ 0.3 & 650 $\pm$ 2 & 41 $\pm$ 1 & 52 $\pm$ 3 & 10.52 $\pm$ 0.09 & 13.44 $\pm$ 0.11 \\
NVSS J173806-262443   & 2.13 $\pm$ 0.26 & 3.5 $\pm$ 0.3 & 9.8 $\pm$ 0.4 & 2100 $\pm$ 4 & 34 $\pm$ 21 & 41 $\pm$ 26 & 13.84 $\pm$ 0.07 & 16.99 $\pm$ 0.09 \\
NVSS J174832-225211   & 1.47 $\pm$ 0.14 & 6.2 $\pm$ 0.2 & 8.4 $\pm$ 0.5 & 1525 $\pm$ 5 & 54 $\pm$ 1 & - $\pm$ - & 12.98 $\pm$ 0.07 & - $\pm$ - \\
NVSS J174931-210847   & 2.36 $\pm$ 0.01 & 7.3 $\pm$ 0.2 & 6.1 $\pm$ 0.2 & 826 $\pm$ 0 & 48 $\pm$ 2 & - $\pm$ - & 13.59 $\pm$ 0.01 & - $\pm$ - \\
NVSS J175233-223012   & 0.17 $\pm$ 0.01 & 0.1 $\pm$ 0.2 & 5.3 $\pm$ 0.2 & 606 $\pm$ 0 & 39 $\pm$ 41 & - $\pm$ - & 0.7 $\pm$ 0.01 & - $\pm$ - \\
NVSS J175233-223012   & 3.2 $\pm$ 0.01 & 5.5 $\pm$ 0.2 & 4.7 $\pm$ 0.2 & 493 $\pm$ 0 & 48 $\pm$ 39 & - $\pm$ - & 14.33 $\pm$ 0.01 & - $\pm$ - \\
NVSS J174915-200033   & 1.6 $\pm$ 0.46 & 3.5 $\pm$ 0.2 & 1.9 $\pm$ 0.2 & 77 $\pm$ 1 & 55 $\pm$ 1 & - $\pm$ - & 3.25 $\pm$ 0.12 & - $\pm$ - \\
NVSS J174915-200033   & 1.46 $\pm$ 0.24 & 0.1 $\pm$ 0.2 & 1.1 $\pm$ 0.2 & 29 $\pm$ 0 & 53 $\pm$ 1 & - $\pm$ - & 1.72 $\pm$ 0.03 & - $\pm$ - \\
NVSS J174915-200033   & 4.29 $\pm$ 1.1 & 7.4 $\pm$ 0.3 & 3.3 $\pm$ 0.4 & 232 $\pm$ 3 & 41 $\pm$ 3 & - $\pm$ - & 11.17 $\pm$ 0.32 & - $\pm$ - \\
NVSS J175218-210508   & 1.79 $\pm$ 0.11 & 6.4 $\pm$ 0.2 & 7.0 $\pm$ 0.4 & 1073 $\pm$ 3 & 44 $\pm$ 1 & - $\pm$ - & 10.74 $\pm$ 0.03 & - $\pm$ - \\
NVSS J175104-235215   & 1.45 $\pm$ 0.01 & 7.9 $\pm$ 0.2 & 4.7 $\pm$ 0.2 & 490 $\pm$ 0 & 56 $\pm$ 1 & 76 $\pm$ 14 & 7.49 $\pm$ 0.01 & 10.14 $\pm$ 0.01 \\
NVSS J175104-235215   & 1.1 $\pm$ 0.01 & 3.4 $\pm$ 0.2 & 2.6 $\pm$ 0.2 & 143 $\pm$ 0 & 77 $\pm$ 1 & 86 $\pm$ 14 & 4.25 $\pm$ 0.01 & 4.72 $\pm$ 0.01\\
NVSS J175157-240425   & 0.27 $\pm$ 0.2 & 8.8 $\pm$ 0.2 & 1.4 $\pm$ 0.8 & 42 $\pm$ 13 & - $\pm$ - & - $\pm$ - & - $\pm$ - & - $\pm$ - \\
NVSS J175157-240425   & 3.65 $\pm$ 0.72 & 4.7 $\pm$ 0.4 & 5.6 $\pm$ 0.4 & 674 $\pm$ 4 & 70 $\pm$ 2 & 64 $\pm$ 4 & 27.87 $\pm$ 0.45 & 25.5 $\pm$ 0.41 \\
NVSS J175548-233322   & 6.28 $\pm$ 1.22 & 6.6 $\pm$ 0.2 & 3.7 $\pm$ 0.3 & 299 $\pm$ 2 & 42 $\pm$ 13 & - $\pm$ - & 19.3 $\pm$ 0.32 & - $\pm$ - \\
NVSS J175151-252359   & 1.49 $\pm$ 0.01 & 0.1 $\pm$ 0.2 & 8.5 $\pm$ 0.2 & 1562 $\pm$ 0 & 118 $\pm$ 6 & 126 $\pm$ 21 & 29.13 $\pm$ 0.01 & 30.9 $\pm$ 0.01 \\
NVSS J175427-235235   & 4.53 $\pm$ 0.73 & 7.2 $\pm$ 0.2 & 4.8 $\pm$ 0.3 & 498 $\pm$ 2 & 46 $\pm$ 5 & - $\pm$ - & 19.34 $\pm$ 0.19 & - $\pm$ - \\
NVSS J175727-223901   & 0.45 $\pm$ 0.4 & 8.3 $\pm$ 0.2 & 1.8 $\pm$ 0.5 & 67 $\pm$ 6 & 8 $\pm$ 2 & - $\pm$ - & 0.13 $\pm$ 0.04 & - $\pm$ - \\
NVSS J175727-223901   & 2.9 $\pm$ 0.5 & 5.1 $\pm$ 0.5 & 4.4 $\pm$ 0.6 & 427 $\pm$ 8 & 50 $\pm$ 2 & - $\pm$ - & 12.5 $\pm$ 0.29 & - $\pm$ - \\
NVSS J174709-295802   & 6.16 $\pm$ 2.41 & 4.7 $\pm$ 0.2 & 6.3 $\pm$ 1.0 & 877 $\pm$ 22 & 98 $\pm$ 7 & 75 $\pm$ 6 & 74.64 $\pm$ 4.67 & 57.2 $\pm$ 3.58 \\
NVSS J174317-305819   & 1.46 $\pm$ 0.01 & 5.6 $\pm$ 0.2 & 7.4 $\pm$ 0.2 & 1183 $\pm$ 0 & 88 $\pm$ 3 & 100 $\pm$ 2 & 18.4 $\pm$ 0.01 & 20.93 $\pm$ 0.01 \\
NVSS J174423-311636   & 3.93 $\pm$ 0.6 & 5.4 $\pm$ 0.4 & 5.4 $\pm$ 0.3 & 638 $\pm$ 3 & 101 $\pm$ 5 & 103 $\pm$ 3 & 41.69 $\pm$ 0.41 & 42.56 $\pm$ 0.41 \\
NVSS J174513-315104   & 1.09 $\pm$ 0.37 & 3.6 $\pm$ 0.2 & 2.6 $\pm$ 0.4 & 152 $\pm$ 3 & 101 $\pm$ 1 & 127 $\pm$ 3 & 5.66 $\pm$ 0.25 & 7.13 $\pm$ 0.32 \\
NVSS J174513-315104   & 1.1 $\pm$ 0.09 & 8.6 $\pm$ 0.7 & 9.1 $\pm$ 1.2 & 1820 $\pm$ 30 & 73 $\pm$ 1 & 78 $\pm$ 3 & 14.41 $\pm$ 0.16 & 15.29 $\pm$ 0.17 \\
NVSS J174748-312315   & 1.92 $\pm$ 0.47 & 8.0 $\pm$ 0.4 & 5.0 $\pm$ 0.3 & 536 $\pm$ 2 & 70 $\pm$ 3 & 91 $\pm$ 5 & 12.99 $\pm$ 0.2 & 16.85 $\pm$ 0.26 \\
NVSS J174748-312315   & 2.08 $\pm$ 0.38 & 0.5 $\pm$ 0.7 & 6.3 $\pm$ 0.6 & 874 $\pm$ 9 & 108 $\pm$ 4 & 128 $\pm$ 7 & 27.6 $\pm$ 0.5 & 32.88 $\pm$ 0.6 \\
NVSS J174831-324102   & 0.91 $\pm$ 2.27 & 4.5 $\pm$ 3.2 & 7.1 $\pm$ 2.6 & 1103 $\pm$ 152 & 46 $\pm$ 1 & 60 $\pm$ 1 & 5.86 $\pm$ 5.44 & 7.6 $\pm$ 7.06 \\
NVSS J175114-323538   & 0.29 $\pm$ 0.27 & 9.4 $\pm$ 0.8 & 9.9 $\pm$ 0.9 & 2152 $\pm$ 19 & 100 $\pm$ 1 & 88 $\pm$ 8 & 5.6 $\pm$ 0.5 & 4.94 $\pm$ 0.44 \\
NVSS J173811-204411   & 1.53 $\pm$ 0.18 & 5.0 $\pm$ 0.2 & 2.3 $\pm$ 0.2 & 112 $\pm$ 1 & 49 $\pm$ 1 & - $\pm$ - & 3.33 $\pm$ 0.03 & - $\pm$ - \\
NVSS J173811-204411   & 2.75 $\pm$ 0.46 & 2.2 $\pm$ 0.2 & 2.1 $\pm$ 0.2 & 97 $\pm$ 0 & 51 $\pm$ 1 & - $\pm$ - & 5.74 $\pm$ 0.04 & - $\pm$ - \\
NVSS J173939-205505   & 1.69 $\pm$ 0.14 & 3.5 $\pm$ 0.2 & 3.1 $\pm$ 0.2 & 206 $\pm$ 0 & 46 $\pm$ 7 & - $\pm$ - & 4.67 $\pm$ 0.01 & - $\pm$ - \\
NVSS J173939-205505   & 0.46 $\pm$ 0.04 & 8.1 $\pm$ 0.3 & 4.4 $\pm$ 0.5 & 415 $\pm$ 6 & 41 $\pm$ 7 & - $\pm$ - & 1.64 $\pm$ 0.02 & - $\pm$ - \\
NVSS J174343-182838   & 0.16 $\pm$ 0.03 & 0.9 $\pm$ 0.2 & 1.2 $\pm$ 0.2 & 33 $\pm$ 1 & 39 $\pm$ 10 & - $\pm$ - & 0.15 $\pm$ 0.0 & - $\pm$ - \\
NVSS J174343-182838   & 0.09 $\pm$ 0.02 & 5.9 $\pm$ 0.2 & 1.1 $\pm$ 0.3 & 27 $\pm$ 2 & - $\pm$ - & - $\pm$ - & - $\pm$ - & - $\pm$ - \\
NVSS J174343-182838   & 0.1 $\pm$ 0.02 & 0.5 $\pm$ 0.9 & 7.8 $\pm$ 1.5 & 1342 $\pm$ 52 & 29 $\pm$ 11 & - $\pm$ - & 0.44 $\pm$ 0.02 & - $\pm$ - \\
NVSS J174343-182838   & 2.91 $\pm$ 0.4 & 4.7 $\pm$ 0.2 & 3.3 $\pm$ 0.2 & 245 $\pm$ 0 & 41 $\pm$ 8 & - $\pm$ - & 7.74 $\pm$ 0.03 & - $\pm$ - \\
		\hline
	\end{tabular}
\end{table*}

\begin{table*}
	\centering
	\contcaption{Line parameters from the Gaussian decomposition in the velocity range of the R-C cloud (0 < $v_{LSR}$ < 10 \kms) and calculated T$_{k,max}$, T$_{s}$ and N(\HI).  }
	\label{tab:continued}
	\begin{tabular}{l c c c c c c c c}
		\hline
		name & $\tau_{HISA}$ & $v_{LSR}$ & FWHM & T$_{k,max}$ & T$_{s,HISA}$ GASS & T$_{s,HISA}$ SGPS & N(\HI) GASS & N(\HI) SGPS   \\
		& & [\kms] & [\kms] & [K] & [K] & [K]  & [10$^{20}$ cm$^{-2}$]  & [10$^{20}$ cm$^{-2}$]  \\
		\hline
NVSS J174637-182629   & 0.56 $\pm$ 0.05 & 8.8 $\pm$ 0.2 & 3.6 $\pm$ 0.4 & 278 $\pm$ 4 & 55 $\pm$ 3 & - $\pm$ - & 2.14 $\pm$ 0.02 & - $\pm$ - \\
NVSS J174637-182629   & 1.2 $\pm$ 0.07 & 3.5 $\pm$ 0.2 & 3.2 $\pm$ 0.2 & 219 $\pm$ 1 & 43 $\pm$ 1 & - $\pm$ - & 3.16 $\pm$ 0.01 & - $\pm$ - \\
NVSS J174716-191954   & 3.07 $\pm$ 0.25 & 5.0 $\pm$ 0.2 & 3.6 $\pm$ 0.2 & 278 $\pm$ 0 & 35 $\pm$ 16 & - $\pm$ - & 7.51 $\pm$ 0.02 & - $\pm$ - \\
NVSS J174716-191954   & 0.33 $\pm$ 0.12 & 7.3 $\pm$ 0.2 & 1.6 $\pm$ 0.2 & 58 $\pm$ 1 & - $\pm$ - & - $\pm$ - & - $\pm$ - & - $\pm$ - \\
		\hline
	\end{tabular}
\end{table*}

\begin{table*}
	\centering
	\caption{\Ts\ calculated at the minimum point ($v_{min}$, T$_{min}$) of the HISA feature in the \HI\ emission spectra for both GASS and SGPS data.}
	\label{tab:Ts_ridge_full}
	\begin{tabular}{l c c c c c c c c}
		\hline
		name & l & b & v$_{min}$ (GASS) & T$_{s,max}$  & T$_{s,HISA,peak}$  & v$_{min}$ (SGPS) & T$_{s,max}$  & T$_{s,HISA,peak}$   \\
		& [$^{\circ}$] & [$^{\circ}$] &  [\kms] & (GASS) [K] & (GASS) [K] & [\kms]  & (SGPS) [K]  & (SGPS)  [K]  \\
		\hline
NVSS J172829-284610   & -1.8738 & 3.2555 & 4.4 & 55.11 & 51 $\pm$ 1 & 4.9 & 66.06 & 57 $\pm$ 7 \\
NVSS J172836-271236   & -0.5563 & 4.0941 & 4.4 & 38.65 & 44 $\pm$ 1 & 4.1 & 44.05 & 49 $\pm$ 2 \\
NVSS J172908-265751   & -0.2845 & 4.1296 & 4.4 & 37.75 & 44 $\pm$ 3 & 4.1 & 44.66 & 53 $\pm$ 16 \\
NVSS J172920-234535   & 2.432 & 5.8467 & 5.3 & 46.58 & 50 $\pm$ 58 & - & - & - $\pm$ - \\
NVSS J173107-245703   & 1.6506 & 4.8577 & 3.6 & 46.52 & 51 $\pm$ 1 & 4.1 & 56.83 & 59 $\pm$ 1 \\
NVSS J173133-264015   & 0.2574 & 3.8392 & 5.3 & 37.96 & 41 $\pm$ 1 & 4.9 & 44.72 & 48 $\pm$ 2 \\
NVSS J173203-285516   & -1.5699 & 2.5169 & 5.3 & 42.06 & 46 $\pm$ 1 & 4.9 & 43.25 & 46 $\pm$ 3 \\
NVSS J173205-242651   & 2.1956 & 4.945 & 4.4 & 46.14 & 50 $\pm$ 4 & 4.1 & 52.35 & 56 $\pm$ 1 \\
NVSS J173252-223511   & 3.8664 & 5.7991 & 5.3 & 50.1 & 53 $\pm$ 4 & - & - & - $\pm$ - \\
NVSS J173524-251036   & 1.9877 & 3.9175 & 4.4 & 42.17 & 45 $\pm$ 10 & 4.1 & 49.03 & 37 $\pm$ 39 \\
NVSS J173620-283552   & -0.7887 & 1.9026 & 5.3 & 29.96 & 31 $\pm$ 1 & 4.9 & 34.21 & 39 $\pm$ 1 \\
NVSS J173713-224734   & 4.2289 & 4.8441 & 5.3 & 47.75 & 51 $\pm$ 15 & 5.8 & 47.34 & 48 $\pm$ 38 \\
NVSS J173718-260426   & 1.4579 & 3.0762 & 5.3 & 43.22 & 51 $\pm$ 17 & 4.9 & 50.54 & 57 $\pm$ 1 \\
NVSS J173722-223000   & 4.498 & 4.968 & 5.3 & 48.19 & 52 $\pm$ 1 & 5.8 & 52.47 & 54 $\pm$ 32 \\
NVSS J173753-254642   & 1.7778 & 3.1242 & 5.3 & 45.22 & 46 $\pm$ 25 & 4.9 & 52.52 & 57 $\pm$ 32 \\
NVSS J173806-262443   & 1.2669 & 2.7455 & 5.3 & 39.83 & 45 $\pm$ 23 & 4.9 & 45.53 & 55 $\pm$ 28 \\
NVSS J173811-204411   & 6.1012 & 5.7415 & 3.6 & 47.13 & 51 $\pm$ 1 & - & - & - $\pm$ - \\
NVSS J173850-221918   & 4.8301 & 4.7751 & 5.3 & 48.56 & 50 $\pm$ 4 & 5.8 & 57.25 & 59 $\pm$ 2 \\
NVSS J173939-205505   & 6.1259 & 5.3569 & 3.6 & 46.58 & 51 $\pm$ 5 & - & - & - $\pm$ - \\
NVSS J174202-271311   & 1.0472 & 1.5731 & 5.3 & 38.11 & 33 $\pm$ 1 & 5.8 & 41.72 & 36 $\pm$ 15 \\
NVSS J174224-203729   & 6.7168 & 4.9604 & 4.4 & 56.06 & 60 $\pm$ 2 & - & - & - $\pm$ - \\
NVSS J174317-305819   & -1.9982 & -0.6363 & 6.9 & 66.0 & 76 $\pm$ 3 & 7.4 & 82.63 & 88 $\pm$ 2 \\
NVSS J174343-182838   & 8.7203 & 5.8132 & 6.1 & 40.23 & 44 $\pm$ 8 & - & - & - $\pm$ - \\
NVSS J174351-261059   & 2.1435 & 1.7718 & 5.3 & 37.92 & 33 $\pm$ 16 & 5.8 & 45.71 & 51 $\pm$ 13 \\
NVSS J174423-311636   & -2.135 & -0.9956 & 6.9 & 77.73 & 86 $\pm$ 4 & 7.4 & 86.47 & 94 $\pm$ 2 \\
NVSS J174513-315104   & -2.533 & -1.446 & 7.7 & 71.66 & 80 $\pm$ 1 & 7.4 & 84.13 & 92 $\pm$ 3 \\
NVSS J174618-193006   & 8.1553 & 4.7619 & 5.3 & 41.91 & 45 $\pm$ 1 & - & - & - $\pm$ - \\
NVSS J174637-182629   & 9.1081 & 5.2404 & 5.3 & 43.51 & 47 $\pm$ 1 & - & - & - $\pm$ - \\
NVSS J174709-295802   & -0.7073 & -0.8216 & 5.3 & 50.32 & 73 $\pm$ 6 & 5.8 & 58.65 & 77 $\pm$ 4 \\
NVSS J174713-192135   & 8.3906 & 4.6467 & 6.9 & 37.53 & 42 $\pm$ 14 & - & - & - $\pm$ - \\
NVSS J174716-191954   & 8.4192 & 4.6536 & 6.1 & 38.29 & 43 $\pm$ 17 & - & - & - $\pm$ - \\
NVSS J174748-312315   & -1.8502 & -1.6766 & 6.1 & 55.64 & 66 $\pm$ 3 & 5.8 & 65.1 & 73 $\pm$ 5 \\
NVSS J174831-324102   & -2.8834 & -2.4753 & 4.4 & 69.29 & 74 $\pm$ 1 & 4.9 & 82.54 & 39 $\pm$ 1 \\
NVSS J174832-225211   & 5.5312 & 2.5801 & 3.6 & 43.06 & 50 $\pm$ 1 & - & - & - $\pm$ - \\
NVSS J174915-200033   & 8.0755 & 3.9058 & 5.3 & 34.23 & 40 $\pm$ 1 & - & - & - $\pm$ - \\
NVSS J174931-210847   & 7.1301 & 3.2676 & 8.6 & 42.84 & 50 $\pm$ 1 & - & - & - $\pm$ - \\
NVSS J175104-235215   & 4.9686 & 1.5693 & 5.3 & 52.62 & 56 $\pm$ 1 & 6.6 & 65.66 & 72 $\pm$ 15 \\
NVSS J175114-323538   & -2.514 & -2.9188 & 5.3 & 71.94 & 69 $\pm$ 1 & 4.9 & 88.18 & 89 $\pm$ 8 \\
NVSS J175151-252359   & 3.745 & 0.6349 & 6.1 & 54.98 & 64 $\pm$ 5 & 5.8 & 60.73 & 69 $\pm$ 23 \\
NVSS J175157-240425   & 4.8976 & 1.2915 & 5.3 & 52.75 & 64 $\pm$ 1 & 5.8 & 60.05 & 67 $\pm$ 3 \\
NVSS J175218-210508   & 7.513 & 2.7416 & 5.3 & 40.48 & 46 $\pm$ 1 & - & - & - $\pm$ - \\
NVSS J175233-223012   & 6.32 & 1.9719 & 6.9 & 49.97 & 56 $\pm$ 42 & - & - & - $\pm$ - \\
NVSS J175427-235235   & 5.358 & 0.8988 & 6.9 & 45.45 & 48 $\pm$ 4 & - & - & - $\pm$ - \\
NVSS J175526-223211   & 6.6287 & 1.3809 & 6.9 & 39.96 & 49 $\pm$ 1 & - & - & - $\pm$ - \\
NVSS J175548-233322   & 5.7906 & 0.7936 & 6.1 & 42.94 & 55 $\pm$ 13 & - & - & - $\pm$ - \\
NVSS J175727-223901   & 6.7646 & 0.9197 & 6.1 & 43.97 & 54 $\pm$ 1 & - & - & - $\pm$ - \\
\hline
\end{tabular}
\end{table*}

\section{HI absorption and emission spectra for all observed sources}
\label{Appendix:HIspectra}
	
\begin{figure*}
	\subfigure{\includegraphics[width=8.5cm]{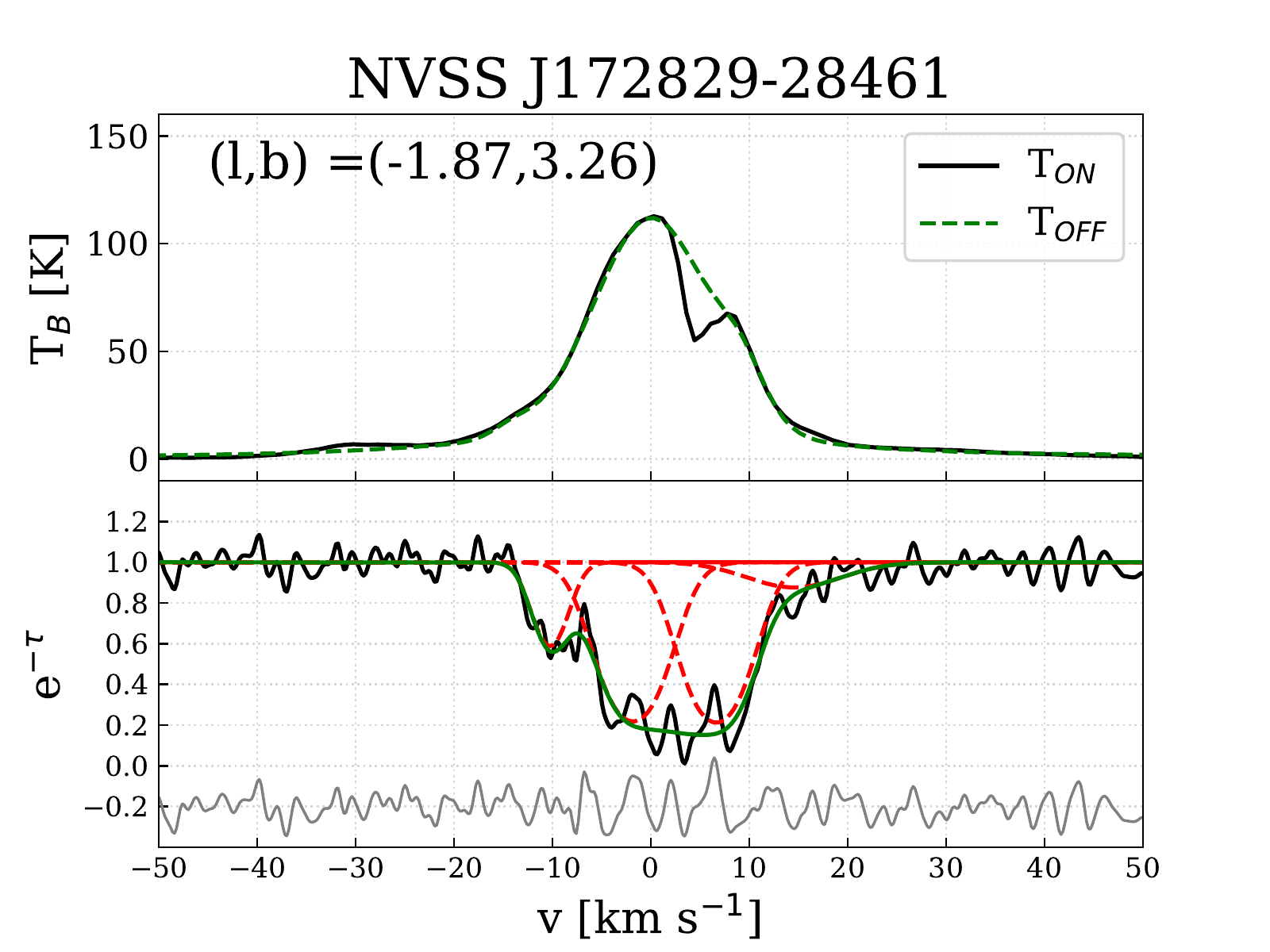}}
	\hfill
	\subfigure{\includegraphics[width=8.5cm]{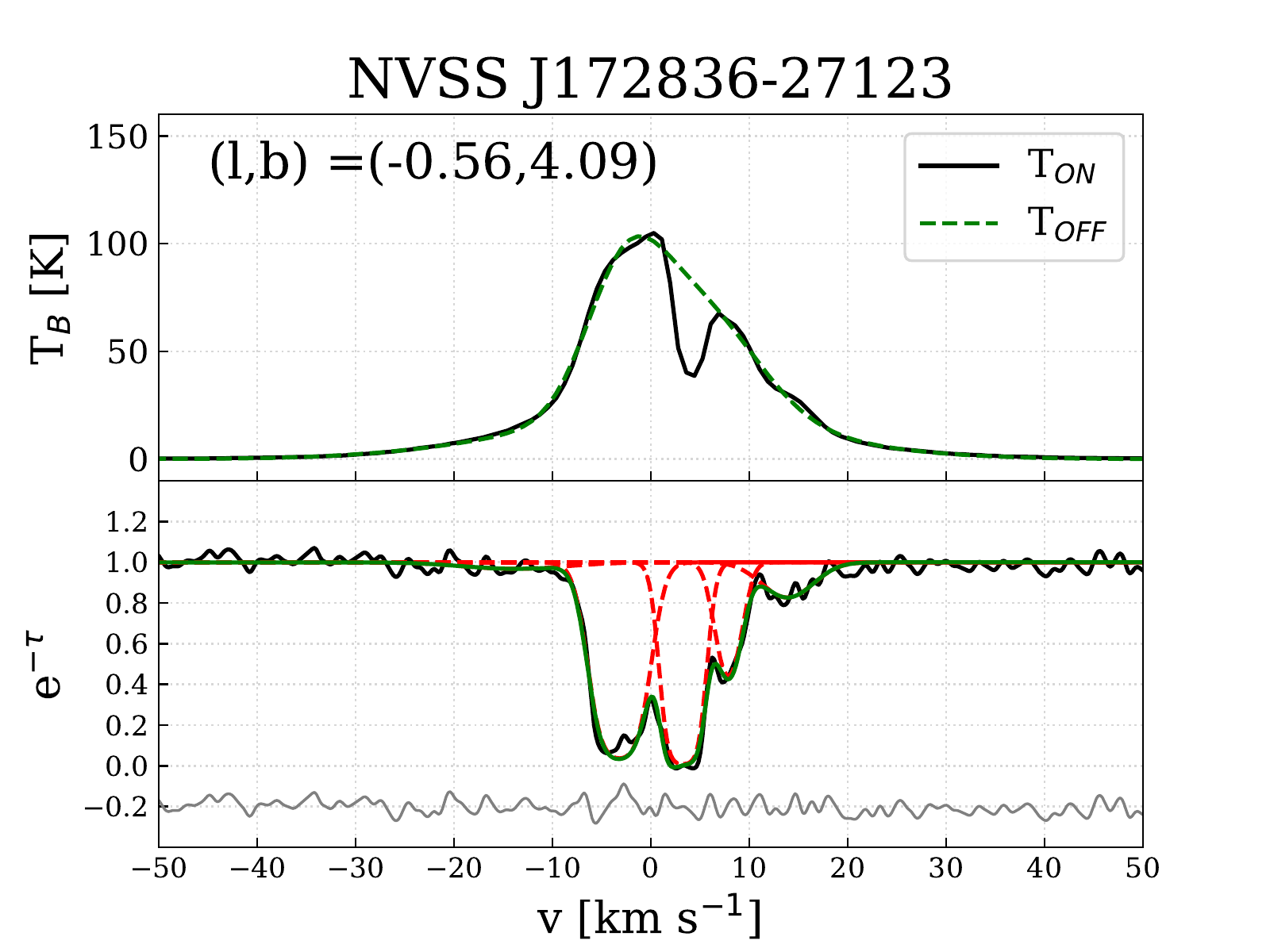}}
	\hfill
	\subfigure{\includegraphics[width=8.5cm]{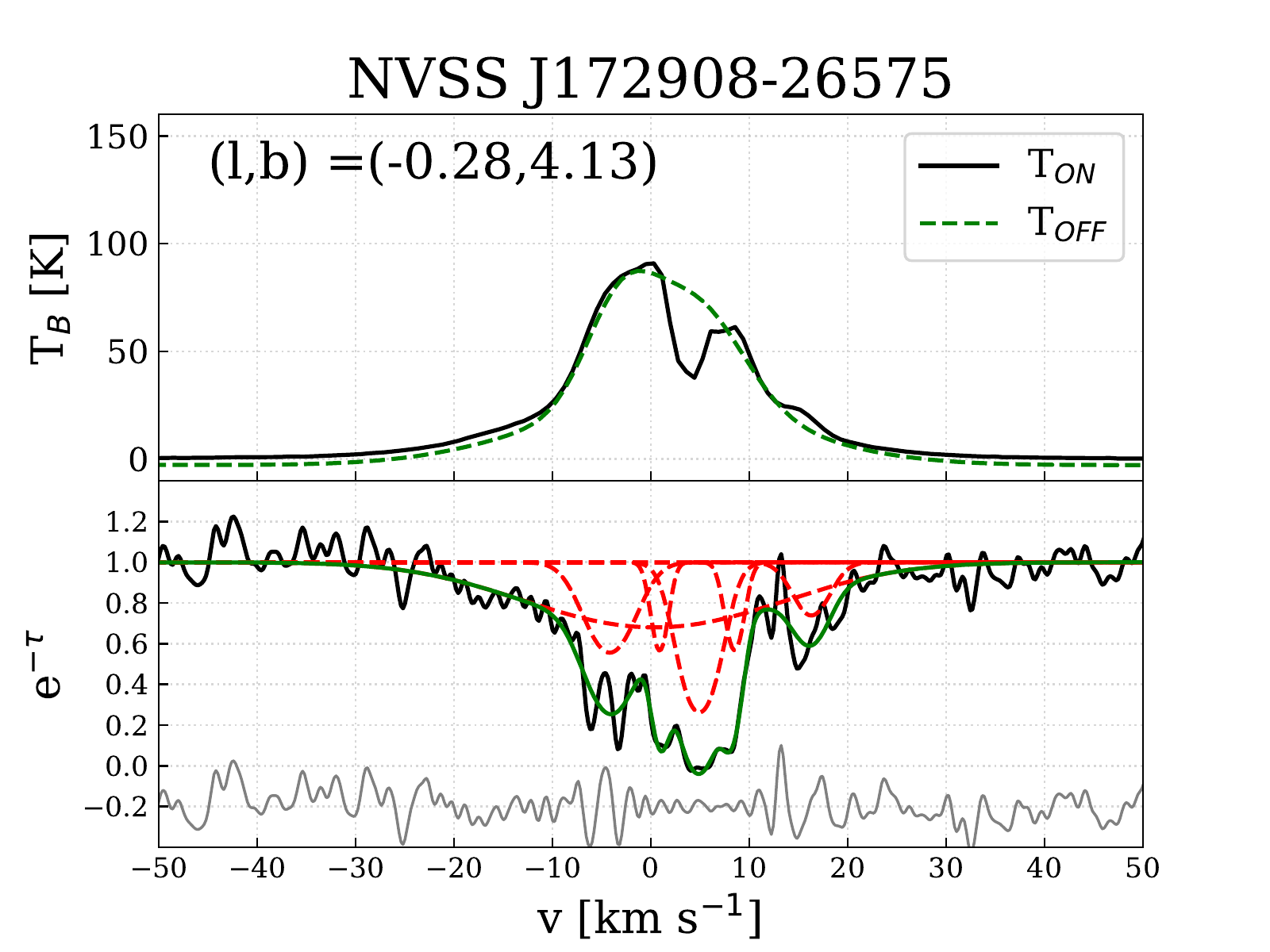}}
	\hfill
	\subfigure{\includegraphics[width=8.5cm]{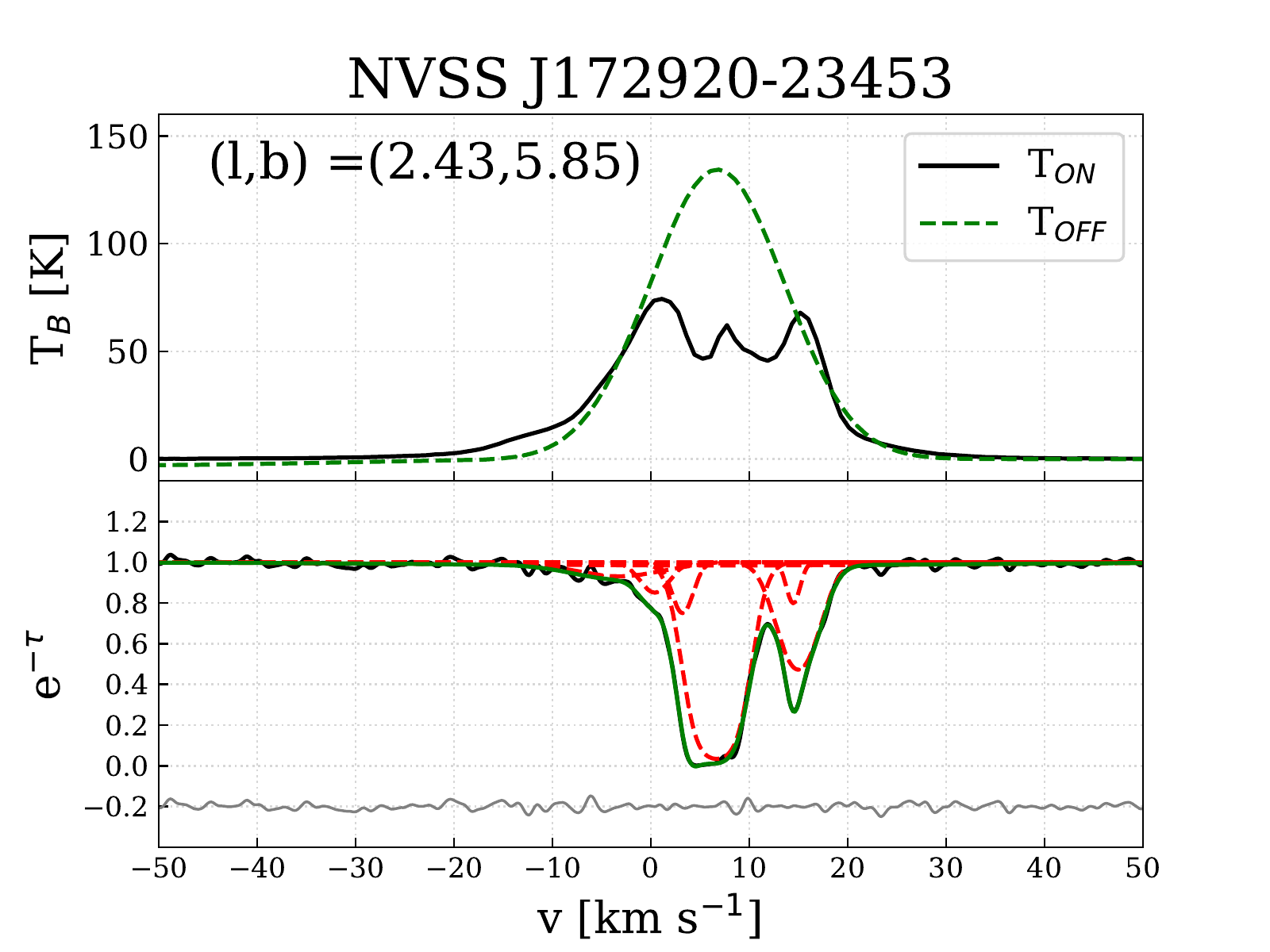}}
	\hfill
	\subfigure{\includegraphics[width=8.5cm]{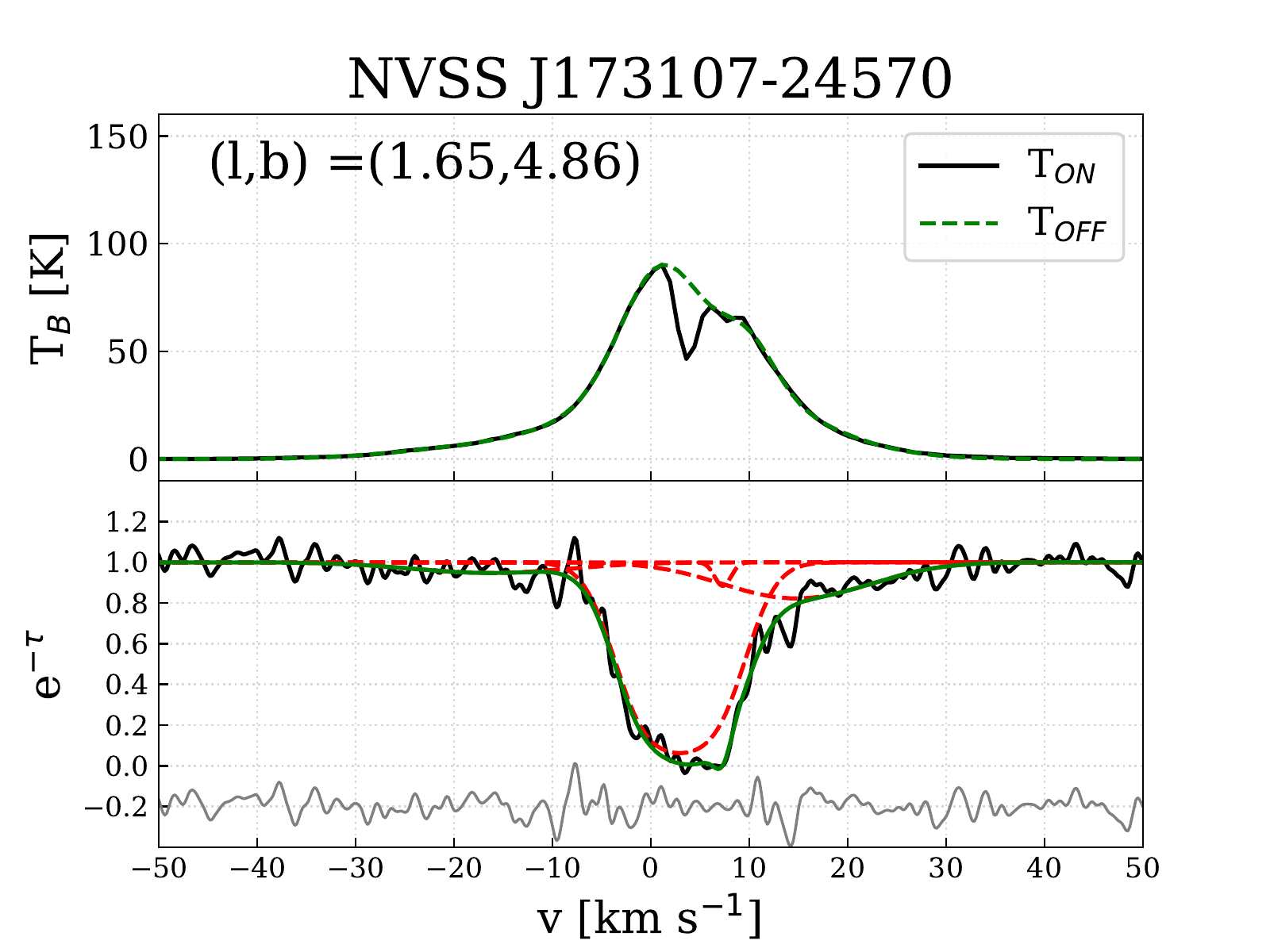}}
	\hfill
	\subfigure{\includegraphics[width=8.5cm]{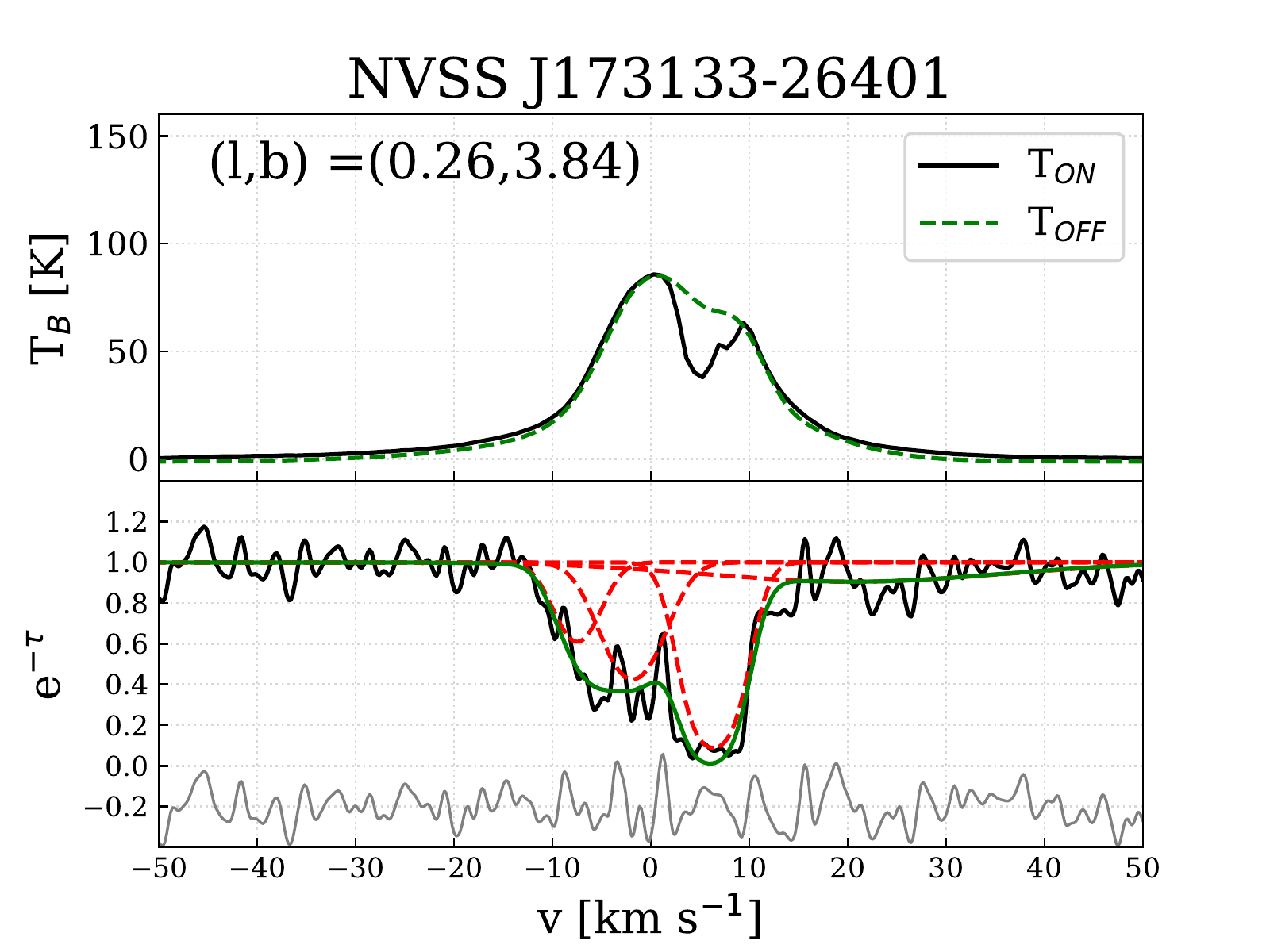}}
	\hfill
	\caption{Top panel: the solid black line shows the emission spectra ($T_{ON}$) extracted from GASS and the dashed line shows the modelled $T_{OFF}$ spectra. Bottom panel: the black line shows the ATCA absorption spectra, the green line is the fitted Gaussian model and the red dashed lines are the individual Gaussian components of the fit. The grey line is the residual from the absorption model shifted to -0.2.}
	\label{fig:spectra1}
\end{figure*}

\begin{figure*}
\subfigure{\includegraphics[width=8.5cm]{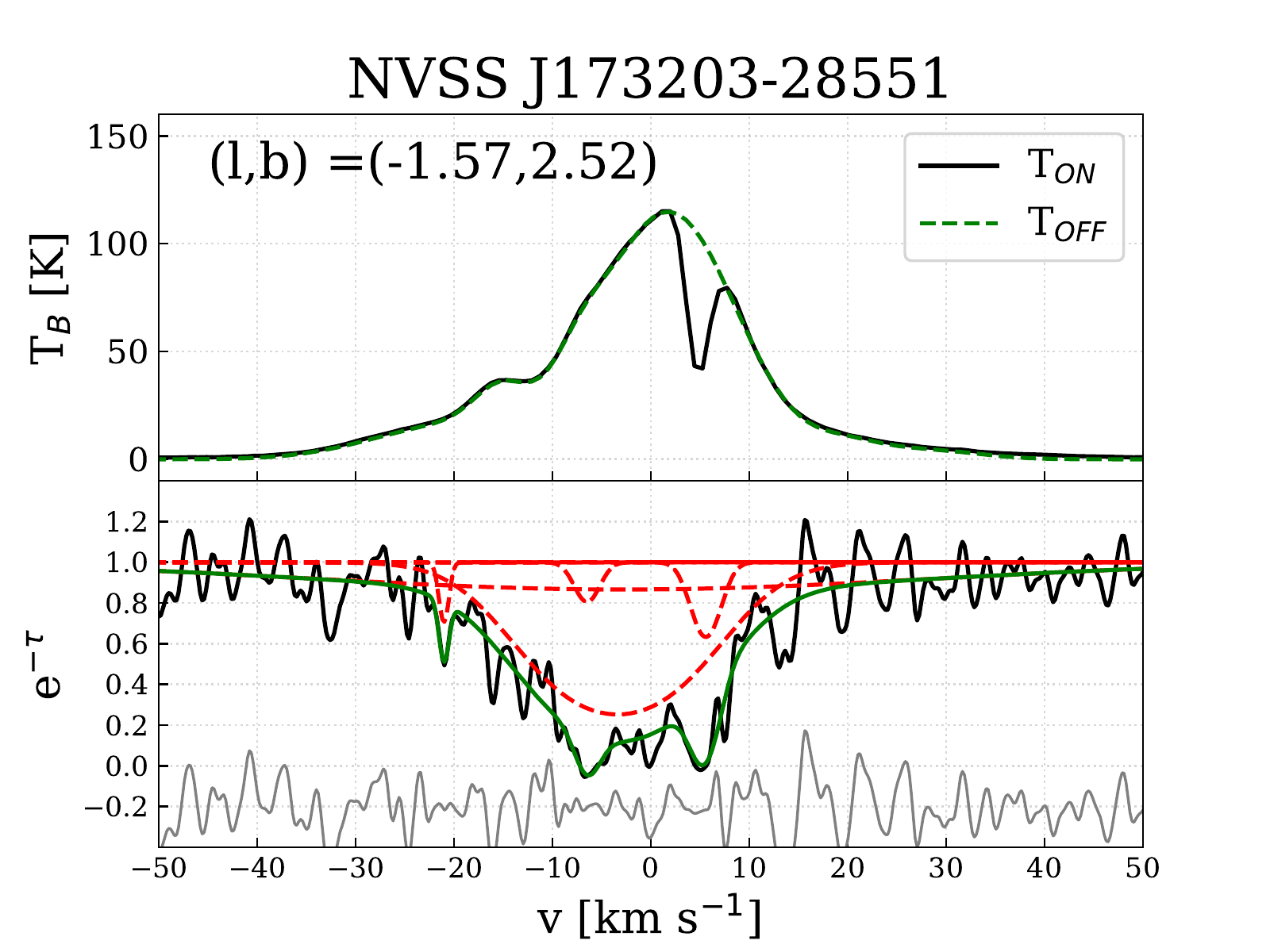}}
\hfill
\subfigure{\includegraphics[width=8.5cm]{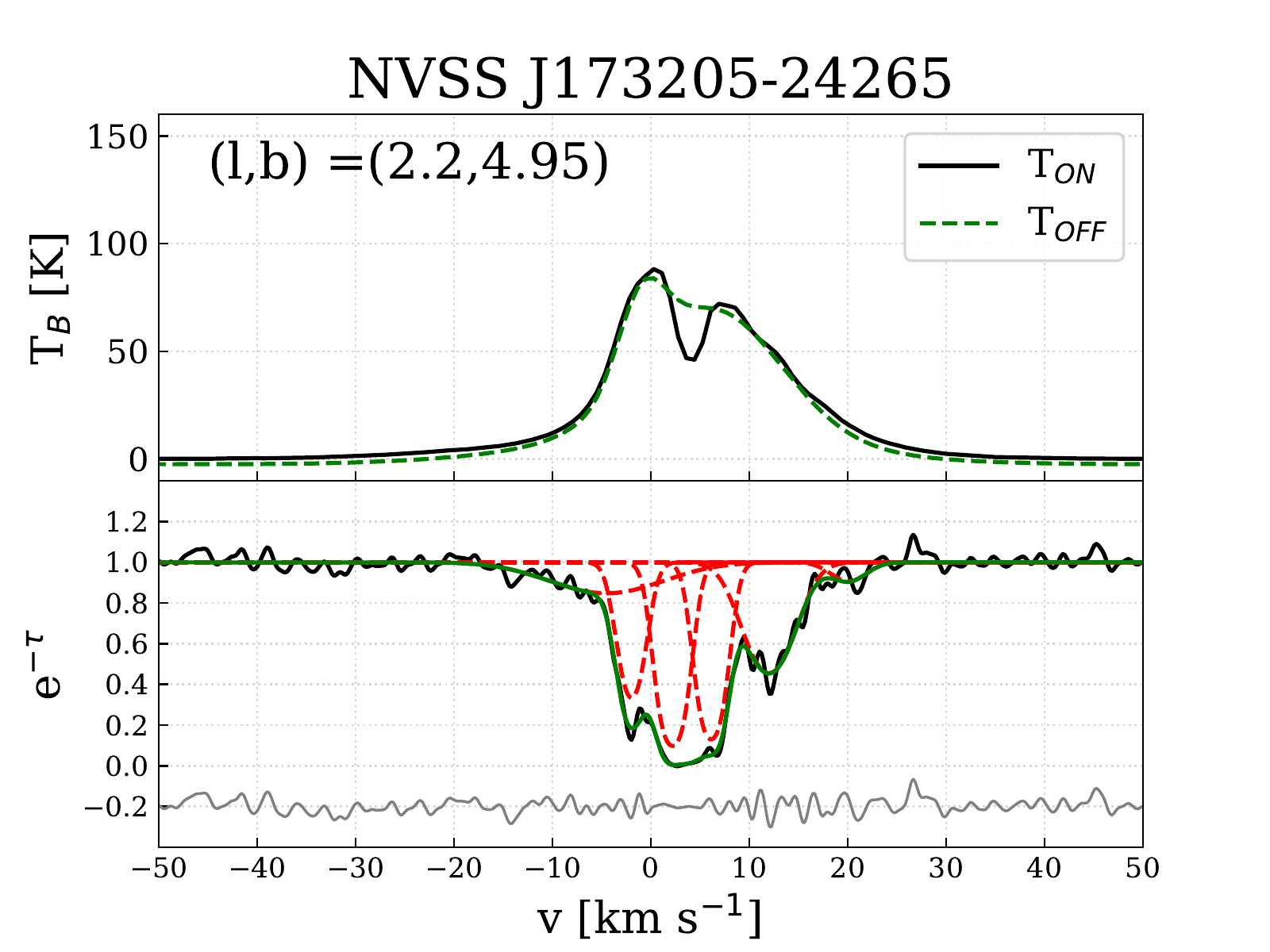}}
\hfill
\subfigure{\includegraphics[width=8.5cm]{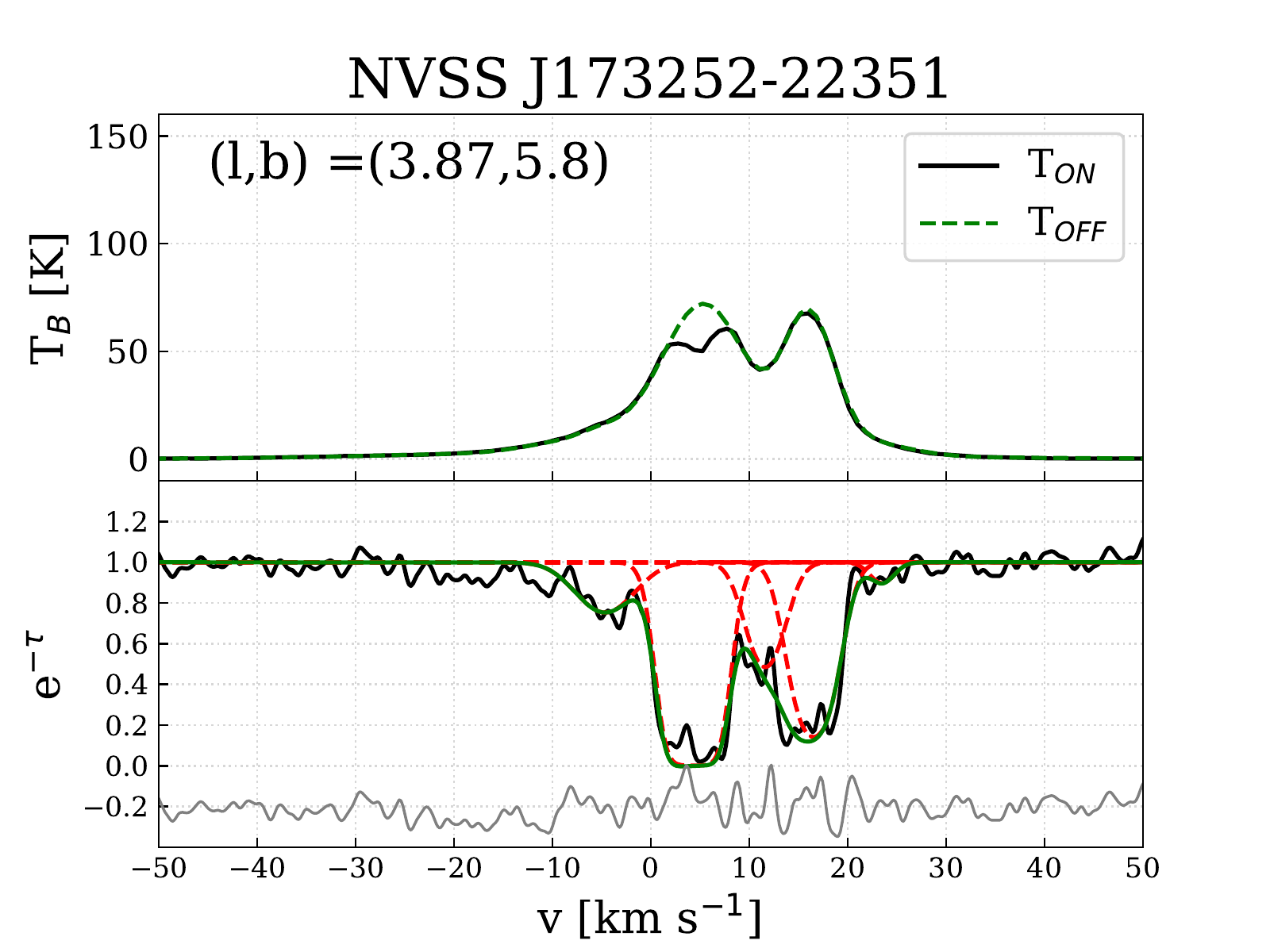}}
\hfill
\subfigure{\includegraphics[width=8.5cm]{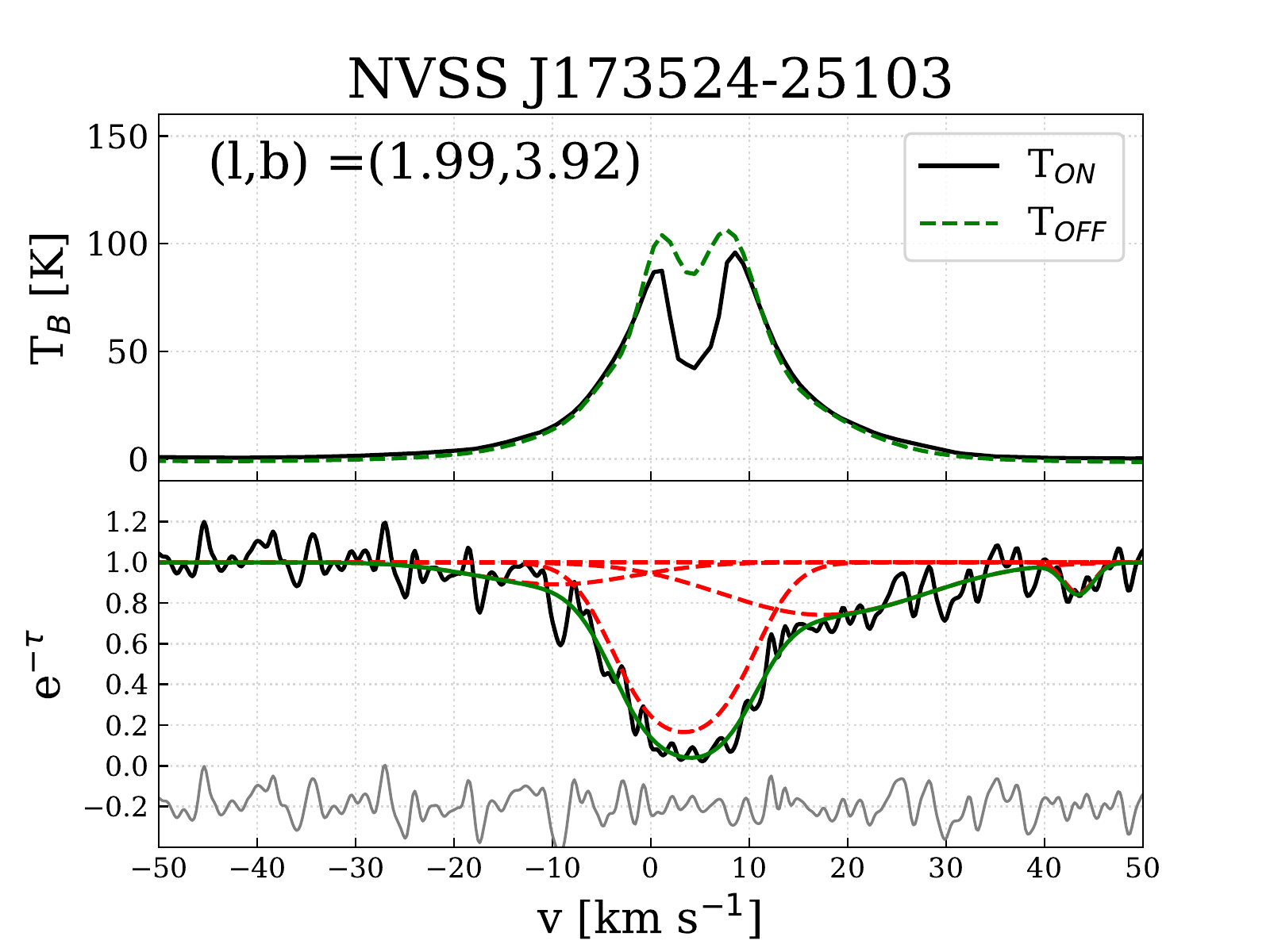}}
\hfill
\subfigure{\includegraphics[width=8.5cm]{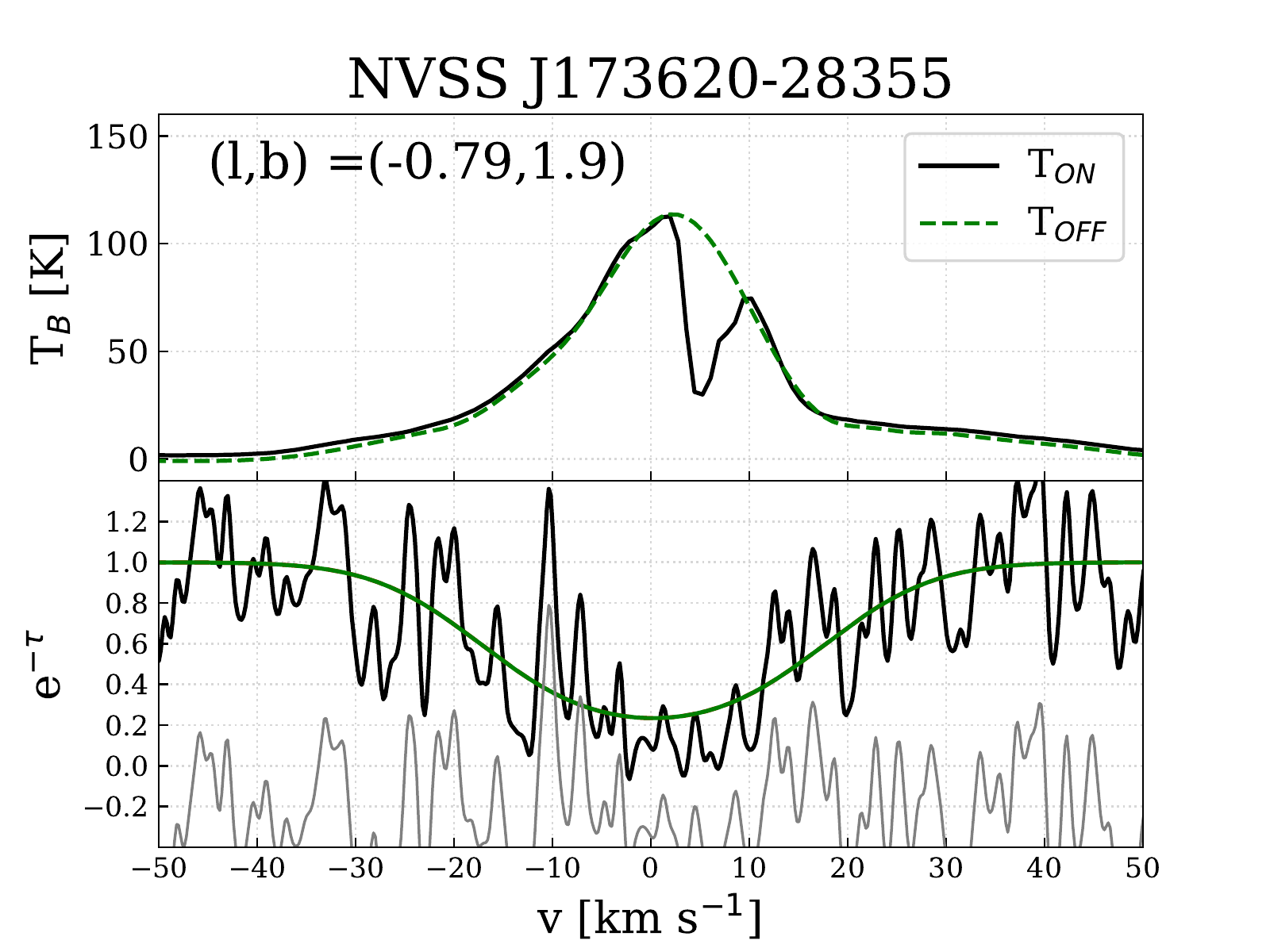}}
\hfill
\subfigure{\includegraphics[width=8.5cm]{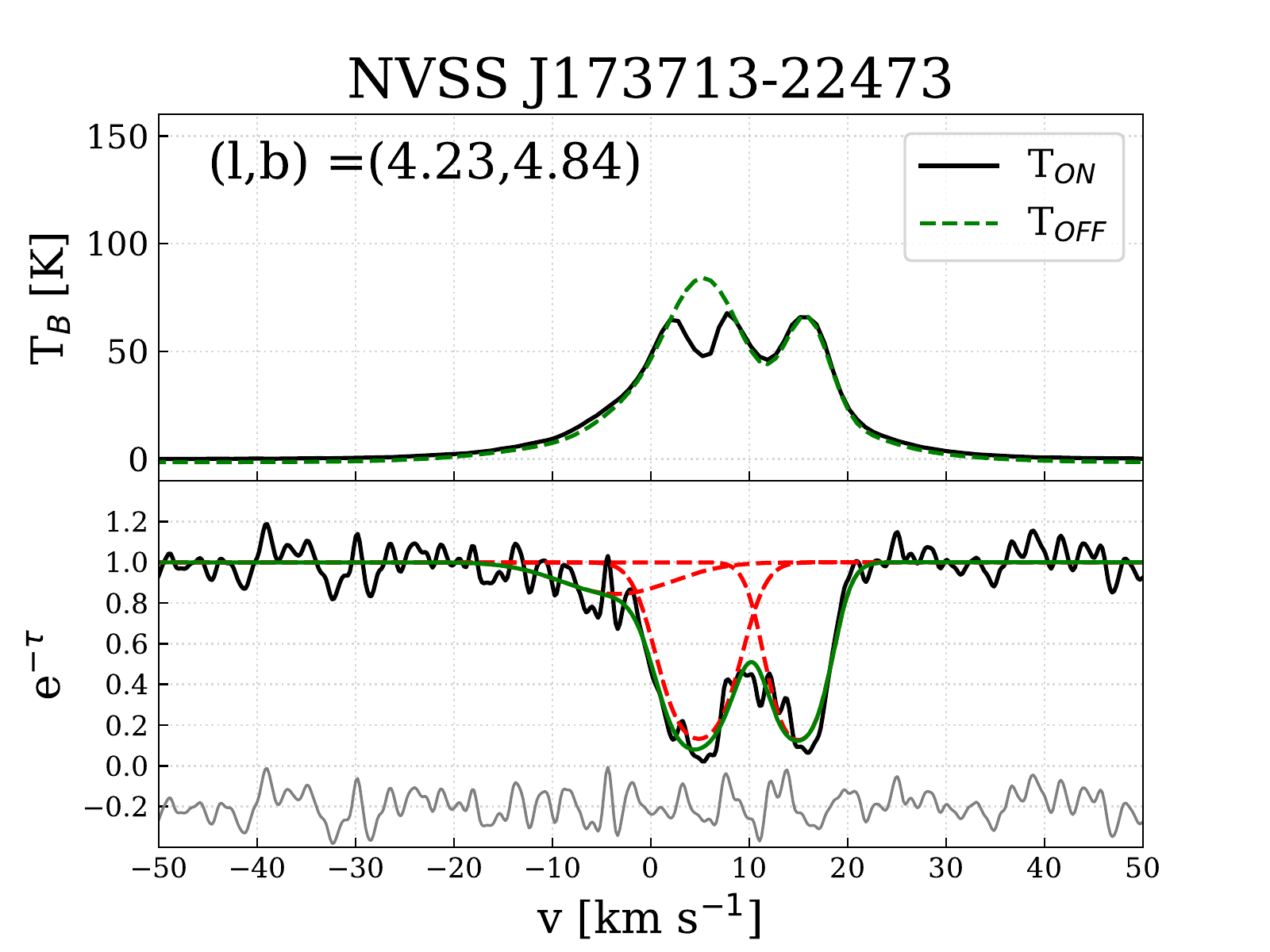}}
\hfill
\caption{Same as \ref{fig:spectra1}}
\label{fig:spectra2}
\end{figure*} 

\begin{figure*}
\subfigure{\includegraphics[width=8.5cm]{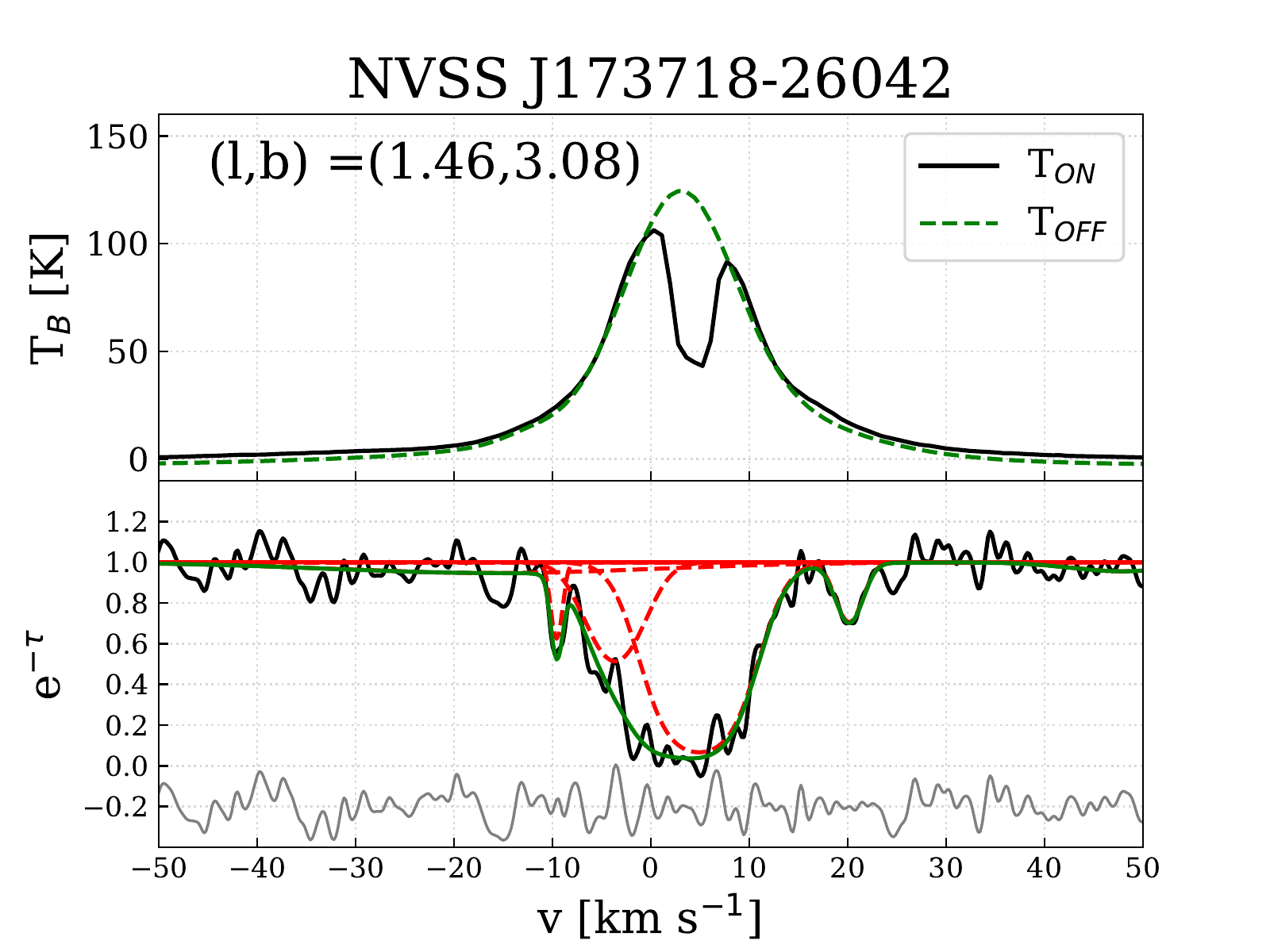}}
\hfill
\subfigure{\includegraphics[width=8.5cm]{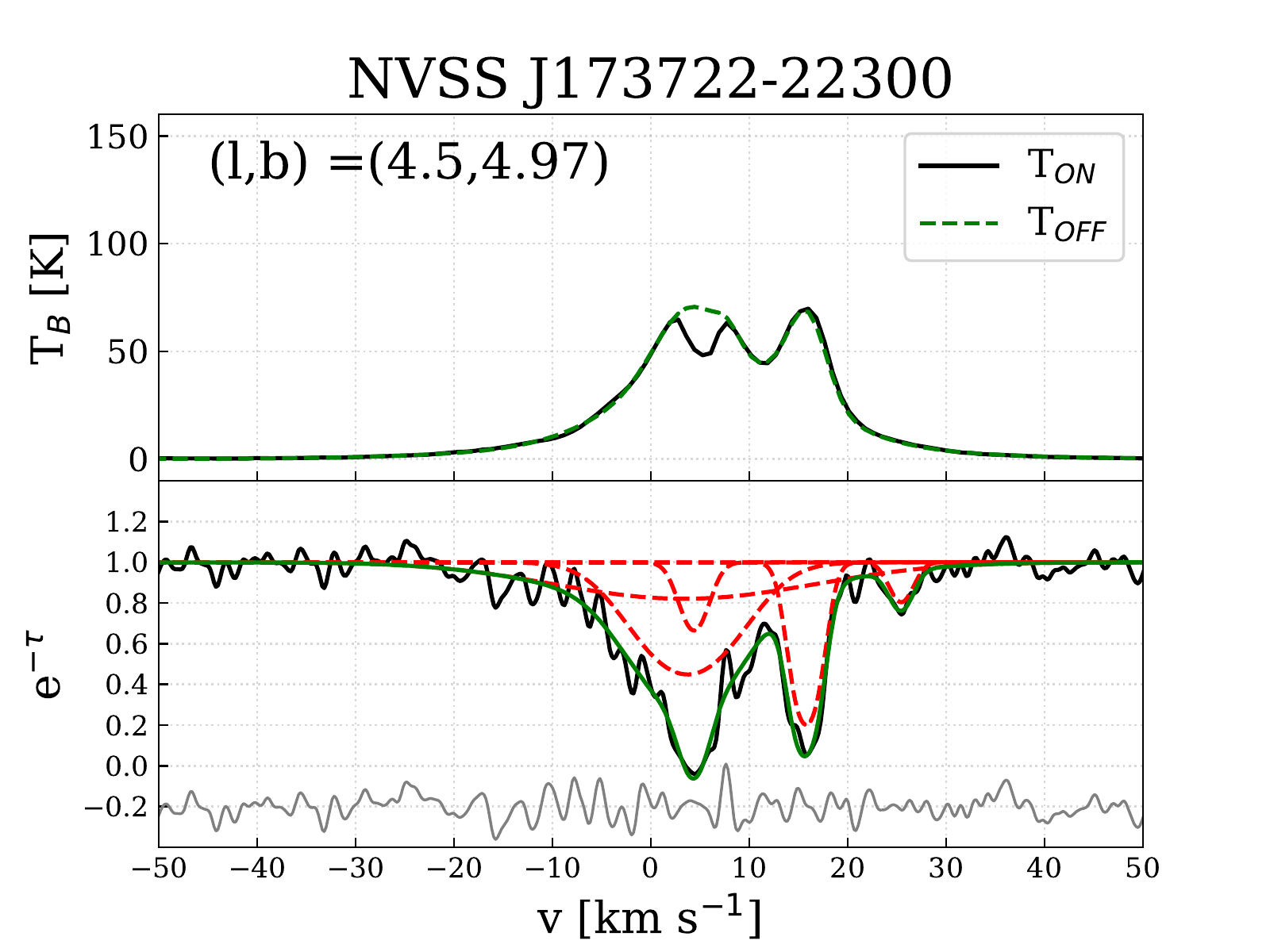}}
\hfill
\subfigure{\includegraphics[width=8.5cm]{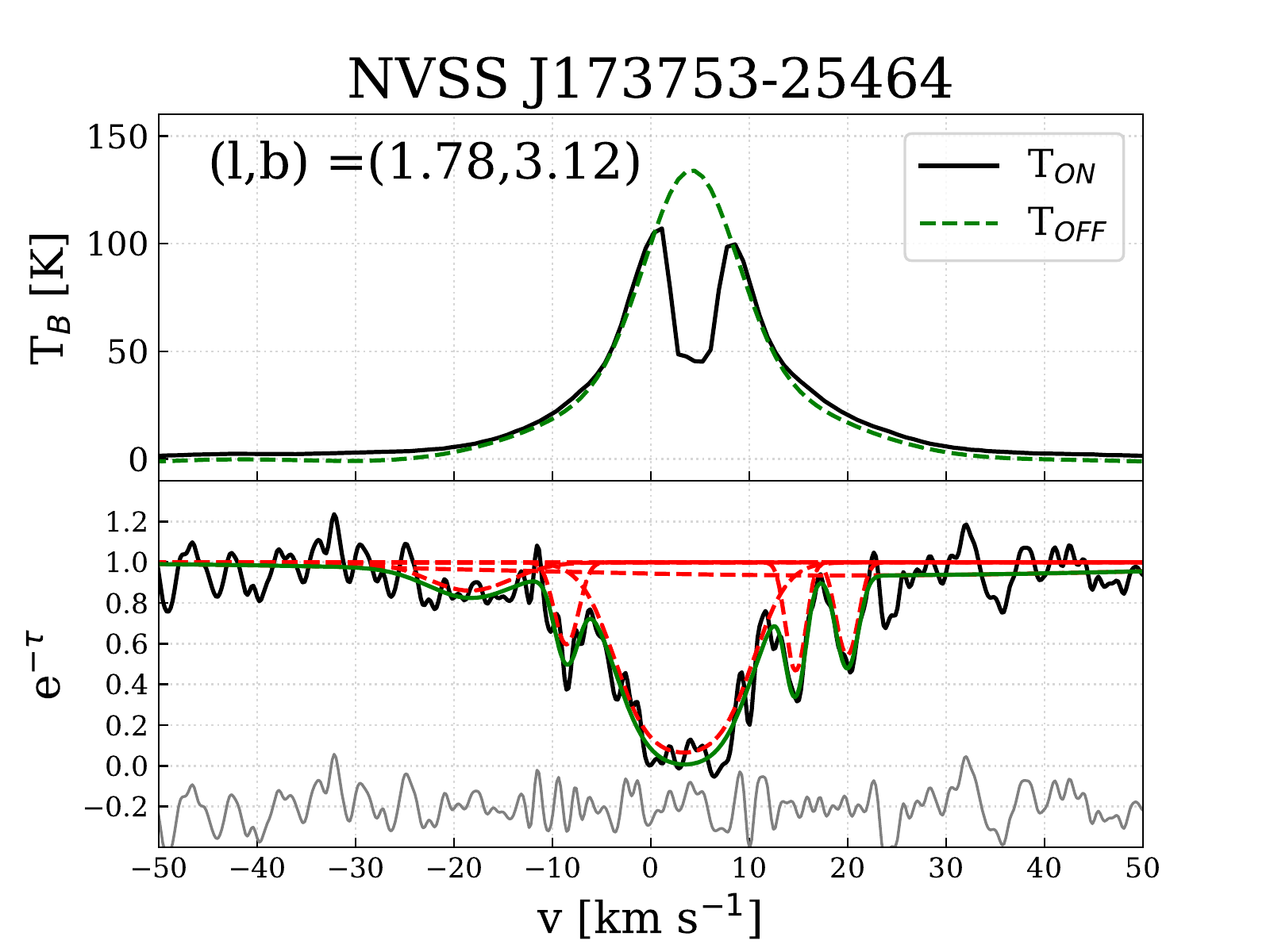}}
\hfill
\subfigure{\includegraphics[width=8.5cm]{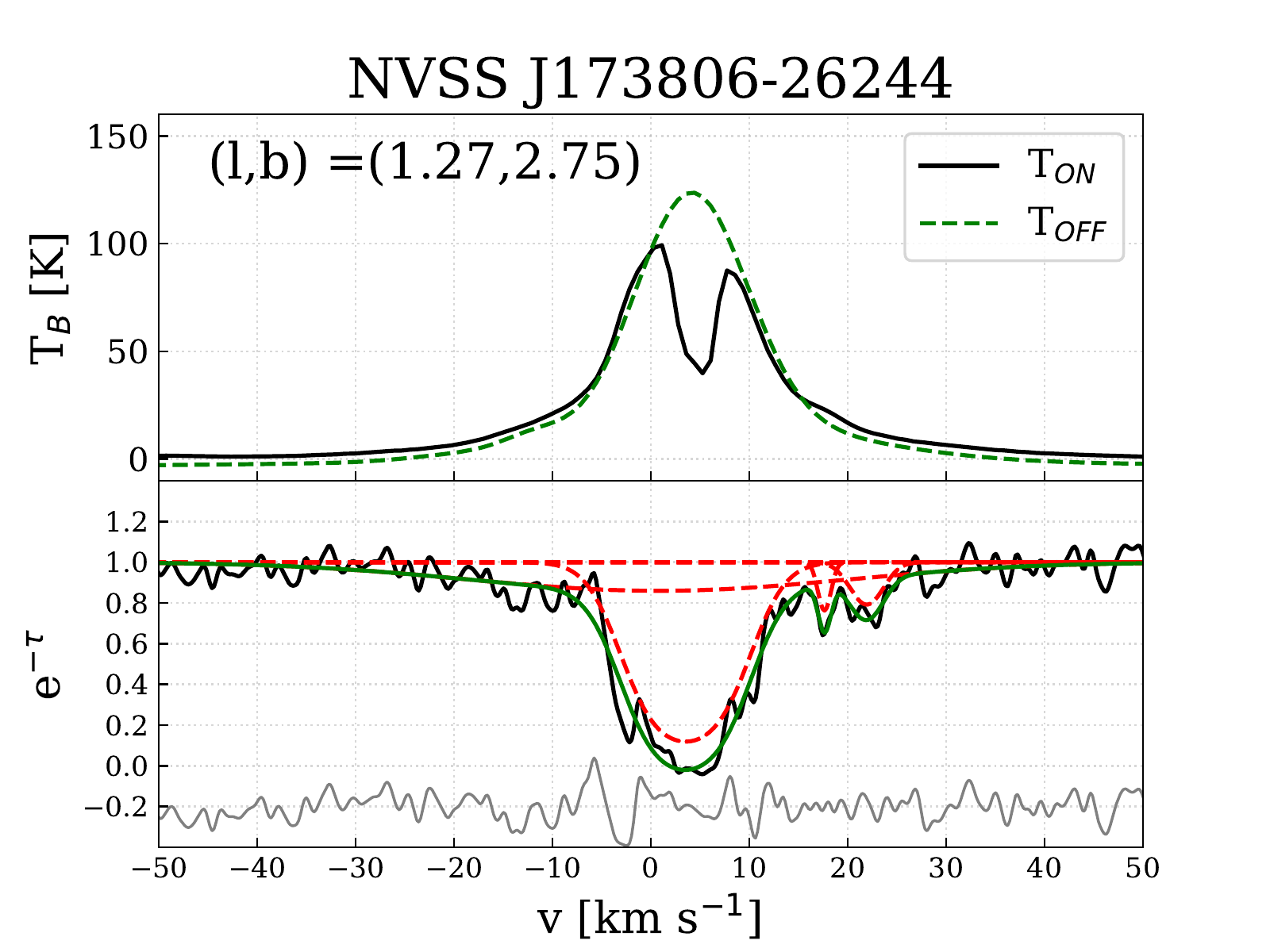}}
\hfill
\subfigure{\includegraphics[width=8.5cm]{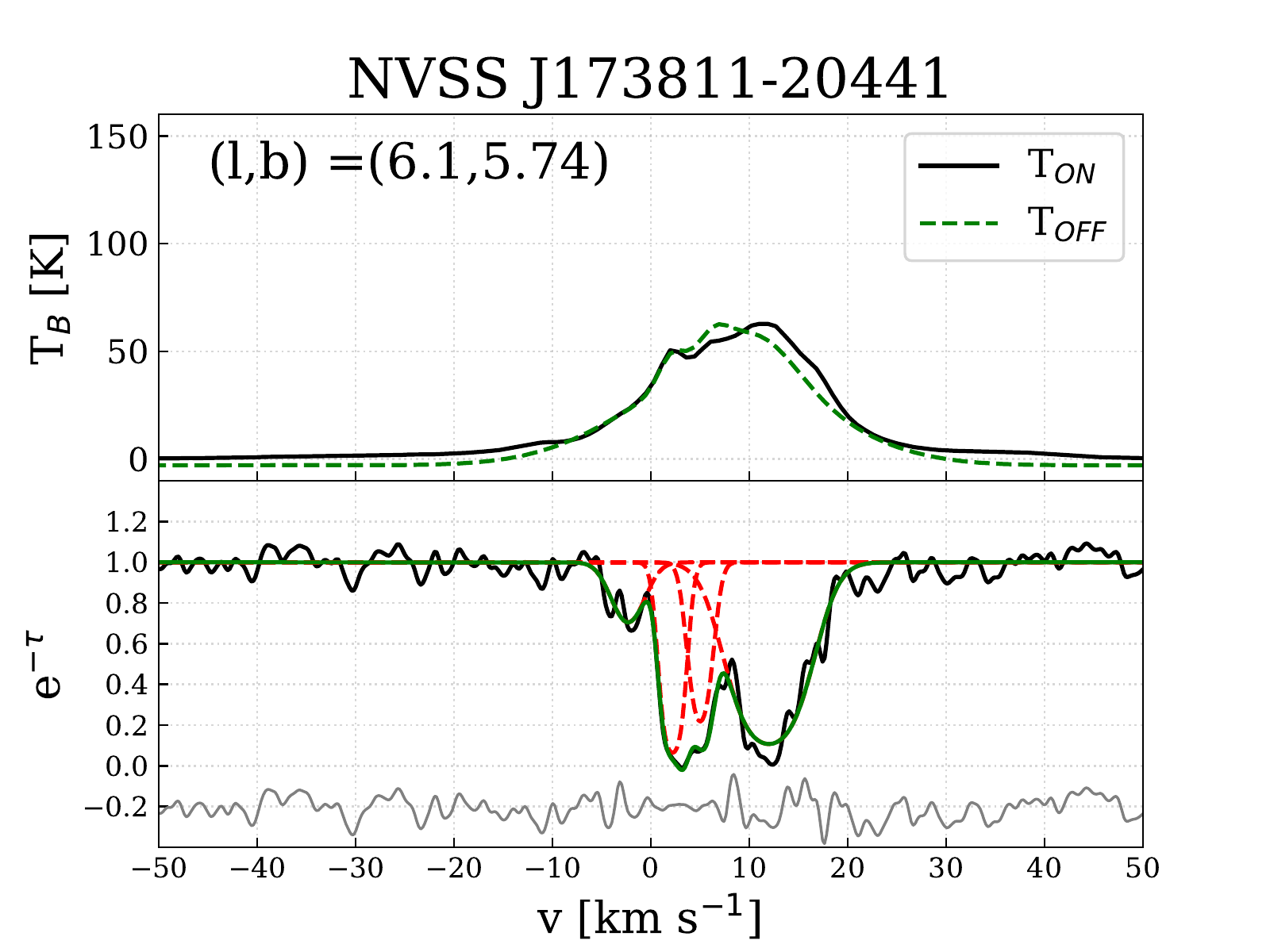}}
\hfill
\subfigure{\includegraphics[width=8.5cm]{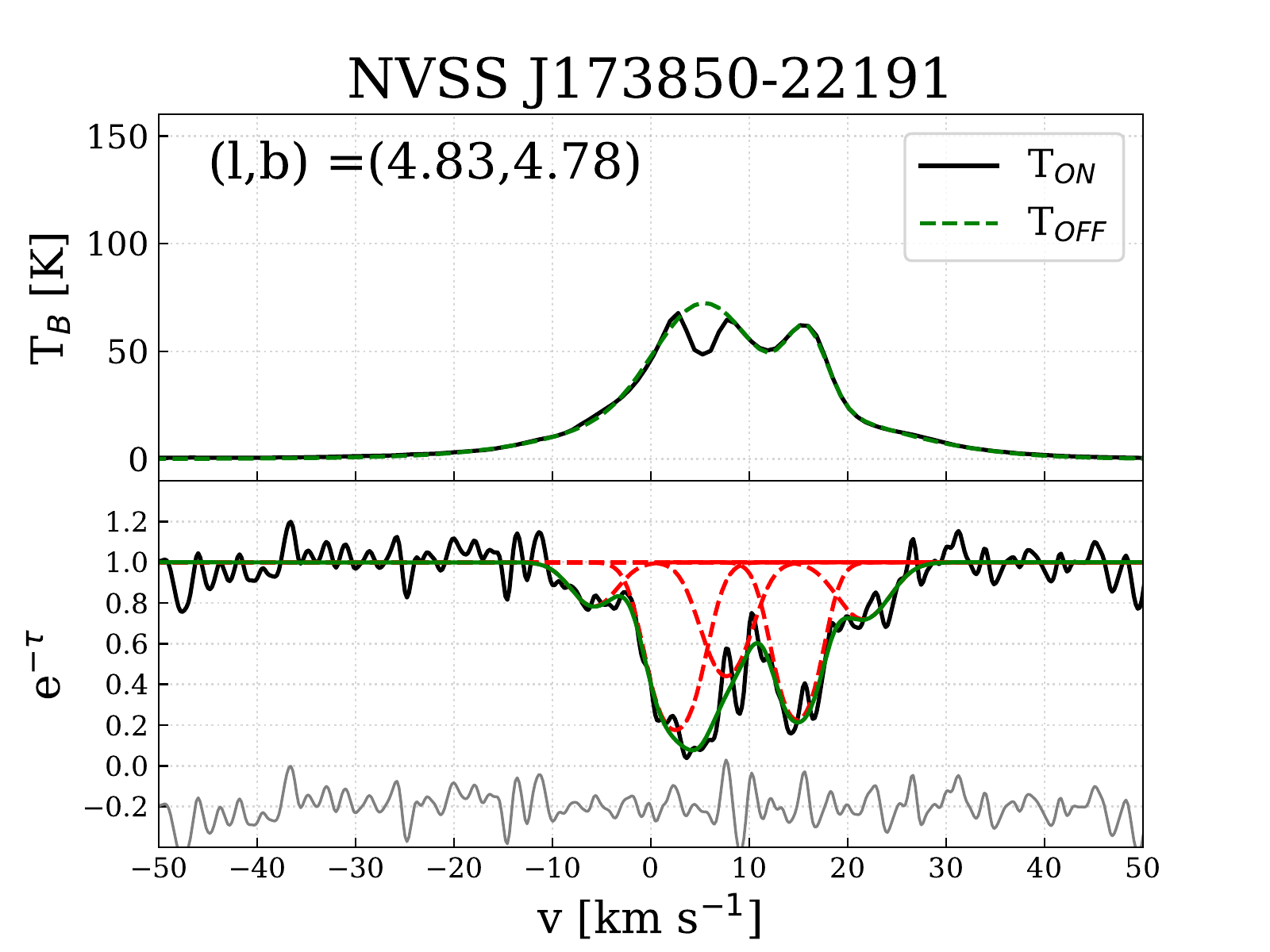}}
\hfill
\caption{Same as \ref{fig:spectra1}}
\label{fig:spectra3}
\end{figure*} 

\begin{figure*}
\subfigure{\includegraphics[width=8.5cm]{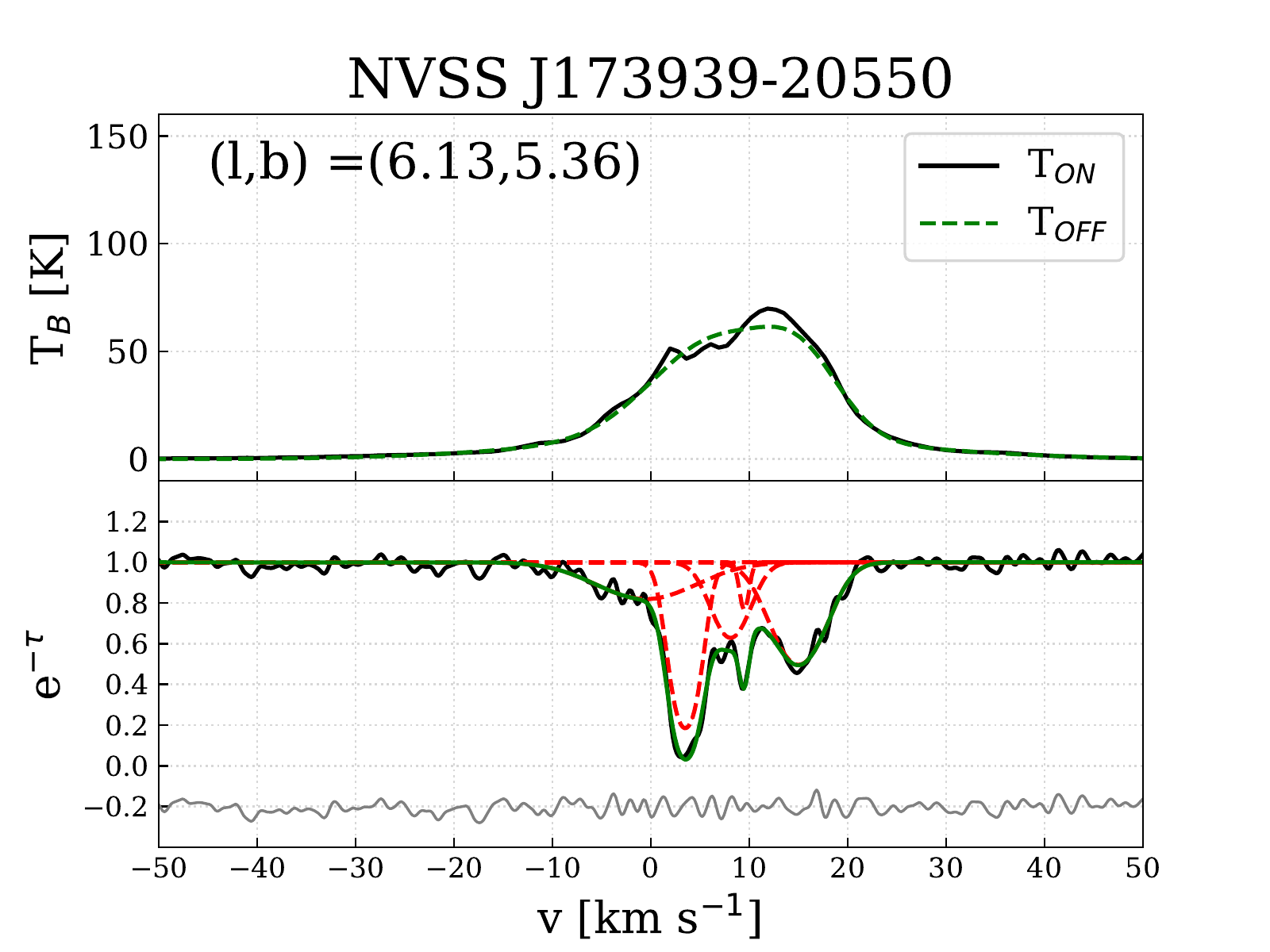}}
\hfill
\subfigure{\includegraphics[width=8.5cm]{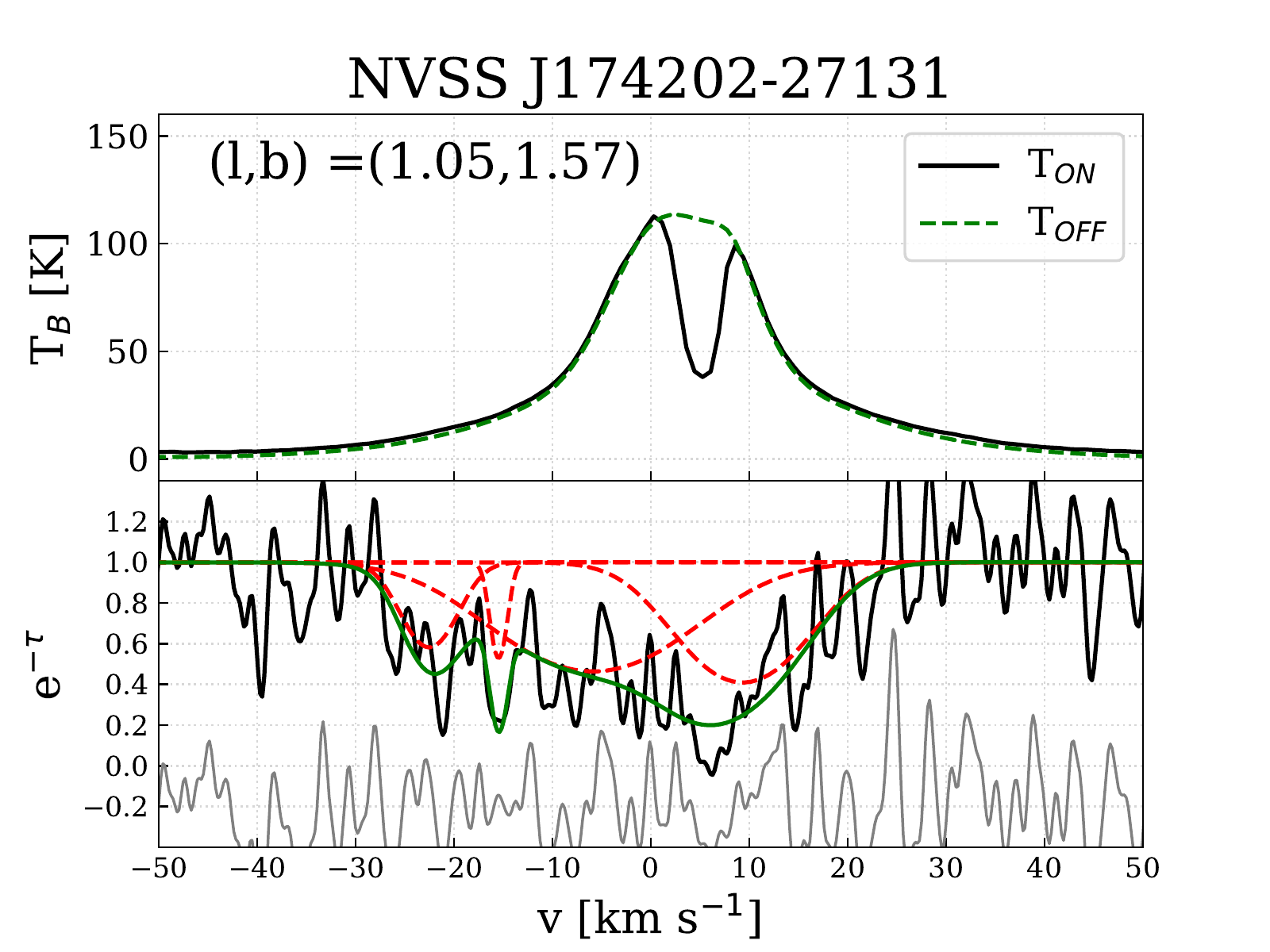}}
\hfill
\subfigure{\includegraphics[width=8.5cm]{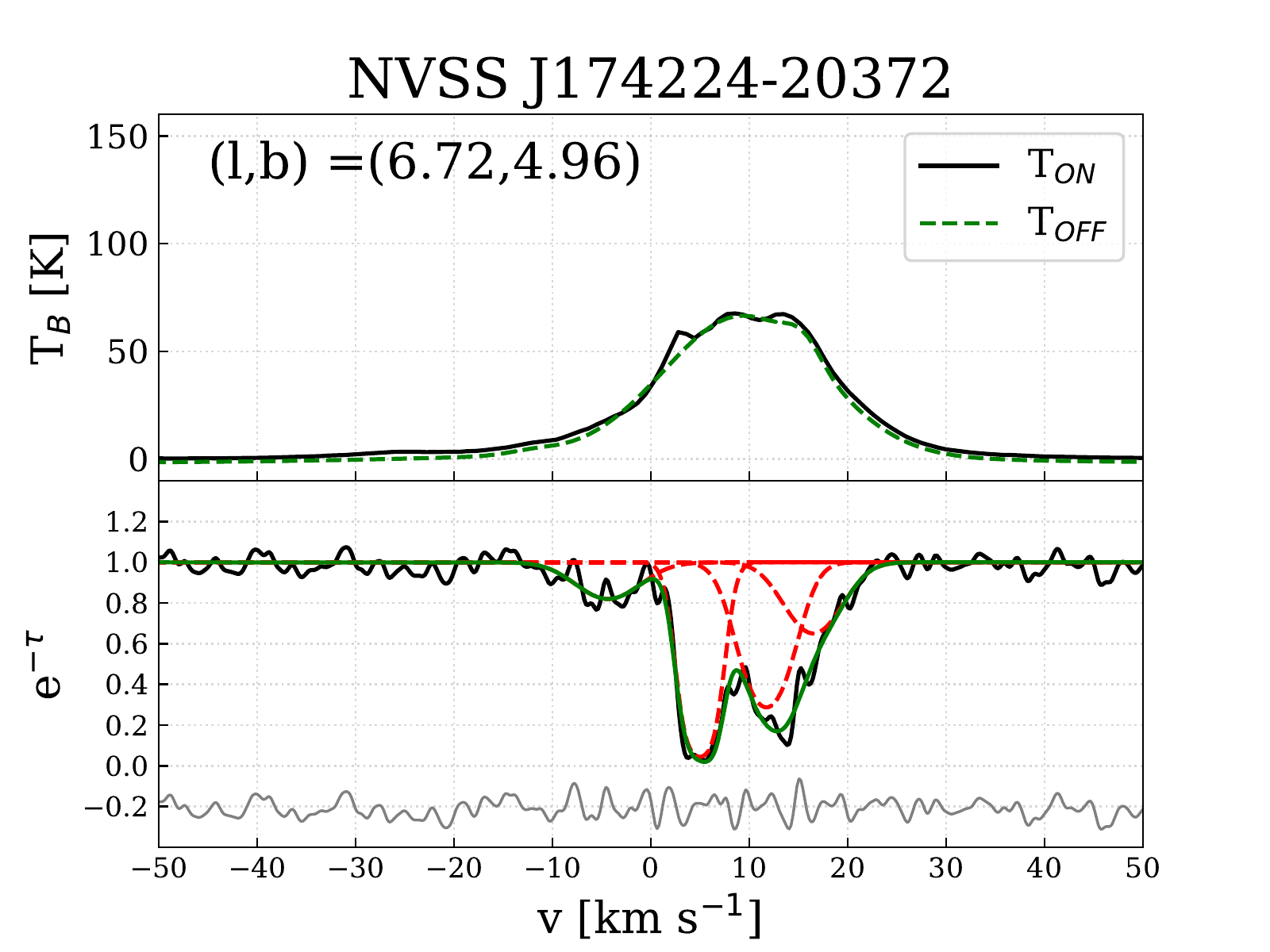}}
\hfill
\subfigure{\includegraphics[width=8.5cm]{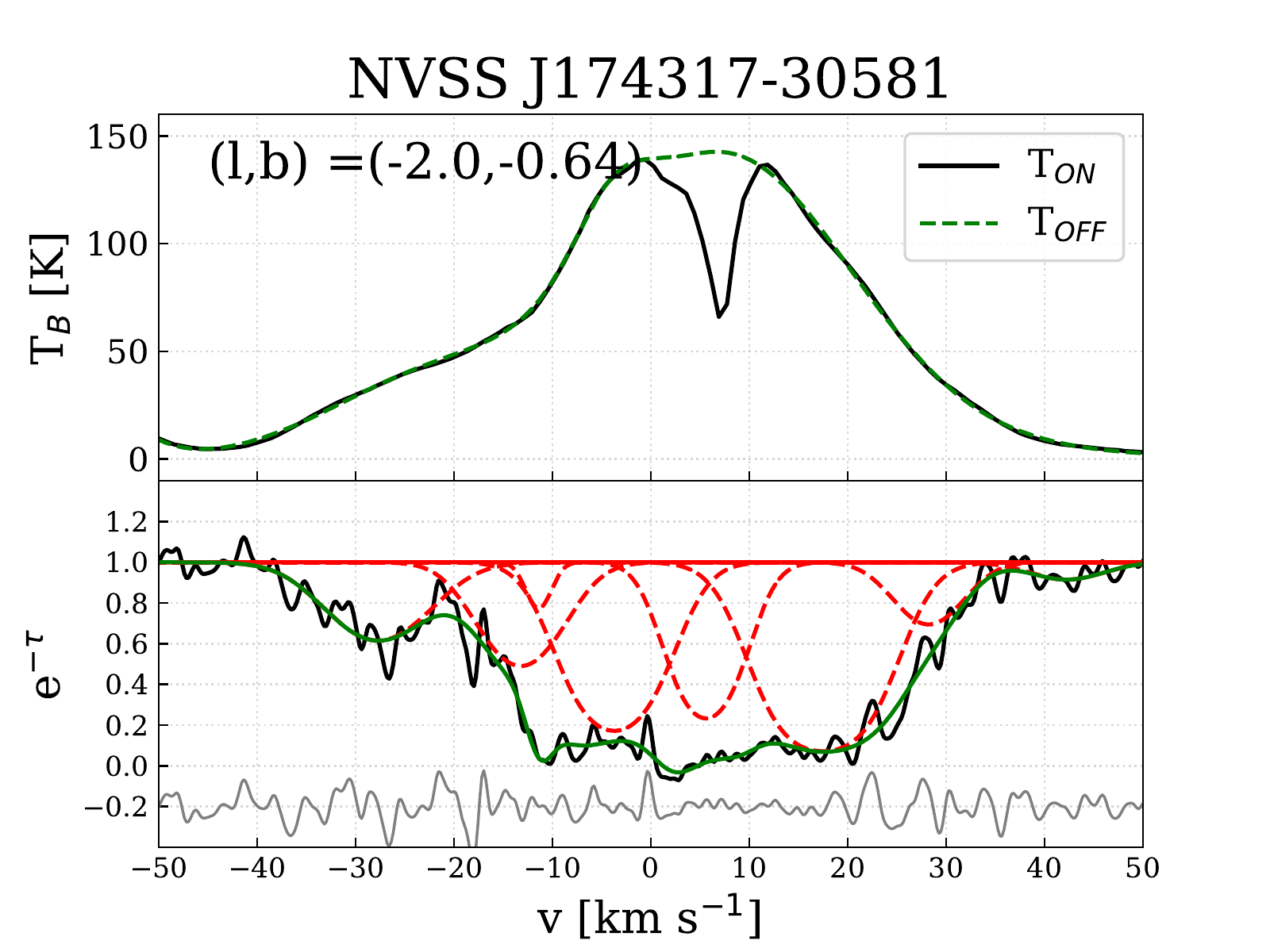}}
\hfill
\subfigure{\includegraphics[width=8.5cm]{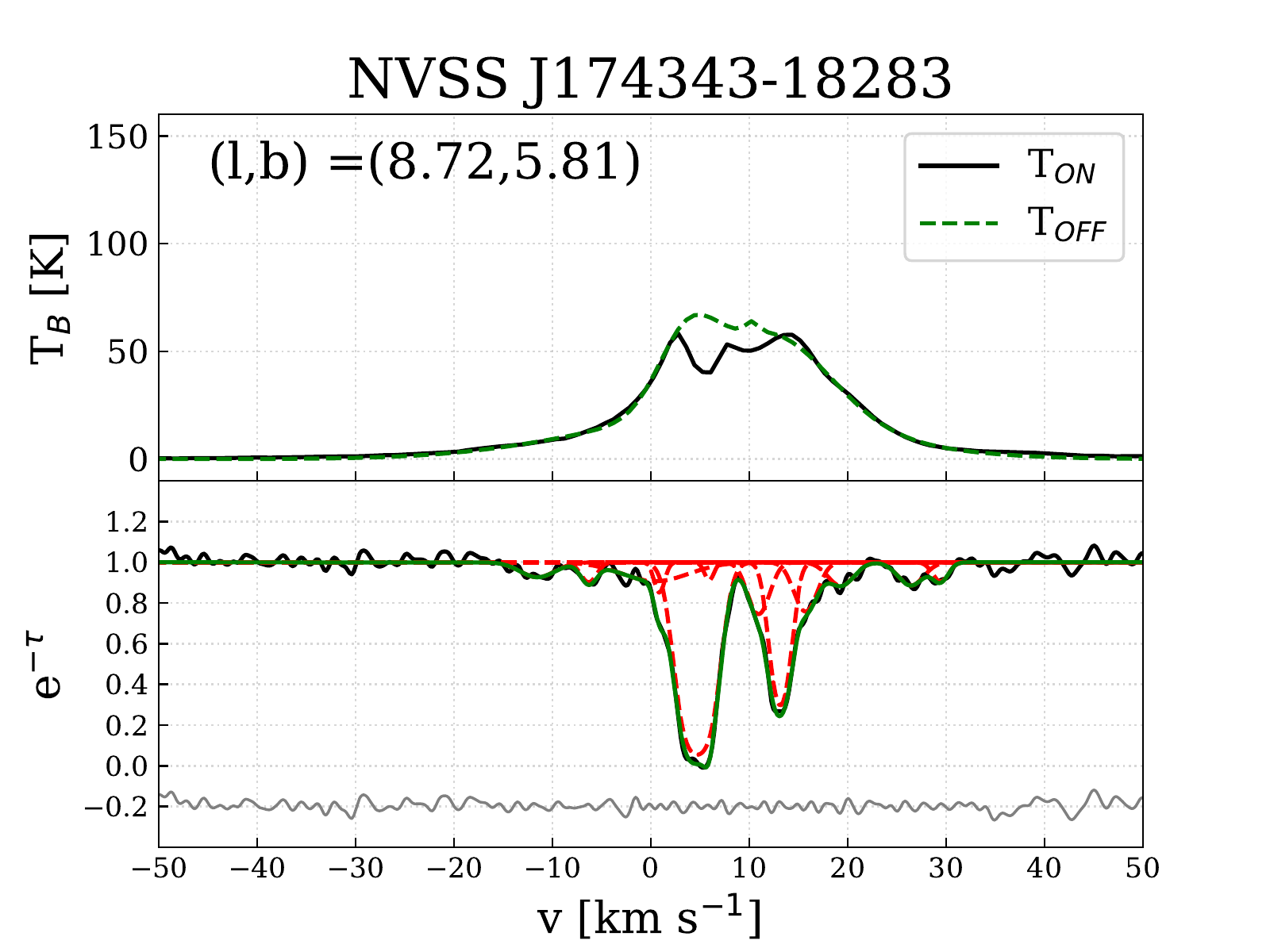}}
\hfill
\subfigure{\includegraphics[width=8.5cm]{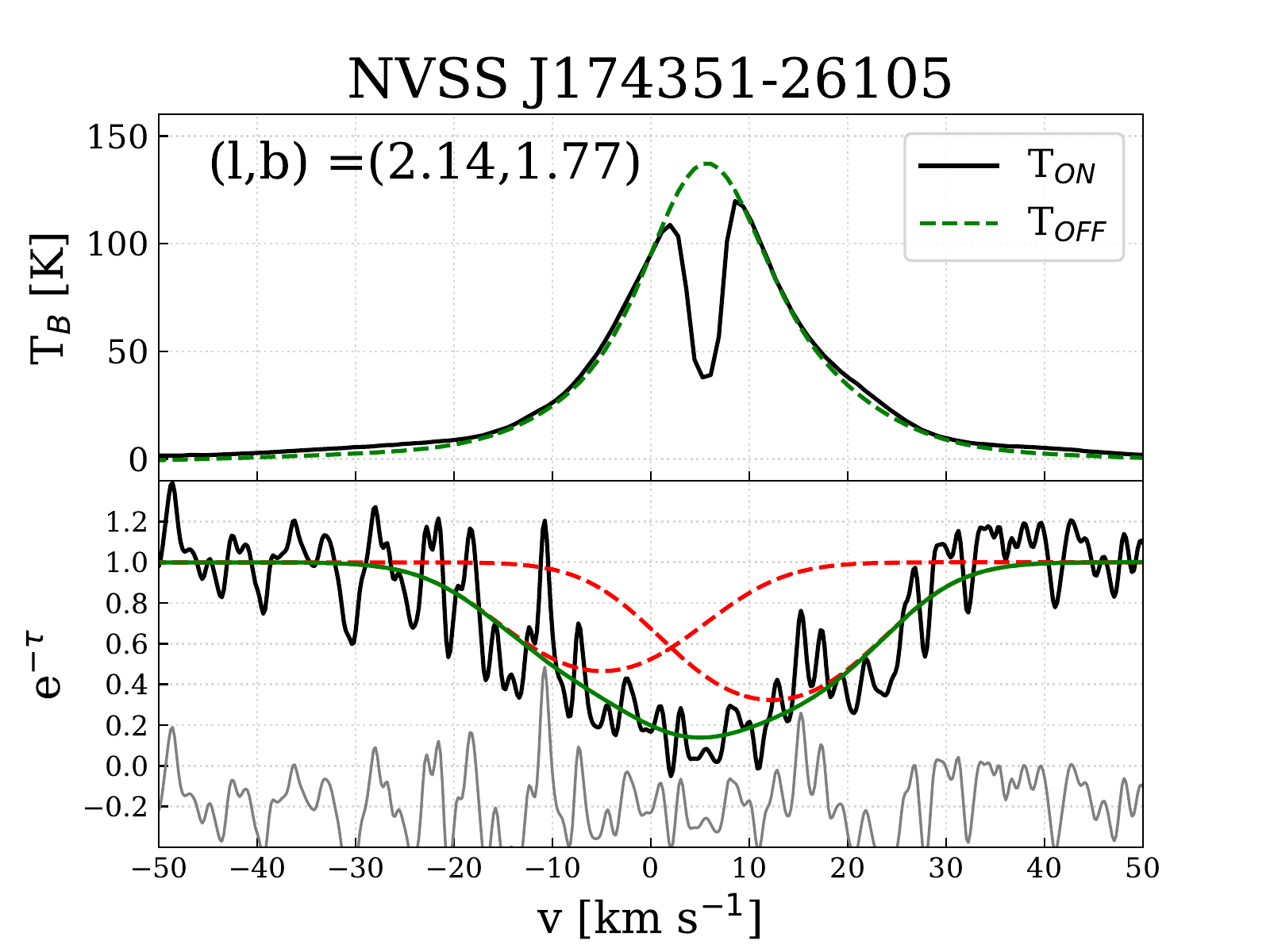}}
\hfill
\caption{Same as \ref{fig:spectra1}}
\label{fig:spectra4}
\end{figure*} 

\begin{figure*}
\subfigure{\includegraphics[width=8.5cm]{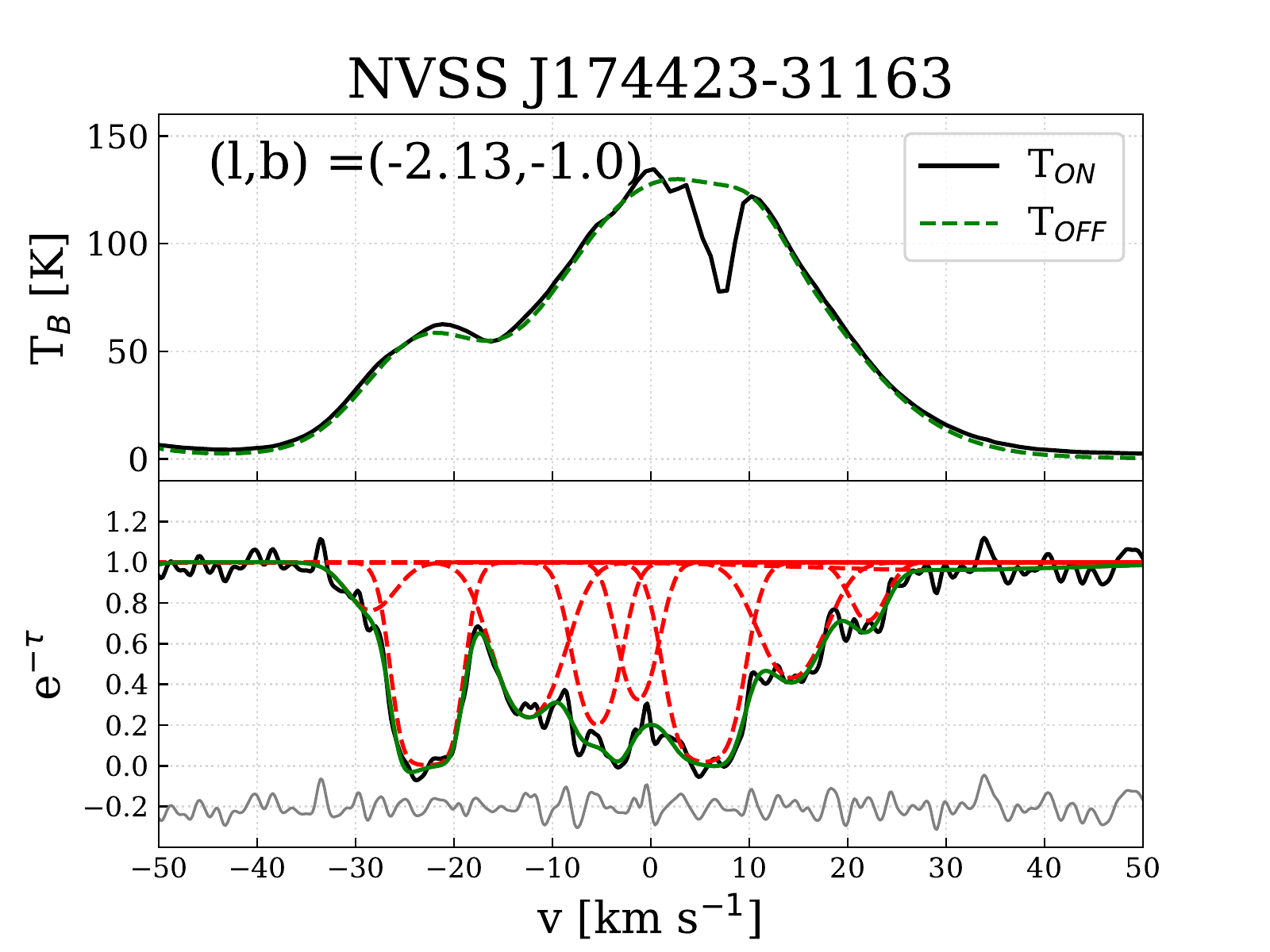}}
\hfill
\subfigure{\includegraphics[width=8.5cm]{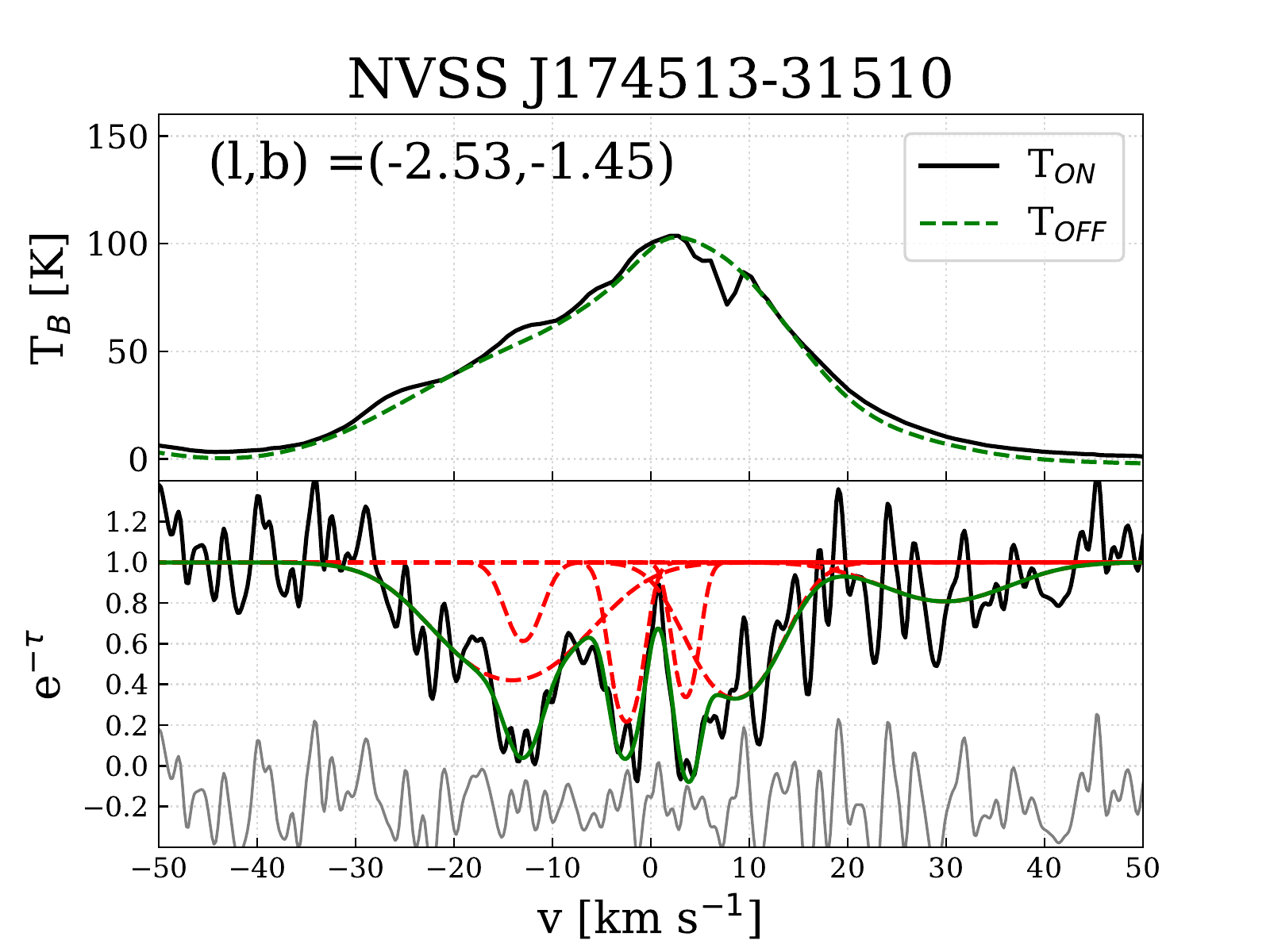}}
\hfill
\subfigure{\includegraphics[width=8.5cm]{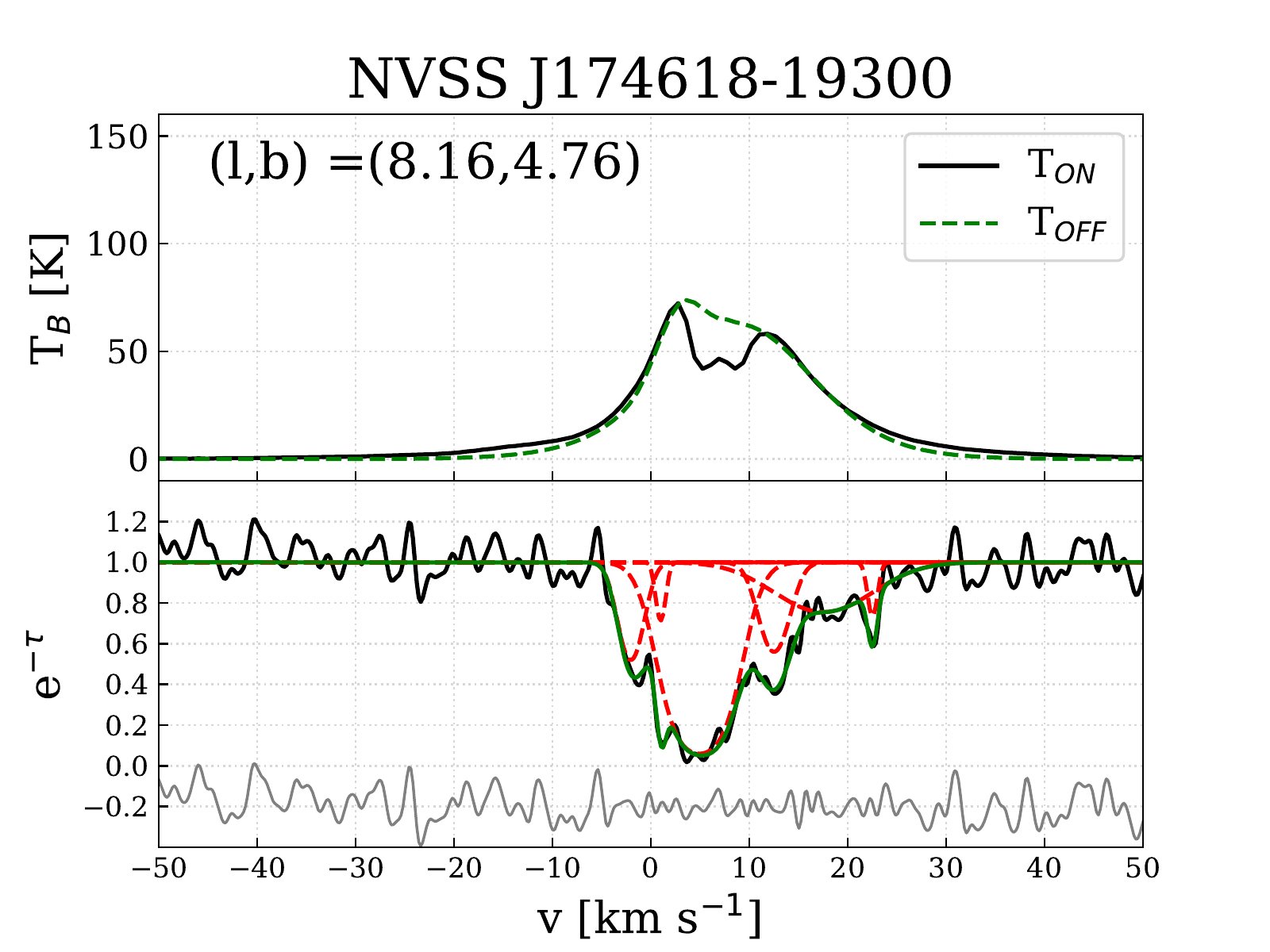}}
\hfill
\subfigure{\includegraphics[width=8.5cm]{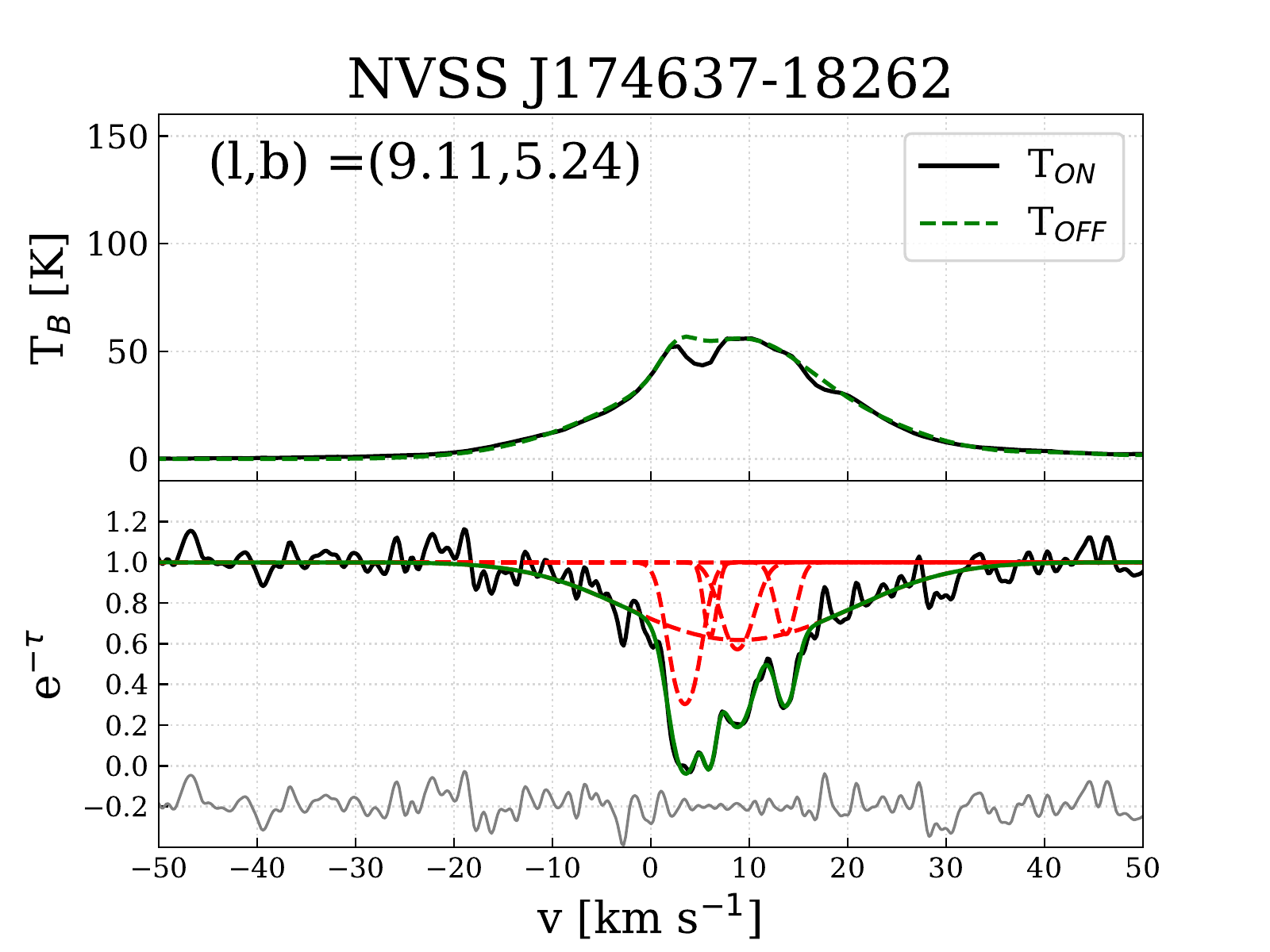}}
\hfill
\subfigure{\includegraphics[width=8.5cm]{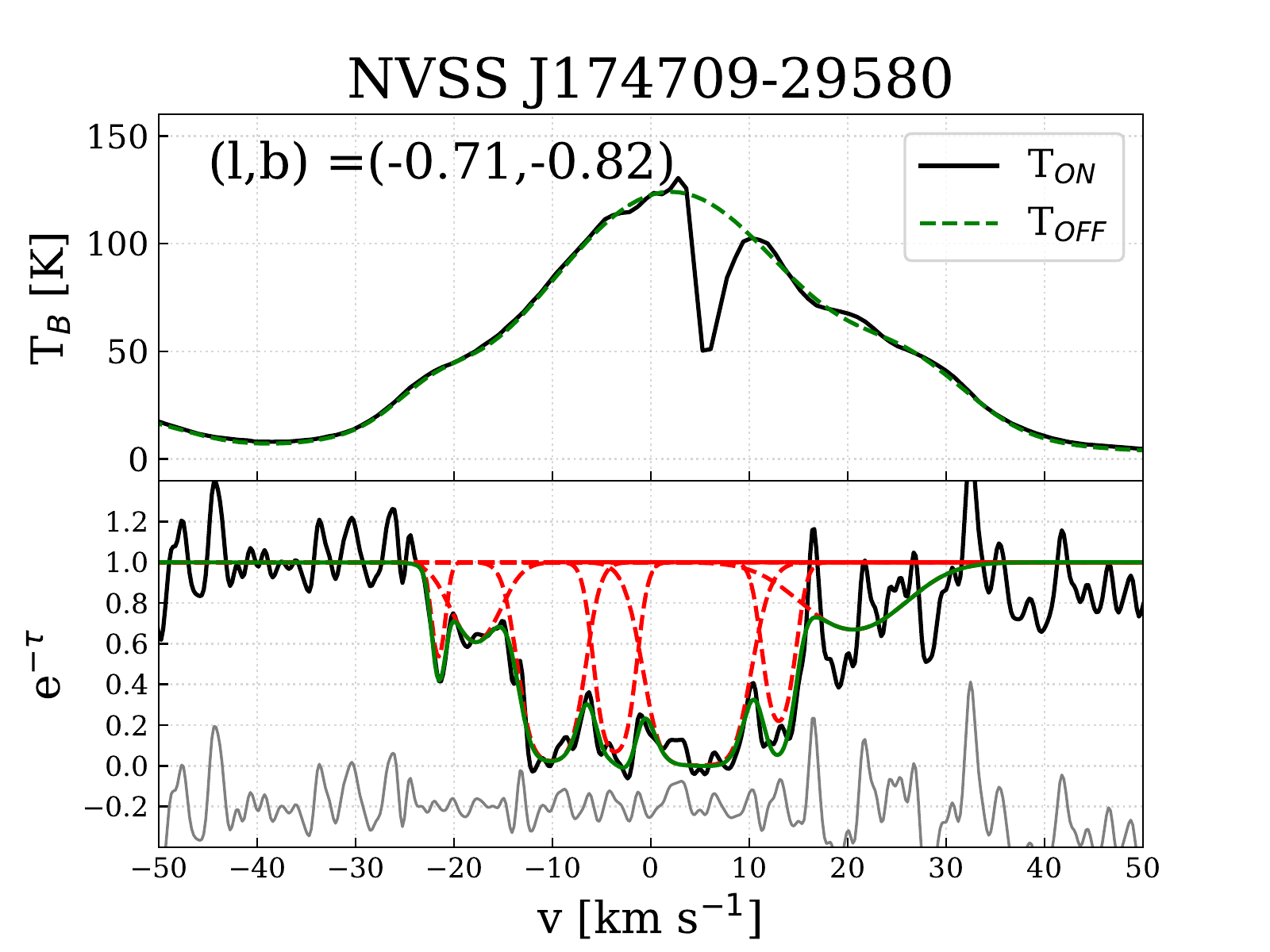}}
\hfill
\subfigure{\includegraphics[width=8.5cm]{174713-19213_spectra_paper.pdf}}
\hfill
\caption{Same as \ref{fig:spectra1}}
\label{fig:spectra5}
\end{figure*} 

\begin{figure*}
\subfigure{\includegraphics[width=8.5cm]{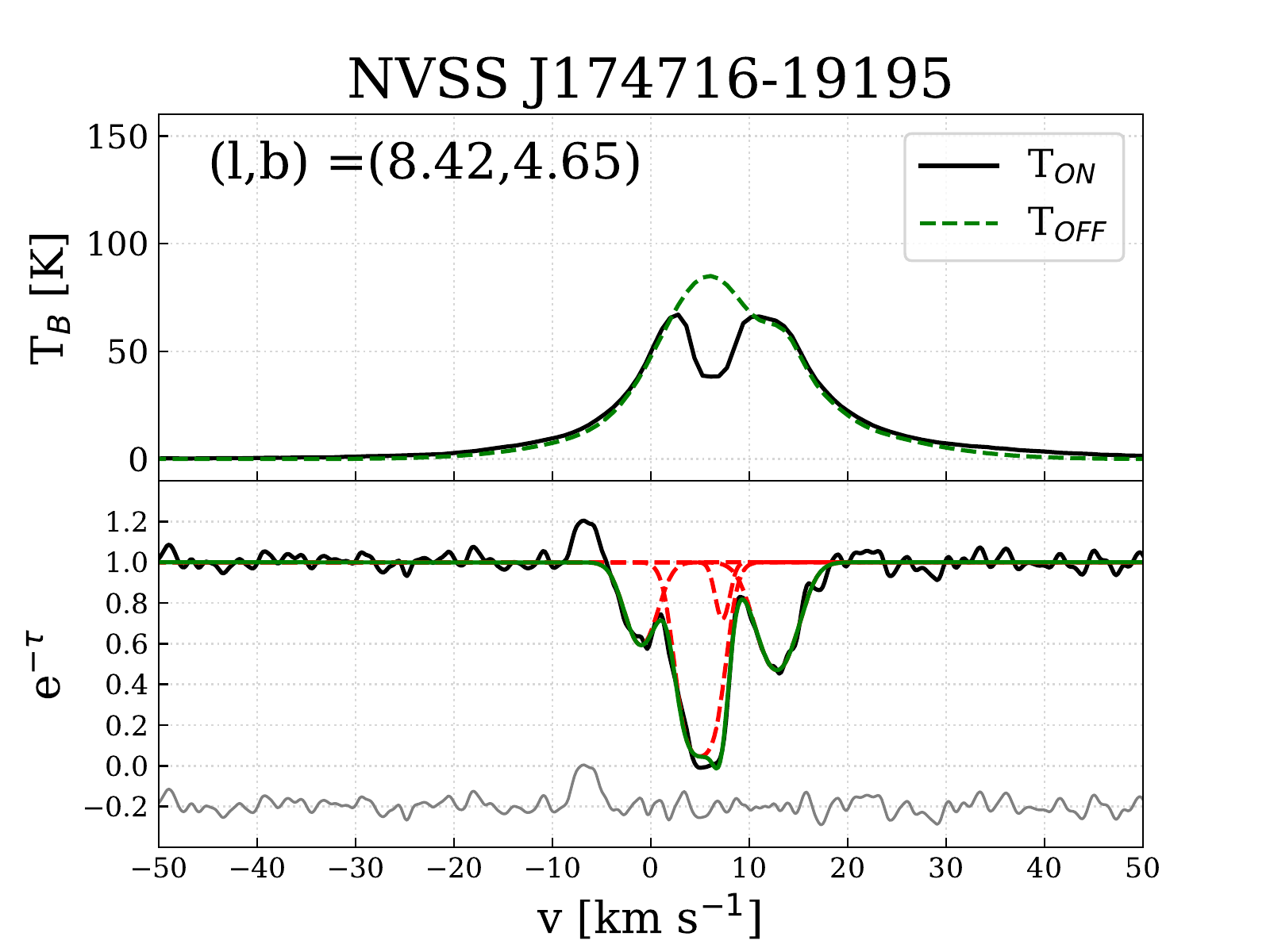}}
\hfill
\subfigure{\includegraphics[width=8.5cm]{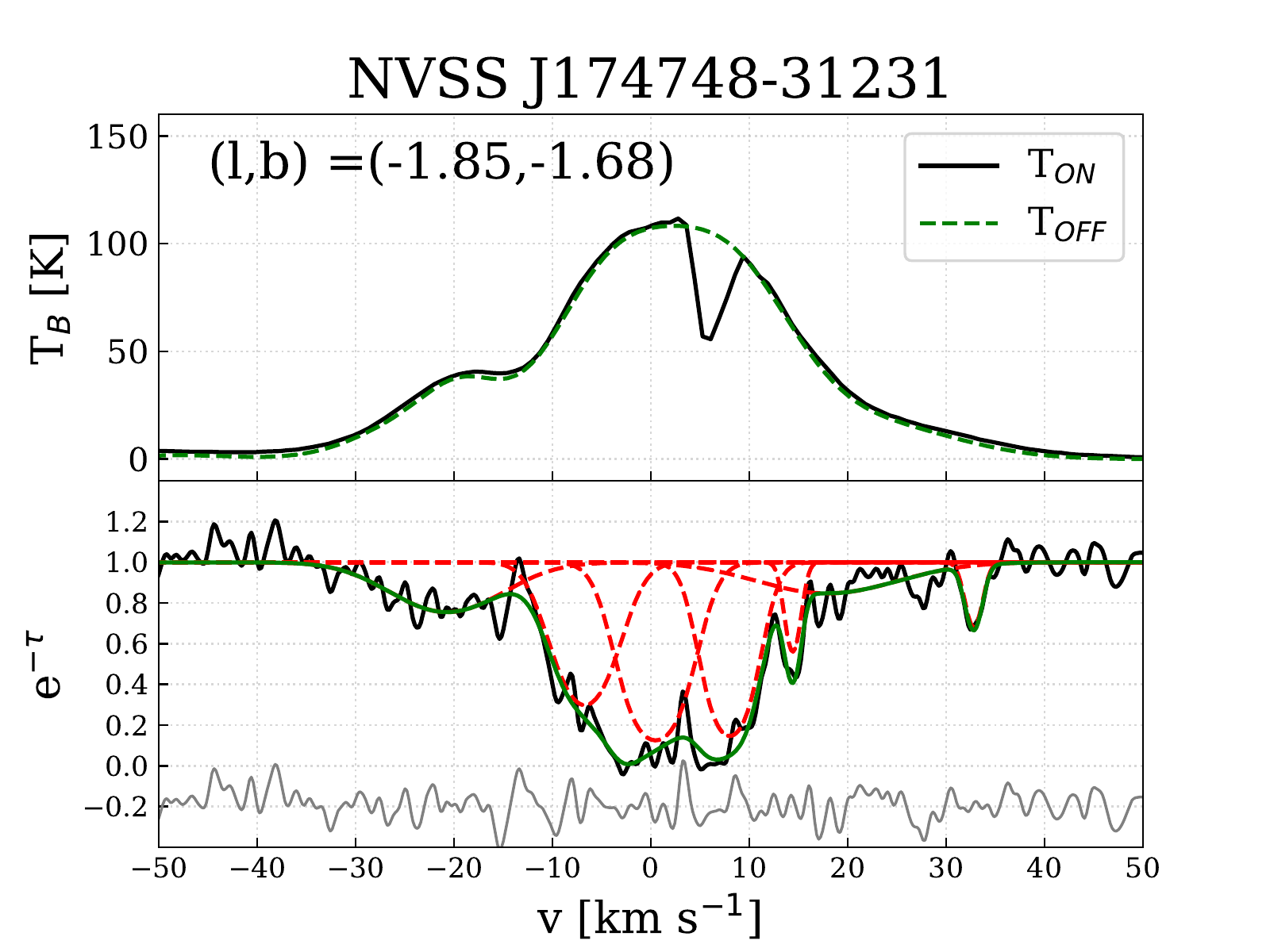}}
\hfill
\subfigure{\includegraphics[width=8.5cm]{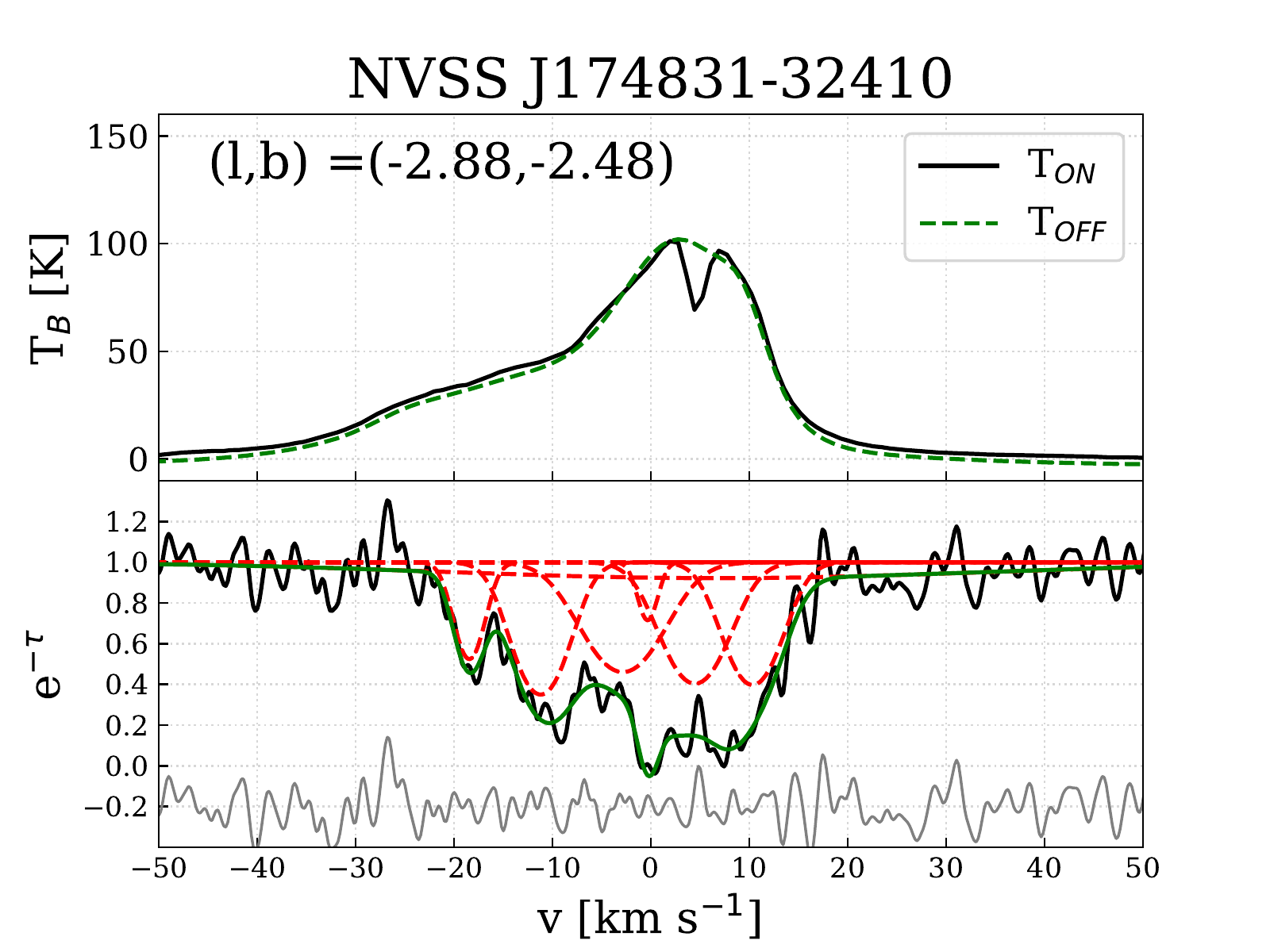}}
\hfill
\subfigure{\includegraphics[width=8.5cm]{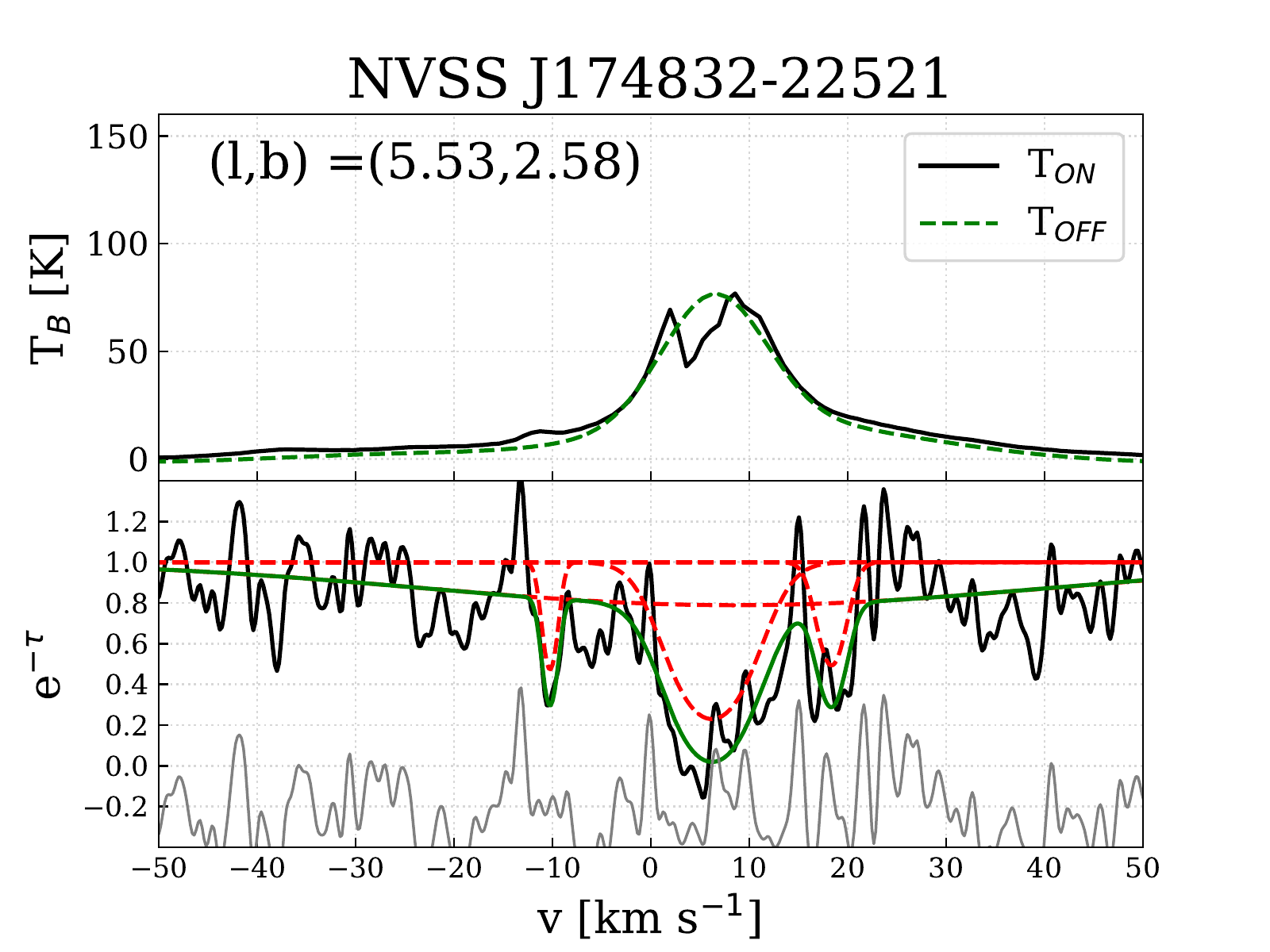}}
\hfill
\subfigure{\includegraphics[width=8.5cm]{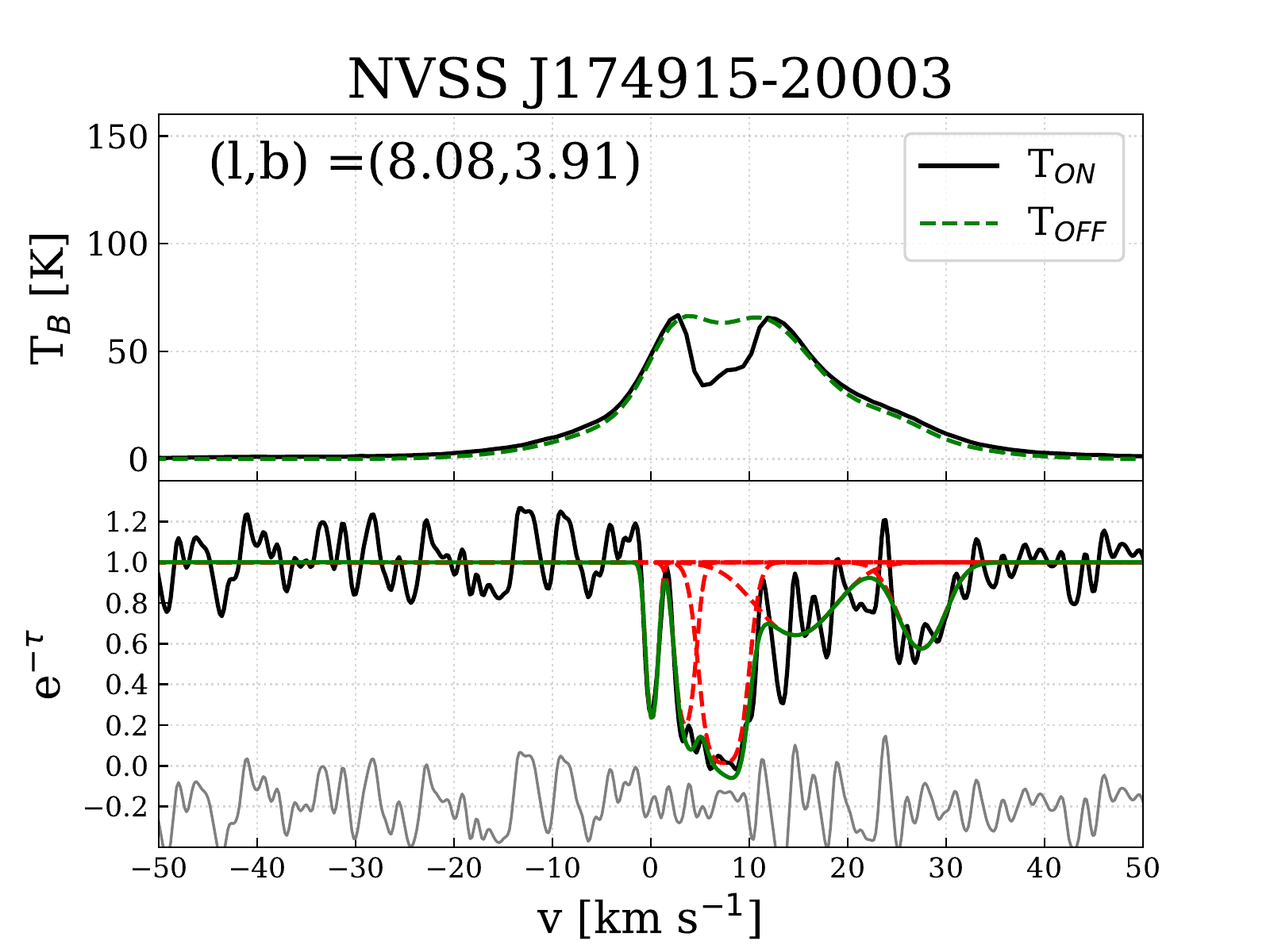}}
\hfill
\subfigure{\includegraphics[width=8.5cm]{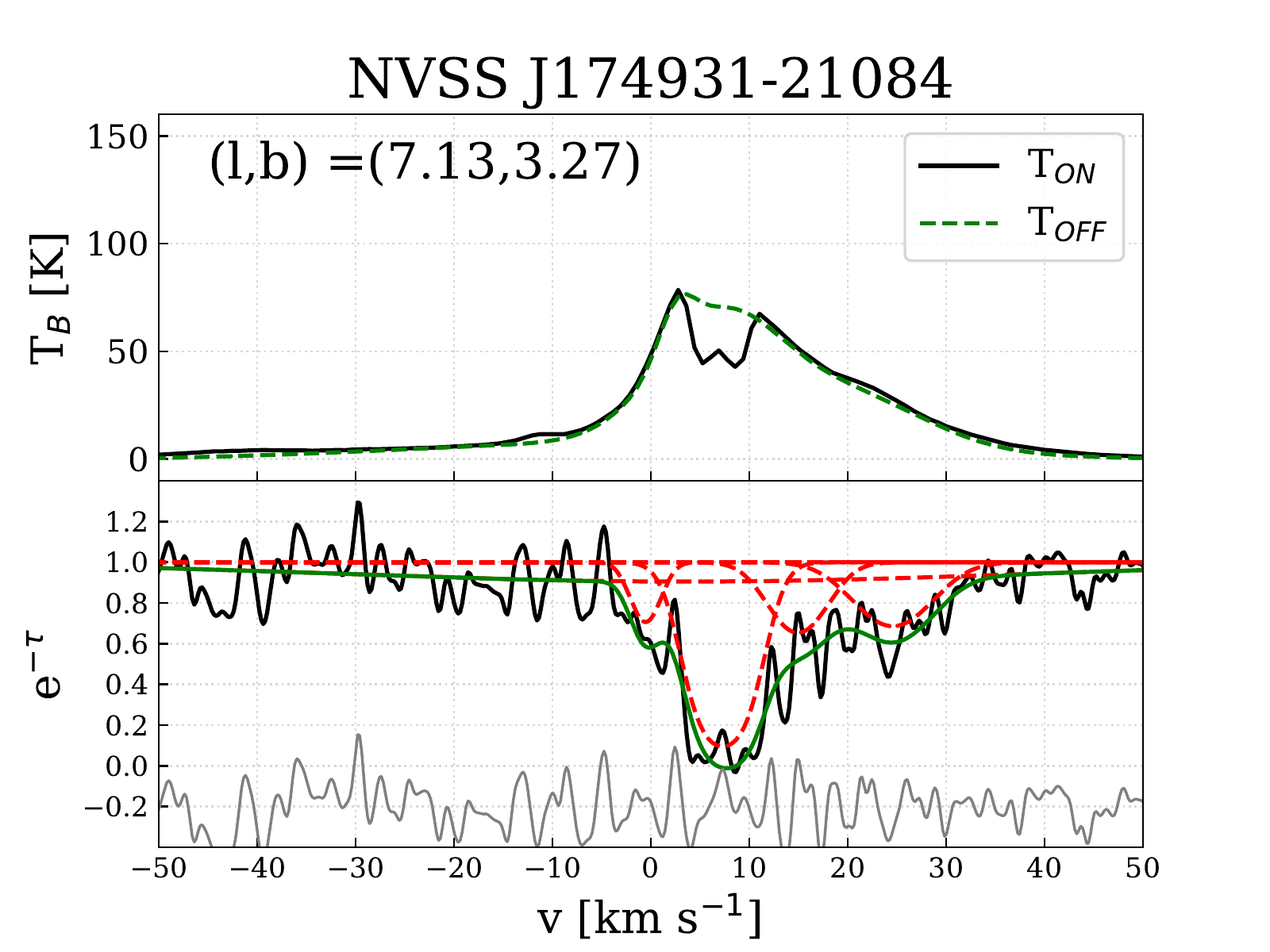}}
\hfill
\caption{Same as \ref{fig:spectra1}}
\label{fig:spectra6}
\end{figure*} 

\begin{figure*}
\subfigure{\includegraphics[width=8.5cm]{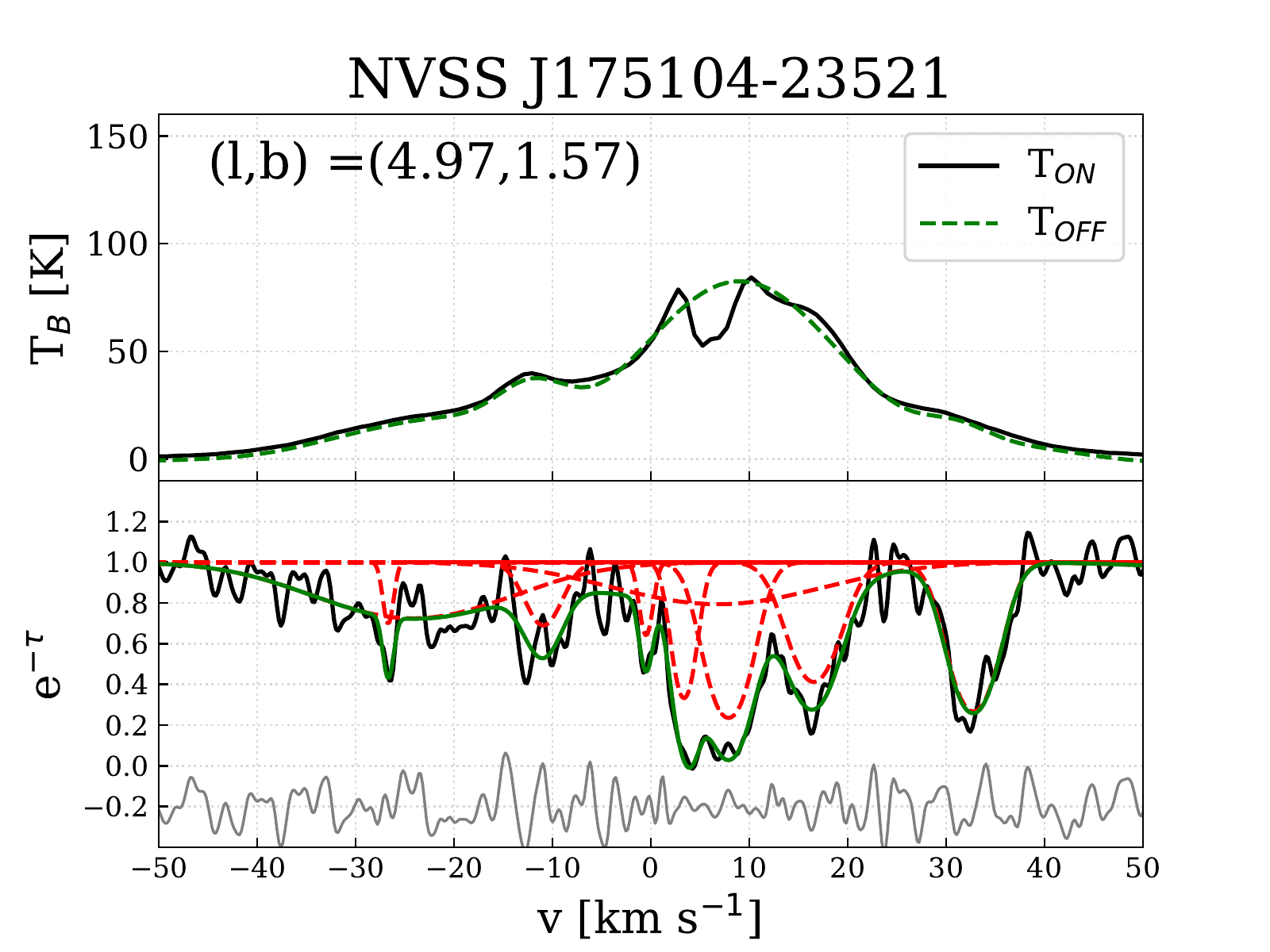}}
\hfill
\subfigure{\includegraphics[width=8.5cm]{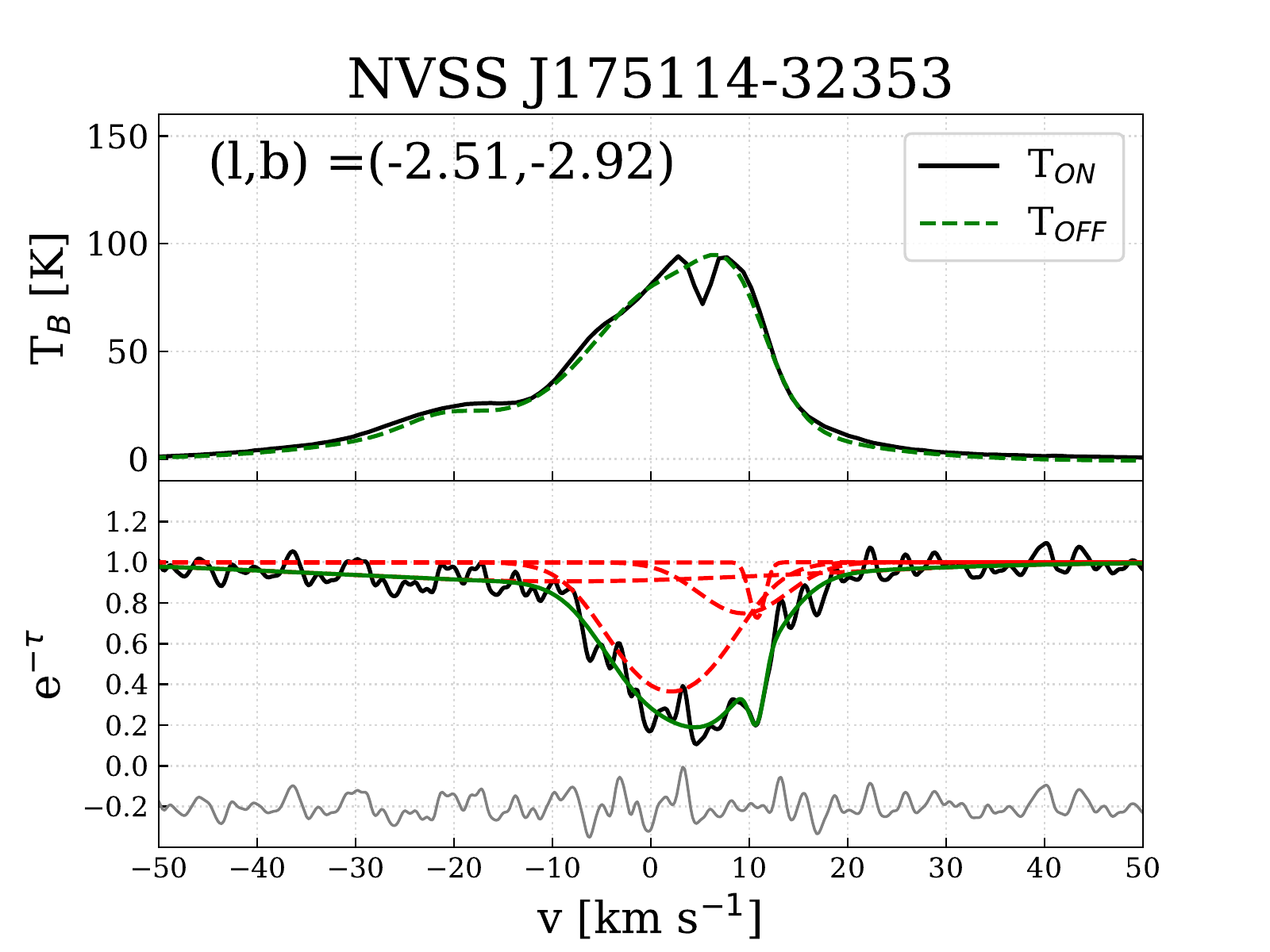}}
\hfill
\subfigure{\includegraphics[width=8.5cm]{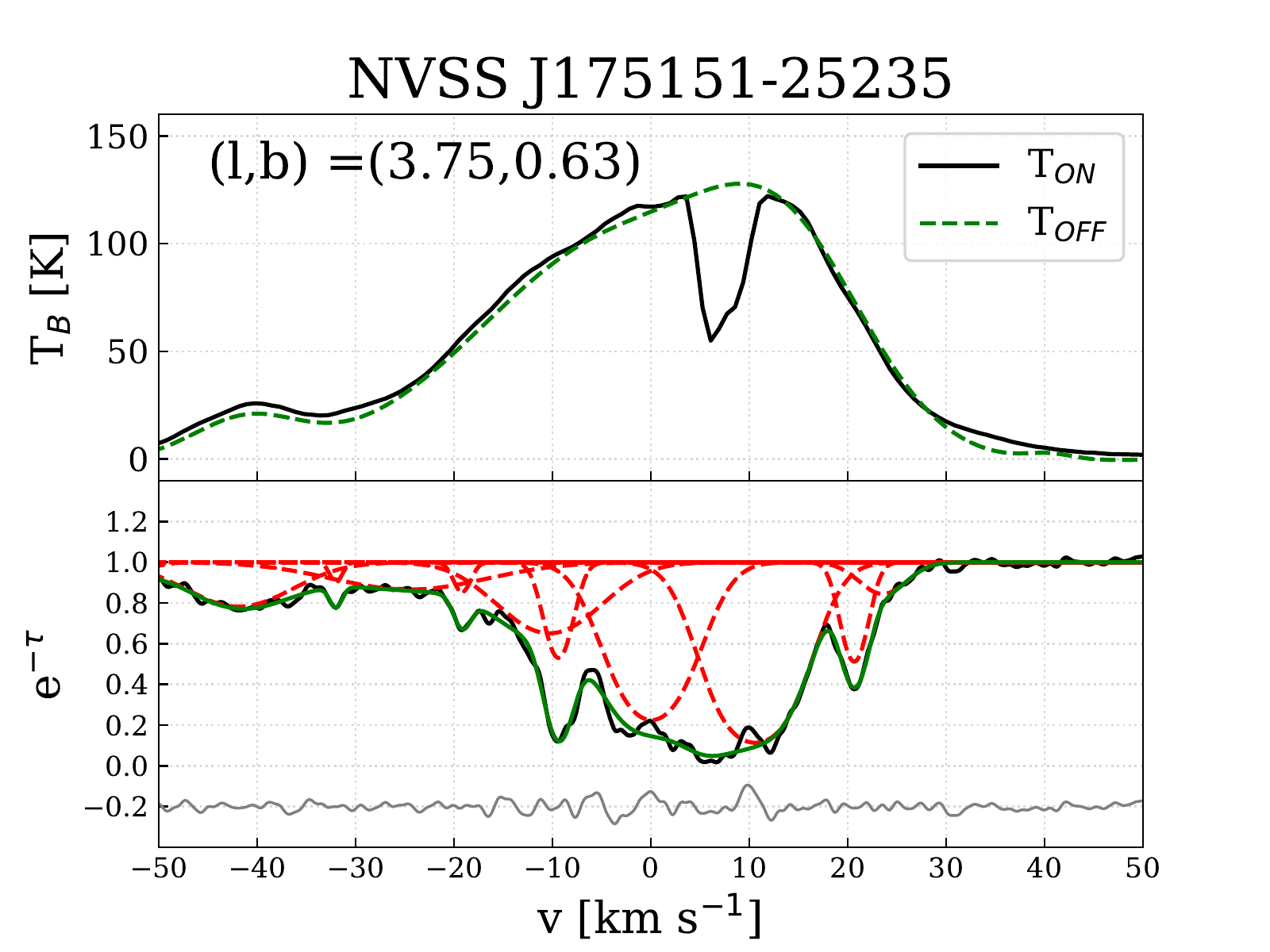}}
\hfill
\subfigure{\includegraphics[width=8.5cm]{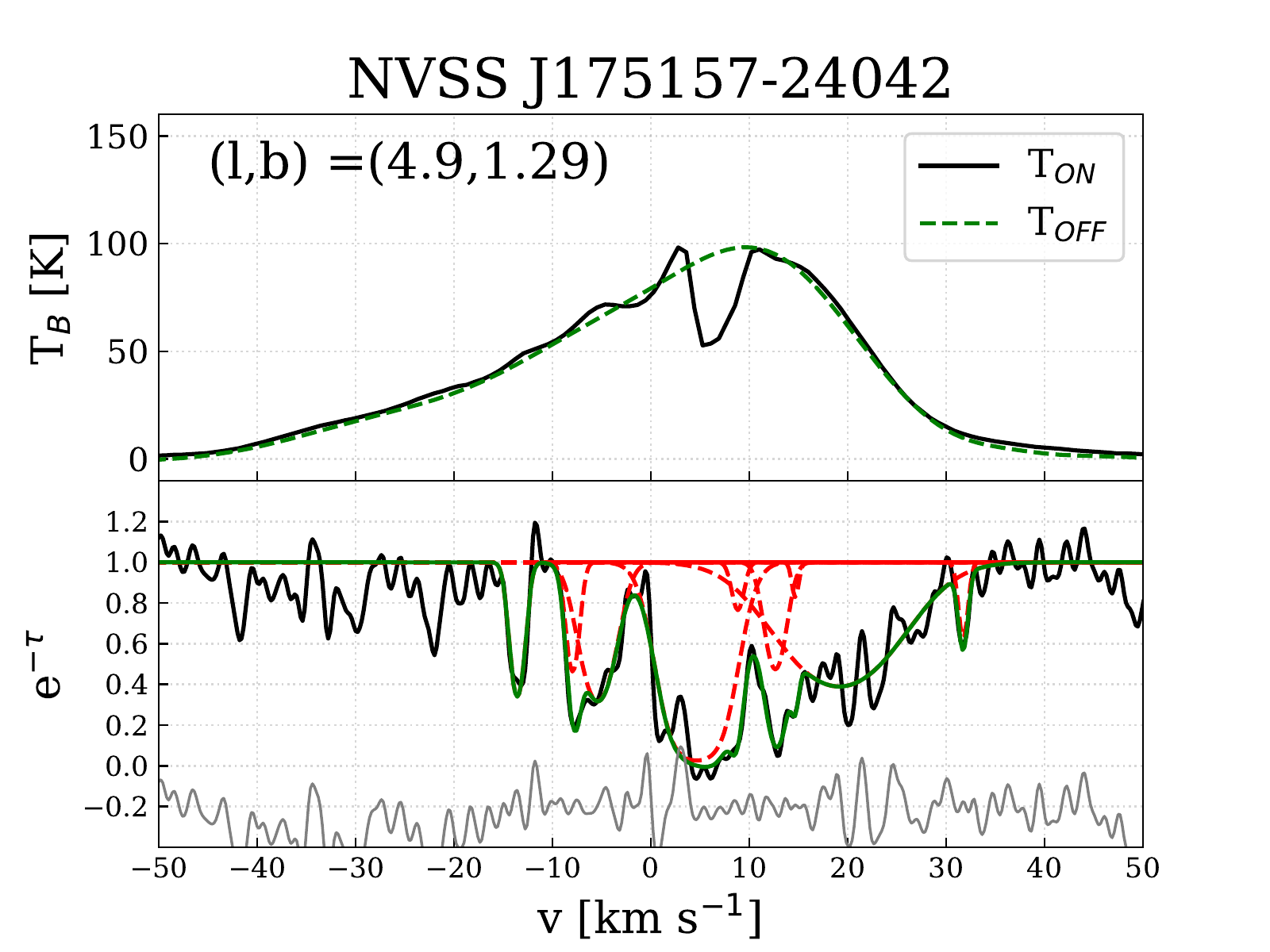}}
\hfill
\subfigure{\includegraphics[width=8.5cm]{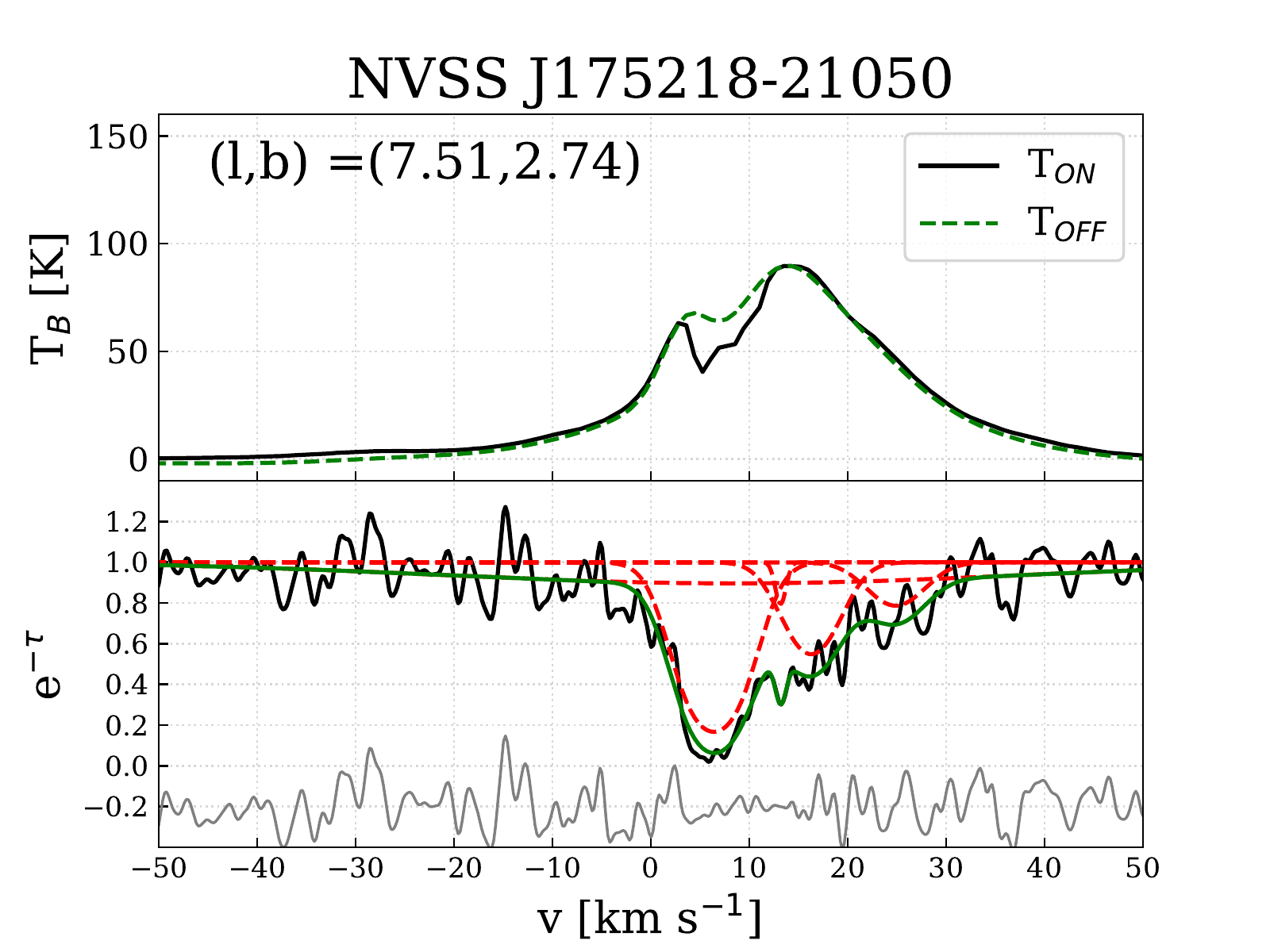}}
\hfill
\subfigure{\includegraphics[width=8.5cm]{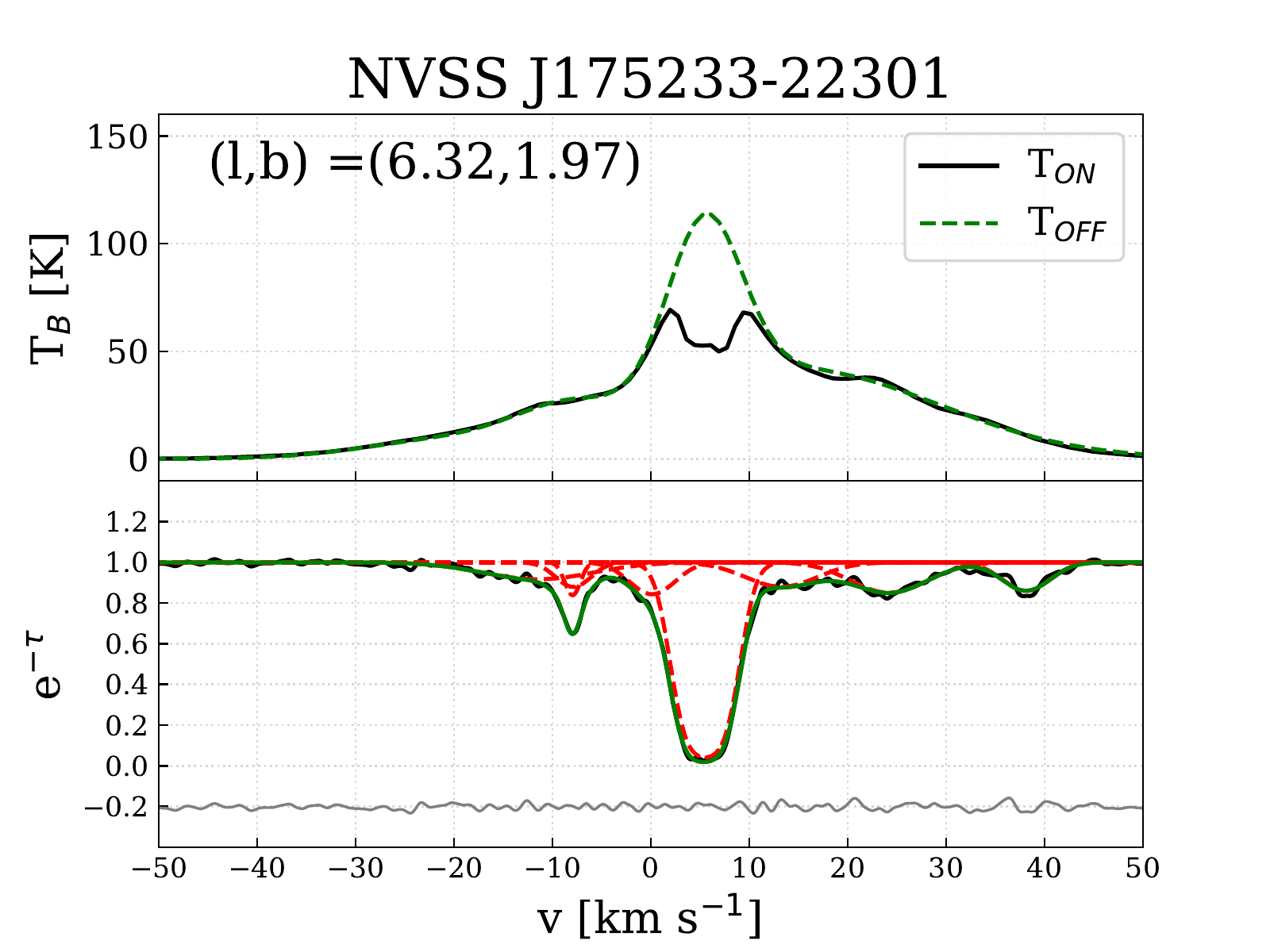}}
\hfill
\caption{Same as \ref{fig:spectra1}}
\label{fig:spectra7}
\end{figure*} 

\begin{figure*}
\subfigure{\includegraphics[width=8.5cm]{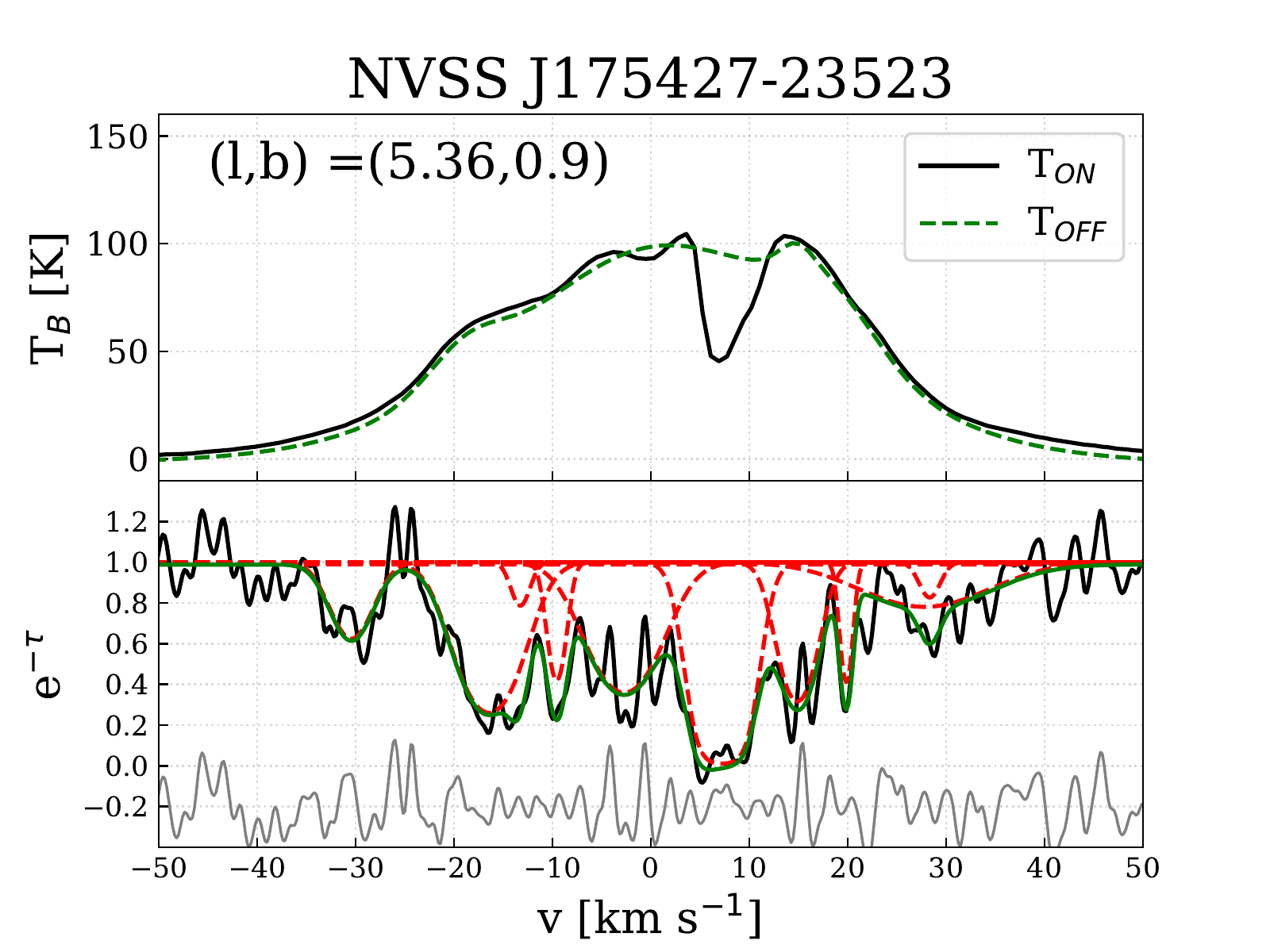}}
\hfill
\subfigure{\includegraphics[width=8.5cm]{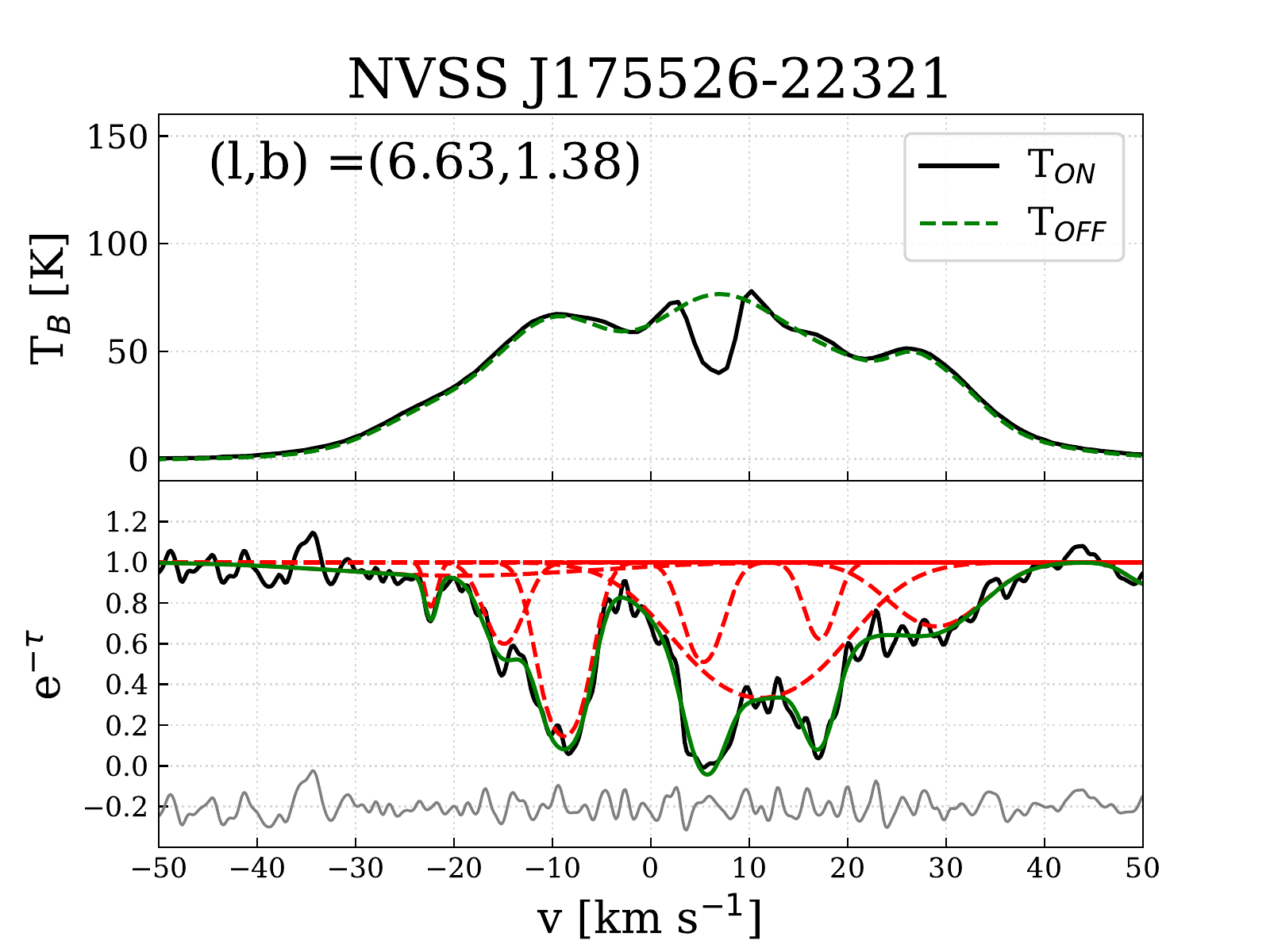}}
\hfill
\subfigure{\includegraphics[width=8.5cm]{175548-23332_spectra_paper.pdf}}
\hfill
\subfigure{\includegraphics[width=8.5cm]{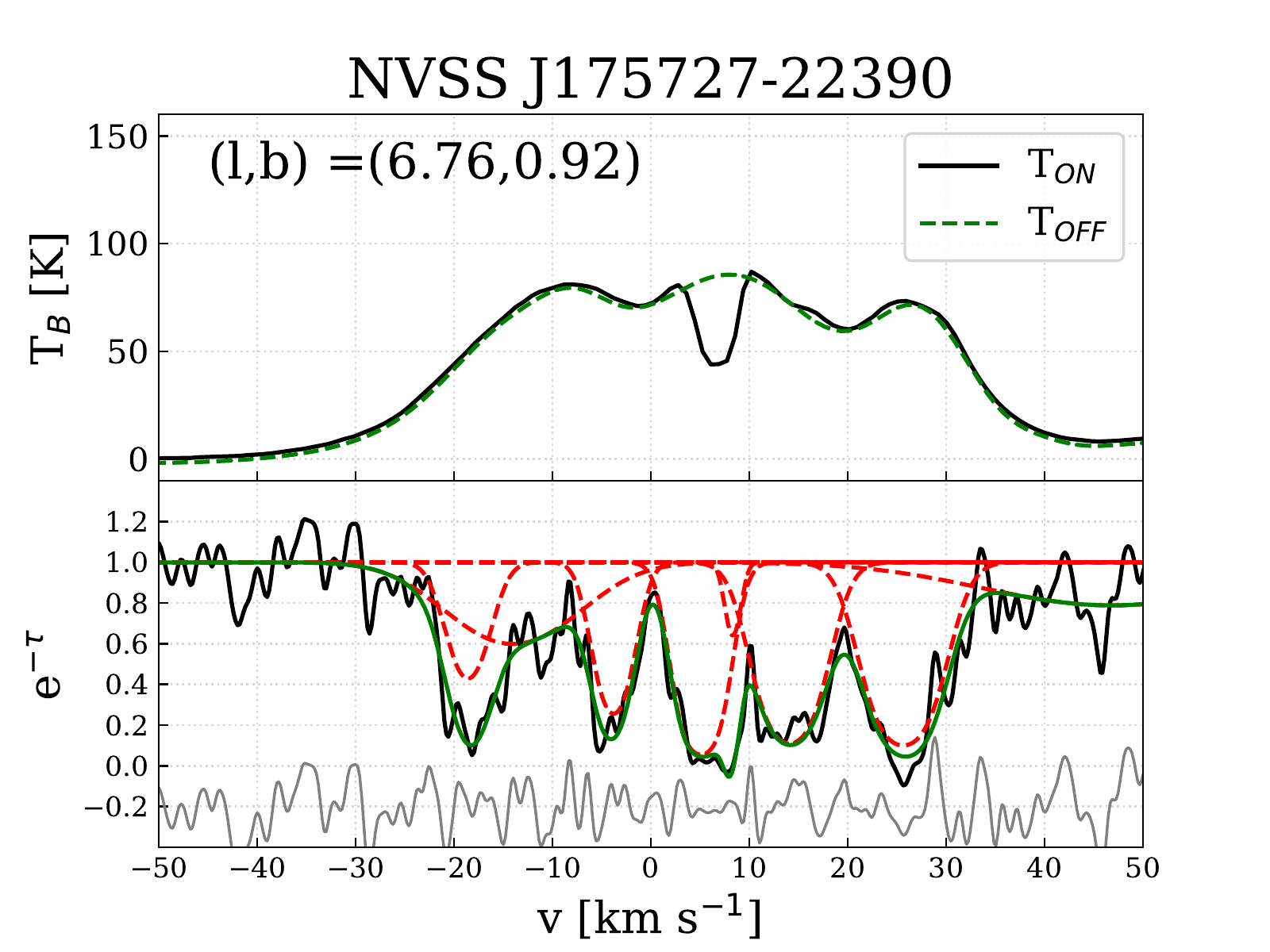}}
\hfill
	\caption{Same as \ref{fig:spectra1}}
	\label{fig:spectra8}
\end{figure*}

\section{Fitted OH spectra}

\begin{figure*}
	\subfigure{\includegraphics[width=8.5cm]{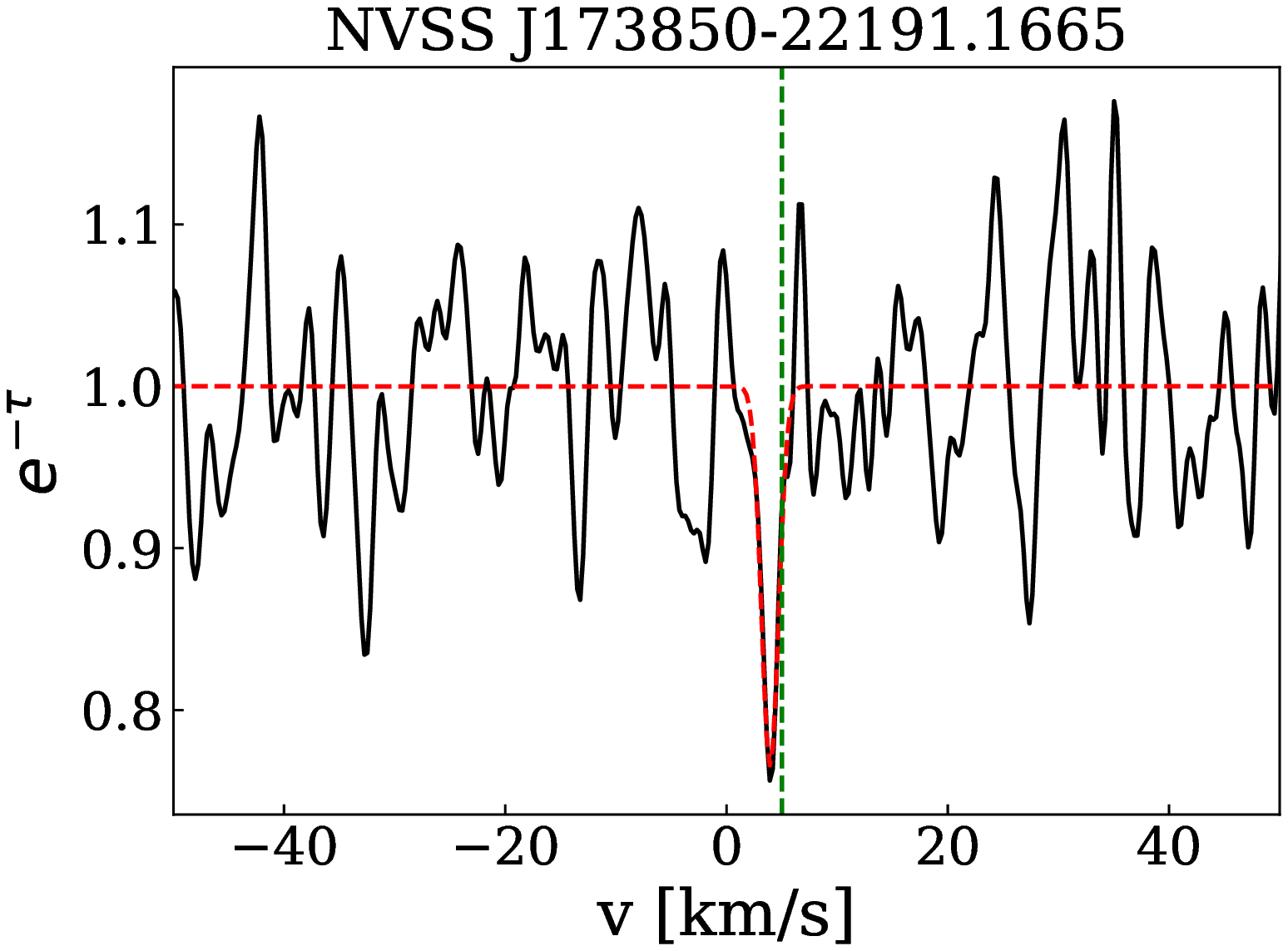}} 
	\hfill
	\subfigure{\includegraphics[width=8.5cm]{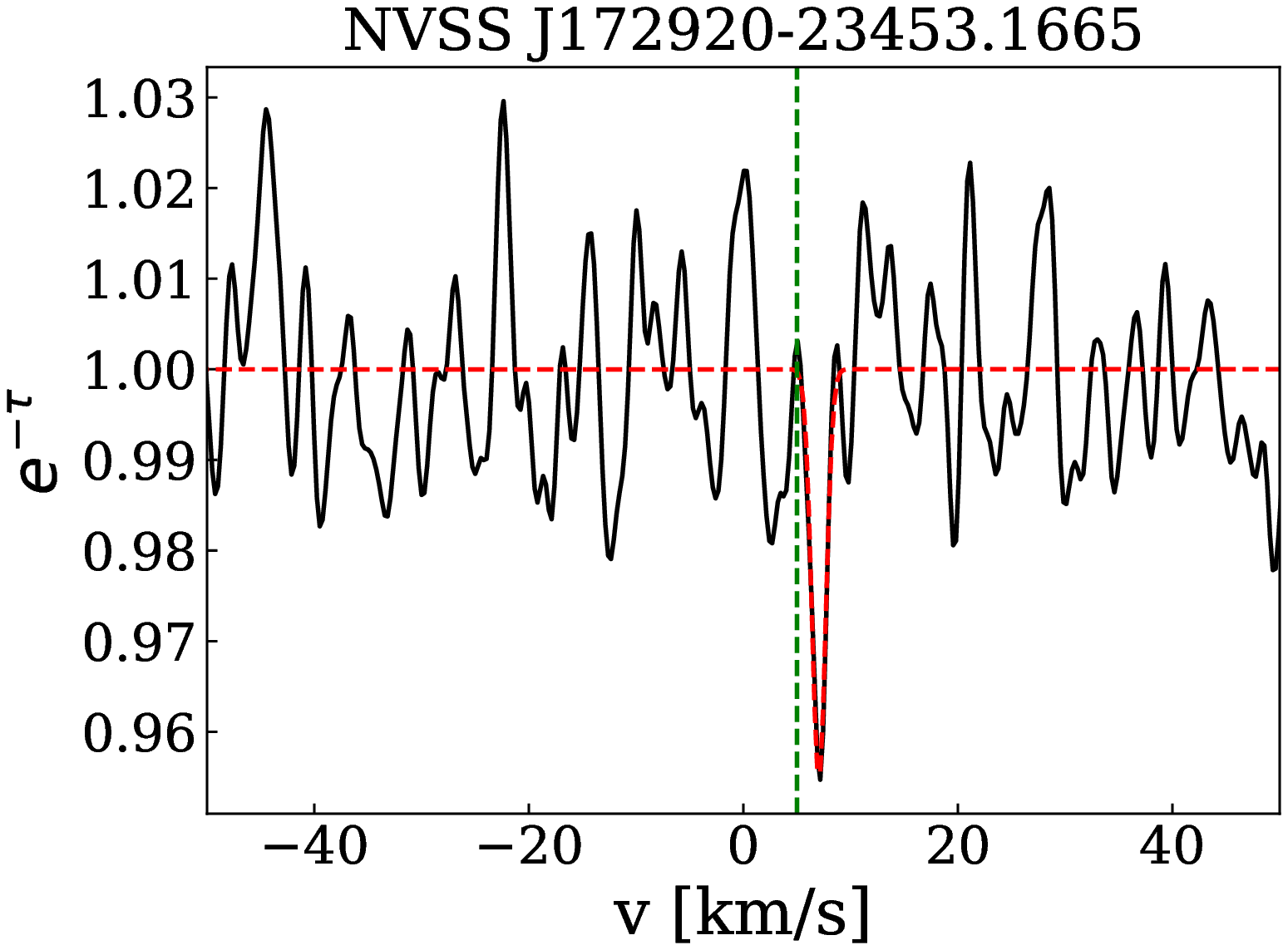}} 
	\hfill
	\caption{1665 MHz OH absorption spectra. The red dashed line shows the fitted Gaussian profile to the line. The green dashed line is $v_{LSR}=5$ \kms.}
	\label{fig:OH_spectra}
\end{figure*}

\begin{figure*}
	\subfigure{\includegraphics[width=8.5cm]{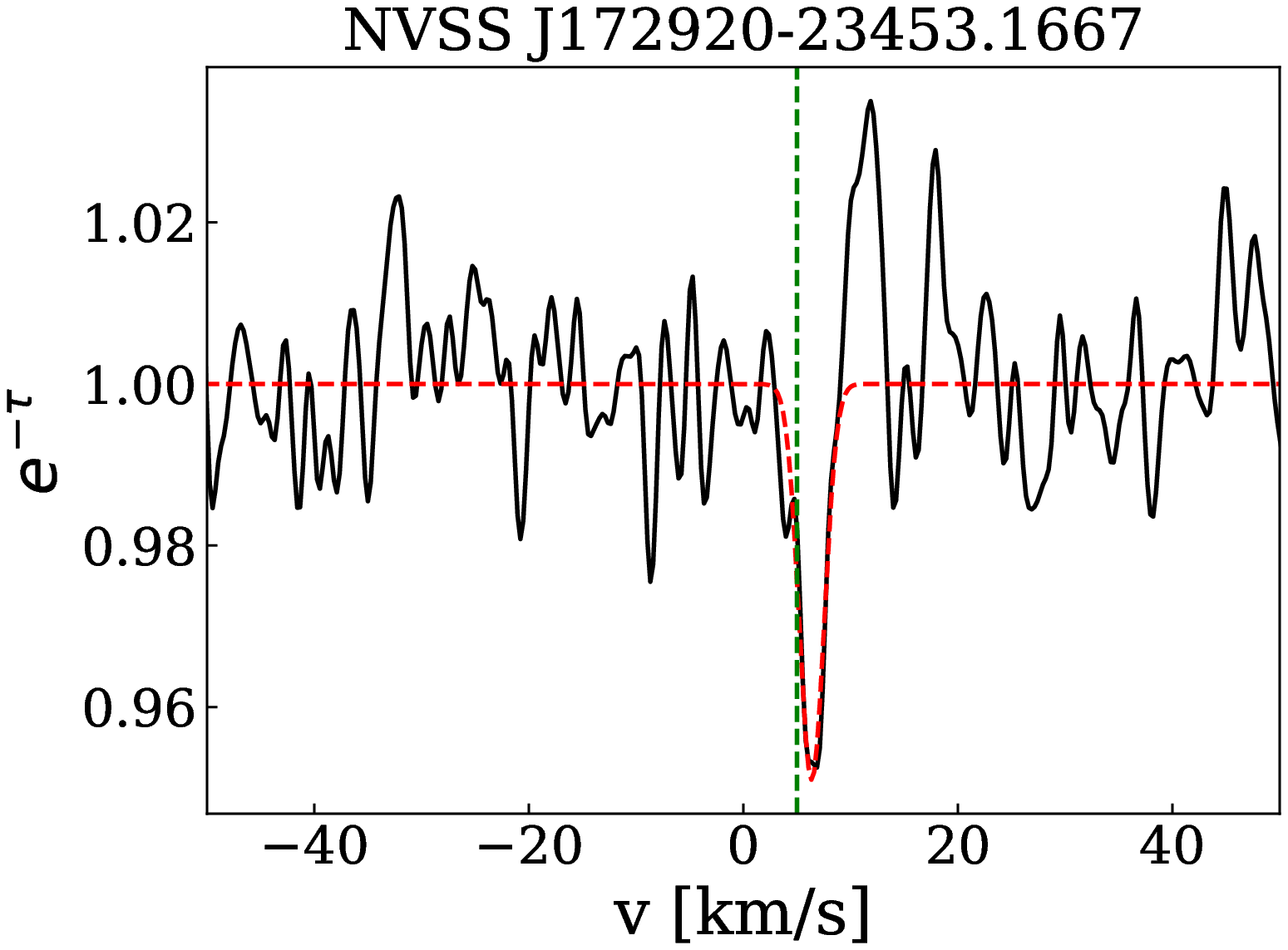}} 
	\hfill
	\subfigure{\includegraphics[width=8.5cm]{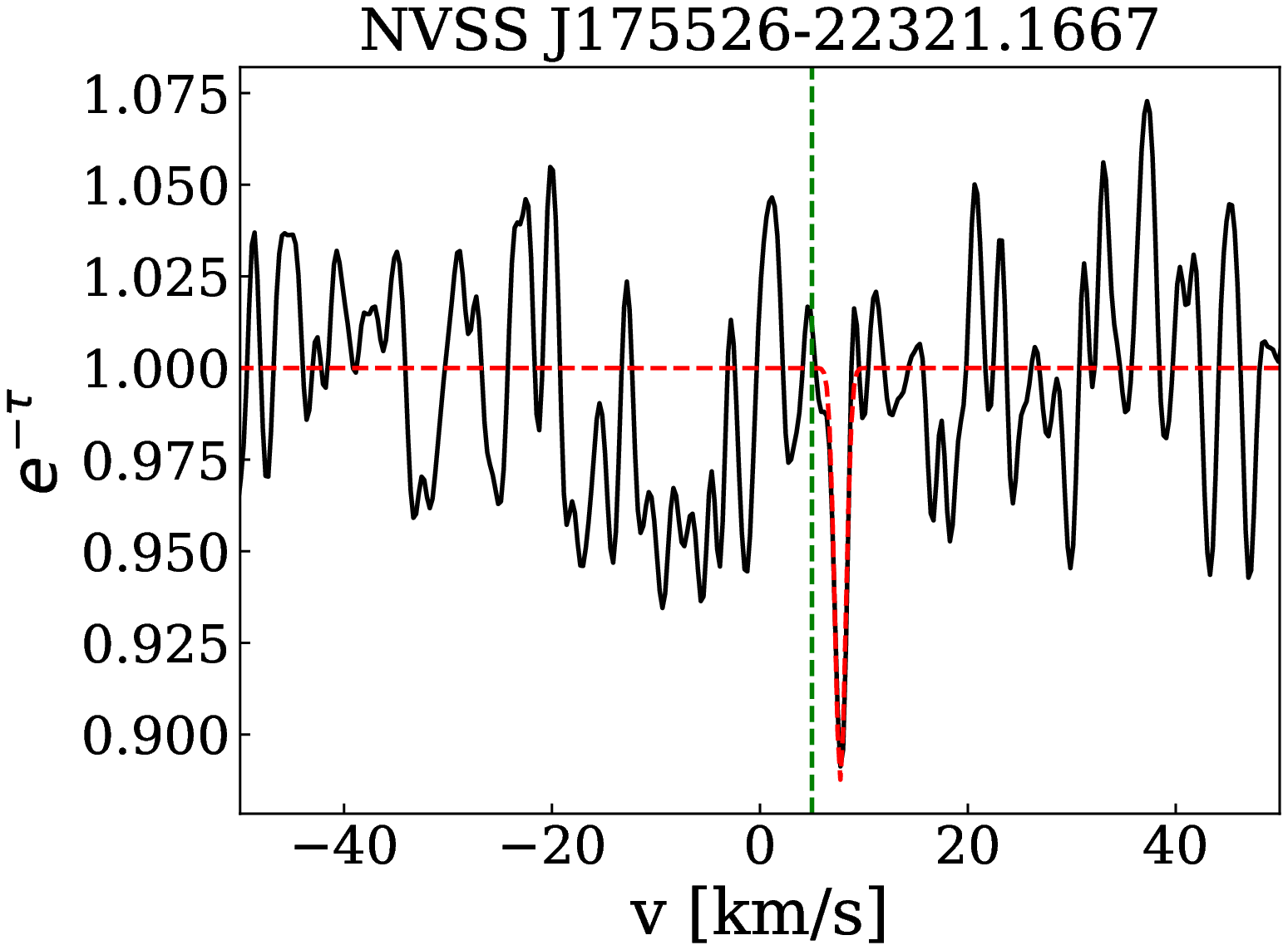}} 
	\hfill
	\caption{1667 MHz OH absorption spectra. The red dashed line shows the fitted Gaussian profile to the line. The green dashed line is $v_{LSR}=5$ \kms.}
	\label{fig:OH_spectra2}
\end{figure*}


\label{lastpage}
\end{document}